\documentclass[useAMS,usenatbib]{mn2e}
\usepackage{graphicx}
\usepackage{amssymb, amsmath}
\usepackage{times}
\usepackage{color}
\usepackage{lscape}
\usepackage[section]{placeins}

\usepackage{lineno}

\newcommand{\hompc}{\,h\,{\rm Mpc}^{-1}}
\newcommand{\mpcoh}{\,h^{-1}\,{\rm Mpc}}

\newcommand\nodata{ ~$\cdots$~ }%

\newcommand{\comment}[1]{\relax}		    

\begin{document}

\title[BAO in SDSS-III BOSS galaxies] {The clustering of
  galaxies in the SDSS-III Baryon Oscillation Spectroscopic Survey:
  Baryon Acoustic Oscillations in the Data Release 10 and 11 Galaxy Samples}

\author[L. Anderson et al.]{\parbox{\textwidth}{\Large
Lauren Anderson$^{1}$,
\'Eric Aubourg$^{2}$,
Stephen Bailey$^{3}$,
Florian Beutler$^{3}$,
Vaishali Bhardwaj$^{1,3}$,
Michael Blanton$^{4}$,
Adam S. Bolton$^{5}$,
J. Brinkmann$^{6}$,
Joel R. Brownstein$^{5}$,
Angela Burden$^{7}$,
Chia-Hsun Chuang$^{8}$,
Antonio J. Cuesta$^{9,10}$,
Kyle S. Dawson$^{5}$,
Daniel J. Eisenstein$^{11}$,
Stephanie Escoffier$^{12}$,
James E. Gunn$^{13}$,
Hong Guo$^{5}$,
Shirley Ho$^{14}$,
Klaus Honscheid$^{15,16}$,
Cullan Howlett$^{7}$,
David Kirkby$^{17}$,
Robert H. Lupton$^{14}$,
Marc Manera$^{7,18}$,
Claudia Maraston$^{7}$,
Cameron K. McBride$^{11}$,
Olga Mena$^{19}$,
Francesco Montesano$^{20}$,
Robert C. Nichol$^{7}$,
Sebasti\'an E. Nuza$^{21}$,
Matthew D. Olmstead$^{5}$,
Nikhil Padmanabhan$^{9}$,
Nathalie Palanque-Delabrouille$^{3,22}$,
John Parejko$^{9}$,
Will J. Percival$^{7}$,
Patrick Petitjean$^{23}$,
Francisco Prada$^{8,24,25}$,
Adrian M. Price-Whelan$^{26}$,
Beth Reid$^{3,27,28}$,
Natalie A. Roe$^{3}$,
Ashley J. Ross$^{7}$,
Nicholas P. Ross$^{3,29}$,
Cristiano G. Sabiu$^{30}$,
Shun Saito$^{31}$,
Lado Samushia$^{7,32}$,
Ariel G. S\'anchez$^{20}$,
David J. Schlegel\thanks{BOSS PI: djschlegel@lbl.gov}$^{3}$,
Donald P. Schneider$^{33,34}$,
Claudia G. Scoccola$^{8,35,36}$,
Hee-Jong Seo$^{16,37}$,
Ramin A. Skibba$^{38}$,
Michael A. Strauss$^{13}$,
Molly E. C. Swanson$^{11}$,
Daniel Thomas$^{7}$,
Jeremy L. Tinker$^{4}$,
Rita Tojeiro$^{7}$,
Mariana Vargas Maga\~na$^{2}$,
Licia Verde$^{39,36}$,
David A. Wake$^{40,41}$,
Benjamin A. Weaver$^{4}$,
David H. Weinberg$^{16,42}$,
Martin White$^{3,28,43}$,
Xiaoying Xu$^{14}$,
Christophe Y\`eche$^{22}$,
Idit Zehavi$^{44}$,
Gong-Bo Zhao$^{7,45}$
 } \vspace*{4pt} \\ 
\scriptsize $^{1}$ Department of Astronomy, University of Washington, Box 351580, Seattle, WA 98195, USA\vspace*{-2pt} \\ 
\scriptsize $^{2}$ APC, Astroparticule et Cosmologie, Universit\'e Paris Diderot, CNRS/IN2P3, CEA/Irfu, Observatoire de Paris, Sorbonne Paris Cit\'e, 10, rue Alice Domon \& L\'eonie Duquet, 75205 Paris Cedex 13, France\vspace*{-2pt} \\ 
\scriptsize $^{3}$ Lawrence Berkeley National Laboratory, 1 Cyclotron Road, Berkeley, CA 94720, USA\vspace*{-2pt} \\ 
\scriptsize $^{4}$ Center for Cosmology and Particle Physics, New York University, New York, NY 10003, USA\vspace*{-2pt} \\ 
\scriptsize $^{5}$ Department Physics and Astronomy, University of Utah, UT 84112, USA\vspace*{-2pt} \\ 
\scriptsize $^{6}$ Apache Point Observatory, P.O. Box 59, Sunspot, NM 88349-0059, USA\vspace*{-2pt} \\ 
\scriptsize $^{7}$ Institute of Cosmology \& Gravitation, Dennis Sciama Building, University of Portsmouth, Portsmouth, PO1 3FX, UK\vspace*{-2pt} \\ 
\scriptsize $^{8}$ Instituto de Fisica Teorica (UAM/CSIC), Universidad Autonoma de Madrid, Cantoblanco, E-28049 Madrid, Spain\vspace*{-2pt} \\ 
\scriptsize $^{9}$ Department of Physics, Yale University, 260 Whitney Ave, New Haven, CT 06520, USA\vspace*{-2pt} \\ 
\scriptsize $^{10}$ Institut de Ci\`encies del Cosmos, Universitat de Barcelona, IEEC-UB, Mart\'\i~i Franqu\`es 1, E08028 Barcelona, Spain\vspace*{-2pt} \\ 
\scriptsize $^{11}$ Harvard-Smithsonian Center for Astrophysics, 60 Garden St., Cambridge, MA 02138, USA\vspace*{-2pt} \\ 
\scriptsize $^{12}$ CPPM, Aix-Marseille Universit\'e, CNRS/IN2P3, Marseille, France\vspace*{-2pt} \\ 
\scriptsize $^{13}$ Department of Astrophysical Sciences, Princeton University, Ivy Lane, Princeton, NJ 08544, USA\vspace*{-2pt} \\ 
\scriptsize $^{14}$ Department of Physics, Carnegie Mellon University, 5000 Forbes Avenue, Pittsburgh, PA 15213, USA\vspace*{-2pt} \\ 
\scriptsize $^{15}$ Department of Physics, Ohio State University, Columbus, Ohio 43210, USA\vspace*{-2pt} \\ 
\scriptsize $^{16}$ Center for Cosmology and Astro-Particle Physics, Ohio State University, Columbus, Ohio, USA\vspace*{-2pt} \\ 
\scriptsize $^{17}$ Department of Physics and Astronomy, UC Irvine, 4129 Frederick Reines Hall, Irvine, CA 92697, USA\vspace*{-2pt} \\ 
\scriptsize $^{18}$ University College London, Gower Street, London WC1E 6BT, UK\vspace*{-2pt} \\ 
\scriptsize $^{19}$ IFIC, Universidad de Valencia-CSIC, 46071, Spain\vspace*{-2pt} \\ 
\scriptsize $^{20}$ Max-Planck-Institut f\"ur extraterrestrische Physik, Postfach 1312, Giessenbachstr., 85748 Garching, Germany\vspace*{-2pt} \\ 
\scriptsize $^{21}$ Leibniz-Institut f\"{u}r Astrophysik Potsdam (AIP), An der Sternwarte 16, 14482 Potsdam, Germany\vspace*{-2pt} \\ 
\scriptsize $^{22}$ CEA, Centre de Saclay, IRFU, 91191 Gif-sur-Yvette, France\vspace*{-2pt} \\ 
\scriptsize $^{23}$ Universit\'e Paris 6, Institut d'Astrophysique de Paris, UMR7095-CNRS, 98bis Boulevard Arago, 75014 Paris, France\vspace*{-2pt} \\ 
\scriptsize $^{24}$ Campus of International Excellence UAM+CSIC, Cantoblanco, E-28049 Madrid, Spain\vspace*{-2pt} \\ 
\scriptsize $^{25}$ Instituto de Astrof\'isica de Andaluc\'ia (CSIC), E-18080 Granada, Spain\vspace*{-2pt} \\ 
\scriptsize $^{26}$ Department of Astronomy, Columbia University, New York, NY, 10027, USA\vspace*{-2pt} \\ 
\scriptsize $^{27}$ Hubble Fellow\vspace*{-2pt} \\ 
\scriptsize $^{28}$ Department of Physics, University of California, 366 LeConte Hall, Berkeley, CA 94720, USA\vspace*{-2pt} \\ 
\scriptsize $^{29}$ Department of Physics, Drexel University, 3141 Chestnut Street, Philadelphia, PA 19104, USA \vspace*{-2pt} \\ 
\scriptsize $^{30}$ Korea Institute for Advanced Study, Dongdaemun-gu, Seoul 130-722, Korea\vspace*{-2pt} \\ 
\scriptsize $^{31}$ Kavli Institute for the Physics and Mathematics of the Universe (WPI), Todai Institues for Advanced Study, The University of Tokyo, Chiba 277-8582, Japan\vspace*{-2pt} \\ 
\scriptsize $^{32}$ National Abastumani Astrophysical Observatory, Ilia State University, 2A Kazbegi Ave., GE-1060 Tbilisi, Georgia\vspace*{-2pt} \\ 
\scriptsize $^{33}$ Department of Astronomy and Astrophysics, The Pennsylvania State University, University Park, PA 16802, USA\vspace*{-2pt} \\ 
\scriptsize $^{34}$ Institute for Gravitation and the Cosmos, The Pennsylvania State University, University Park, PA 16802, USA\vspace*{-2pt} \\ 
\scriptsize $^{35}$ Instituto de Astrof{\'\i}sica de Canarias (IAC), C/V{\'\i}a L\'actea, s/n, E-38200, La Laguna, Tenerife, Spain\vspace*{-2pt} \\ 
\scriptsize $^{36}$ Departamento de F\'isica Te\'orica, Universidad Aut\'onoma de Madrid, E-28049 Cantoblanco, Madrid, Spain\vspace*{-2pt} \\ 
\scriptsize $^{37}$ Berkeley Center for Cosmological Physics, LBL and Department of Physics, University of California, Berkeley, CA 94720, USA\vspace*{-2pt} \\ 
\scriptsize $^{38}$ Center for Astrophysics and Space Sciences, Department of Physics, University of California, 9500 Gilman Dr., San Diego, CA 92093 USA\vspace*{-2pt} \\ 
\scriptsize $^{39}$ ICREA \& ICC-UB University of Barcelona, Marti i Franques 1, 08028 Barcelona, Spain\vspace*{-2pt} \\ 
\scriptsize $^{40}$ Department of Astronomy, University of Wisconsin-Madison, 475 N. Charter Street, Madison, WI, 53706, USA\vspace*{-2pt} \\ 
\scriptsize $^{41}$ Department of Physical Sciences, The Open University, Milton Keynes, MK7 6AA, UK\vspace*{-2pt} \\ 
\scriptsize $^{42}$ Department of Astronomy, Ohio State University, Columbus, Ohio, USA\vspace*{-2pt} \\ 
\scriptsize $^{43}$ Department of Astronomy, University of California at Berkeley, Berkeley, CA 94720, USA\vspace*{-2pt} \\ 
\scriptsize $^{44}$ Department of Astronomy, Case Western Reserve University, Cleveland, Ohio 44106, USA\vspace*{-2pt} \\ 
\scriptsize $^{45}$ National Astronomy Observatories, Chinese Academy of Science, Beijing, 100012, P.R. China\vspace*{-2pt} \\ 
\scriptsize $^{46}$ Institute of theoretical astrophysics, University of Oslo, 0315 Oslo, Norway\vspace*{-2pt} \\ 
}

\date{\today} 
\pagerange{\pageref{firstpage}--\pageref{lastpage}} \pubyear{2014}
\maketitle
\label{firstpage}

\begin{abstract}
We present a one per cent measurement of the cosmic distance scale from
the detections of the baryon acoustic oscillations in the clustering of
galaxies from the Baryon Oscillation Spectroscopic Survey (BOSS), 
which is part of the Sloan Digital Sky Survey III (SDSS-III). 
Our results come from the Data Release 11 (DR11) sample, 
containing nearly one million galaxies and covering approximately
$8\,500$ square degrees and the redshift range $0.2<z<0.7$.
We also compare these results with those from the publicly released DR9
and DR10 samples.  Assuming a concordance $\Lambda$CDM cosmological
model, the DR11 sample covers a volume of 13\,Gpc${}^3$ and is the largest
region of the Universe ever surveyed at this density.
We measure the correlation function and power spectrum, including
density-field reconstruction of the baryon acoustic oscillation (BAO) feature.
The acoustic features are detected at a significance of over
$7\,\sigma$ in both the correlation function and power spectrum.
Fitting for the position of the acoustic features measures the distance
relative to the sound horizon at the drag epoch, $r_d$, which 
has a value of $r_{d,{\rm fid}}=149.28\,$Mpc in our fiducial cosmology.
We find
$D_V=(1264\pm25\,{\rm Mpc})(r_d/r_{d,{\rm fid}})$ at $z=0.32$ and
$D_V=(2056\pm20\,{\rm Mpc})(r_d/r_{d,{\rm fid}})$ at $z=0.57$.
At 1.0 per cent, this latter measure is the most precise distance constraint
ever obtained from a galaxy survey.
Separating the clustering along and transverse to the line-of-sight
yields measurements at $z=0.57$ of
$D_A=(1421\pm20\,{\rm Mpc})(r_d/r_{d,{\rm fid}})$
and 
$H=(96.8\pm3.4\,{\rm km/s/Mpc})(r_{d,{\rm fid}}/r_d)$.
Our measurements of the distance scale are in good agreement with previous
BAO measurements and with the predictions from cosmic microwave background data
for a spatially flat cold dark matter model with a cosmological constant.

\end{abstract}


\section{Introduction}
\label{sec:intro}
Measuring the expansion history of the Universe has been one of the
key goals of observational cosmology since its founding.  To date the
best constraints come from measuring the distance-redshift relation
over as wide a range of redshifts as possible \citep{WeinbergReview}, and
imply that the expansion rate of the Universe has recently
transitioned from a deceleration to an acceleration phase
\citep{riess98, Perl99}. While the flat $\Lambda$CDM model provides a
simple mathematical description of expansion that matches current
observations \citep{PlanckXVI}, it is physically perplexing given the
small vacuum energy density measured, when compared with the high
densities that traditionally correspond to new physics
\citep[see e.g.][for recent reviews]{WeinbergReview,MWW14}. Understanding
the physical cause of the accelerating expansion rate remains one of the
most interesting problems in modern physics.

One of the most robust methods for measuring the distance-redshift
relation is to use the Baryon Acoustic Oscillation (BAO) feature(s) in
the clustering of galaxies as a ``standard ruler''.  The acoustic
oscillations arise from the tight coupling of baryons and photons in
the early Universe: the propagation of sound waves through this medium
gives rise to a characteristic scale in the distribution of
perturbations corresponding to the distance travelled by the wave
before recombination (\citealt{Pee70,Sun70,Dor78}; a description of
the physics leading to the features can be found in \citealt{Eis98} or
Appendix A of \citealt{MeiWhiPea} and a discussion of the acoustic
signal in configuration space can be found in
\citealt{EisSeoWhi07}). This signal is imprinted in the distribution
of both the matter and the radiation. The latter are seen as
anisotropies in the cosmic microwave background (CMB) radiation while
the former are the signal of interest here.  The distance that sound
waves travel before the coupling between baryons and radiation breaks
down, known as the acoustic scale, is quite large, $r_d \approx 150\,$Mpc.
The signal therefore relies on simple, linear, well-understood physics
that can be well calibrated by CMB anisotropy measurements and is quite
insensitive to non-linear or astrophysical processing that typically
occurs on much smaller scales. This makes experiments using the BAO
signal relatively free of systematic errors.

A number of experiments have used the BAO technique to measure the
distance-redshift relationship.  The strongest early detections were with
galaxies at low-redshift \citep{Col05,Eis05,Hut06,Teg06,Per07},
though BAO have now also been detected in the distribution of clusters
\citep{Veropalumbo14}, and at higher redshift using the Lyman $\alpha$ forest
in quasar spectra \citep{Busca13,Slosar13,Kirkby13} and cross-correlation
betwen quasars and the Lyman $\alpha$ forest \citep{Fon13}.
A review of BAO measurements was provided in \citet{And12}, which described
recent experiments
\citep[e.g.~the 6dFGRS, WiggleZ and SDSS;][]{Beutler11,Bla11a,Pad12},
and presented the first set of analyses of the galaxies in Data Release 9
of the Baryon Oscillation Spectroscopic Survey \citep[BOSS][]{Daw12},
part of the Sloan Digital Sky Survey III \citep[SDSS III][]{Eis11}.

In \citet{And12}, we used reconstruction to provide a 1.7 per cent
distance measurement from the BOSS DR9 galaxies, the most precise
measurement ever obtained from a galaxy survey. This measurement
benefitted from a simple reconstruction procedure, that used the
phase information within the density field to reconstruct linear
behaviour and sharpen the BAO \citep{Eis07a}. In \citet{And13} we
fitted moments of the anisotropic correlation function measured from
the same data, providing distance constraints split into radial and
anisotropic directions. We now extend and update the BAO measurements
based on the BOSS galaxy samples to the latest dataset from the
ongoing BOSS.

This paper concentrates on the DR11 data set, comprised of SDSS-III
observations through May 2013, which is scheduled for public release
in December 2014 together with the final SDSS-III data release (DR12).
The DR10 data set, comprised of observations through June 2012,
is already public \citep{Ahn13}.
We provide the DR10 large scale structure samples, including
the masks, weights, and random catalogs needed for clustering analyses,
through the SDSS-III Science Archive Server.
To facilitate community comparisons to our results, in this
paper we also present several of our key analyses for the DR10
subset of our data sample.

Five companion papers present extensions to the methodology, testing,
and data sets beyond those applied previously to the DR9 data:
\begin{enumerate}
\item \citet{Ross13} split the DR10 CMASS sample (see section \ref{sec:data})
  into red and blue galaxies, showing that consistent cosmological
  measurements result from both data sets.
\item \citet{Vargas13} investigates the different possible systematics
  in the anisotropic fitting methodologies, showing that we achieve
  unbiased results with fiducial fitting methodology.
\item \citet{Man13} describes the production of mock catalogues,
  used here to determine errors and test our analysis methods.  
\item \citet{Per13} presents a method to propagate errors in the
  covariance matrices determined from the mocks through to errors on the
  final measurements. 
\item \citet{Tojlowz} presents measurements made at $z=0.32$ from the
  low-redshift ``LOWZ'' BOSS sample of galaxies which we now include in
  our constraints.
\end{enumerate}

We also have produced a series of companion papers presenting
complementary cosmological measurements from the DR10 and DR11 data:
\begin{enumerate}
\item \citet{Beutler13} presents a fit to the CMASS power spectrum monopole
  and quadrupole, measuring Redshift-Space Distortions (RSD).
\item \citet{Sam13} fits the CMASS correlation function monopole
  and quadrupole, measuring Redshift-Space Distortions (RSD) using a
  streaming model.
\item \citet{Chuang13b} fits CMASS correlation function monopole
  and quadrupole using quasi-linear scales (e.g.~above $50\,h^{-1}$Mpc)
  to extract single-probe measurements.
  For the LOWZ sample, they include smaller scales with Finger-of-God modeling.
\item \citet{Sanchez13} fits LOWZ and CMASS correlation function monopole and
  wedges \citep{Kaz12} with a model inspired by renormalised perturbation
  theory.
\end{enumerate}

The layout of this paper is as follows.  We introduce the data and the
catalogue in the next section.  The catalogue construction is similar
to that described in \citet{And12} for DR9, and so we focus primarily
on the differences and improvements in Section~\ref{sec:analysis}.
We present the analysis methods for our isotropic and anisotropic measurements
in Sections~\ref{sec:fit-2pt} and \ref{sec:aniso_fit}, respectively. 
We then present the isotropic results in Section \ref{sec:results_iso}
and the anisotropic results in Section \ref{sec:results_aniso}.
Our systematic error assessment and final distance measurements
are presented in Section~\ref{sec:distance_scale} and these measurements are
placed in a cosmological context in Section~\ref{sec:params}.
We conclude in Section~\ref{sec:discuss}.  

Throughout the paper we
assume a fiducial $\Lambda$CDM+GR, flat cosmological model with
$\Omega_m=0.274$, $h=0.7$, $\Omega_bh^2=0.0224$, $n_s=0.95$ and
$\sigma_8=0.8$, matching that used in \citet{And12,And13}.
Note that this model is different from the current best-fit cosmology;
however these parameters allow us to translate angles and redshifts into
distances and provide a reference against which we measure distances.
The BAO measurement allows us to constrain changes in the distance scale
relative to that predicted by this fiducial model.

\section{The Data}
\label{sec:data}
\subsection{SDSS-III BOSS}

\begin{figure}
  \centering
\includegraphics[width=84mm]{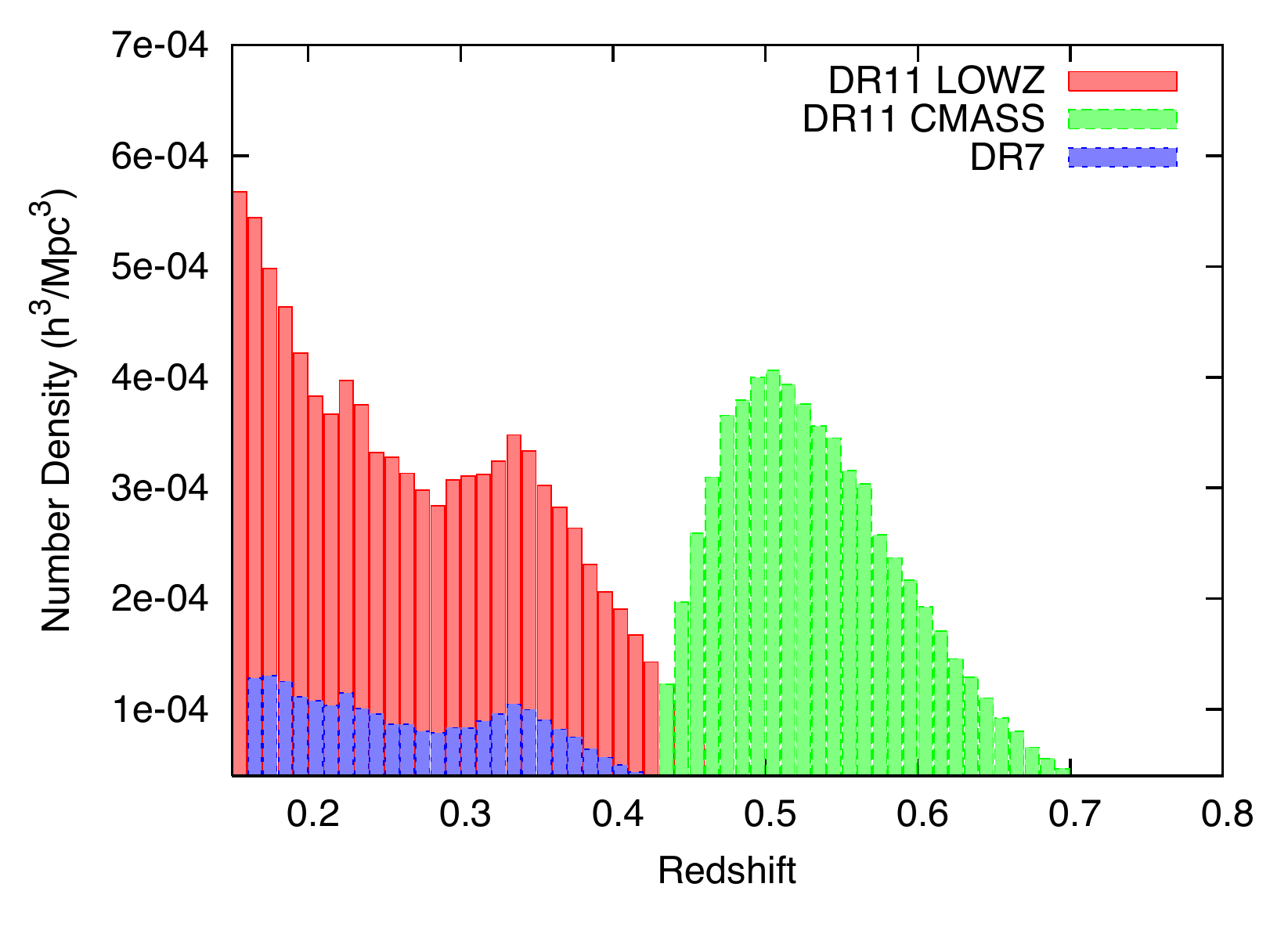}
\caption{Histograms of the galaxy number density as a function of
  redshift for LOWZ (red) and CMASS (green) samples we analyse. We
  also display the number density of the SDSS-II DR7 LRG sample in
  order to illustrate the increase in sample size provided by BOSS LOWZ galaxies.  }
\label{fig:redshift}
\end{figure}

\begin{figure*}
  \centering
  \resizebox{\textwidth}{!}{\includegraphics{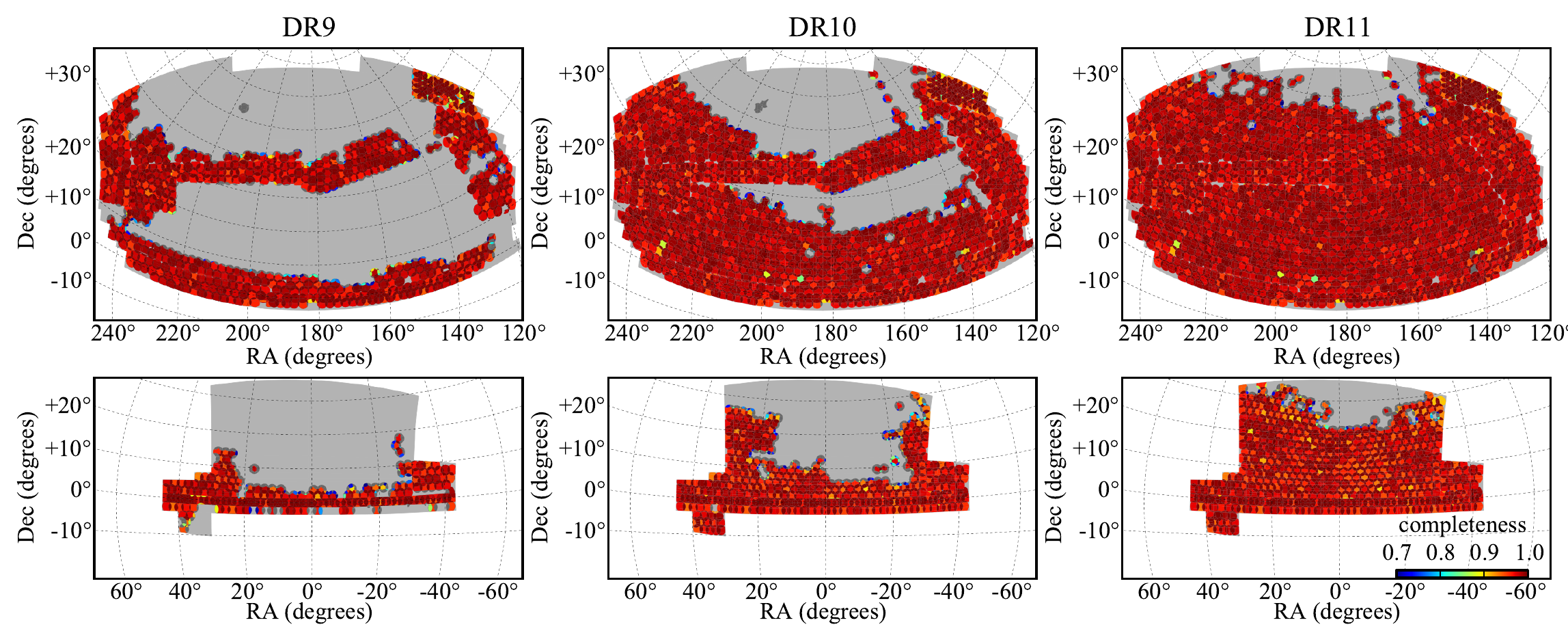}}
  \caption{ Evolution of the BOSS sky coverage from DR9 to DR11. Top
    panels show our observations in the North Galactic Cap (NGC) while lower
    panels show observations in the South Galactic Cap (SGC). Colors
    indicate the spectroscopic completeness within each sector as
    indicated in the key in the lower right panel. Gray areas indicate
    our expected footprint upon completion of the survey.  The total
    sky coverage in DR9, DR10, and DR11 is 3,275 deg$^2$, 6,161
    deg$^2$, and 8,377 deg$^2$, respectively.}
\label{fig:footprints}
\end{figure*}

We use data included in data releases 10 (DR10;\citealt{Ahn13}) and 11 (DR11; to be publicly released with the final BOSS data set)
of the Sloan Digital Sky Survey \citep[SDSS;][]{Yor00}.
Together, SDSS I, II
\citep{Abazajian09}, and III \citep{Eis11} used a drift-scanning
mosaic CCD camera \citep{Gun98} to image over one third of the sky
($14\,555$ square degrees) in five photometric bandpasses
\citep{Fuk96,Smi02,Doi10} to a limiting magnitude of $r\simeq 22.5$
using the dedicated 2.5-m Sloan Telescope \citep{Gun06} located at
Apache Point Observatory in New Mexico. The imaging data were
processed through a series of pipelines that perform astrometric
calibration \citep{Pie03}, photometric reduction \citep{Photo}, and
photometric calibration \citep{Pad08}. All of the imaging was re-processed as part
of SDSS Data Release 8 (DR8; \citealt{DR8}).

BOSS is designed to obtain spectra and redshifts for 1.35 million
galaxies over a footprint covering $10\,000$ square
degrees. These galaxies are selected from the SDSS DR8 imaging and are
being observed together with $160\,000$ quasars and approximately
$100\,000$ ancillary targets.  The targets are assigned to tiles of
diameter $3^\circ$ using a tiling algorithm that is adaptive to the
density of targets on the sky \citep{Tiling}. Spectra are obtained
using the double-armed BOSS spectrographs \citep{Smee13}. Each
observation is performed in a series of $900$-second exposures,
integrating until a minimum signal-to-noise ratio is achieved for the
faint galaxy targets. This ensures a homogeneous data set with a high
redshift completeness of more than $97$ per cent over the full survey
footprint. Redshifts are extracted from the spectra using the methods
described in \cite{Bolton12}. A summary of the survey design appears
in \citet{Eis11}, and a full description, including a discussion of the motivation for the targeting criteria, is provided in \citet{Daw12}.

\subsection{Galaxy Catalogues}

BOSS selects two classes of galaxies to be targeted for spectroscopy
using SDSS DR8 imaging. The `LOWZ' algorithm is designed to select red
galaxies at $z<0.45$ from the SDSS DR8 imaging data via
\begin{eqnarray}
r_{\rm cmod} < 13.5 + c_{\parallel}/0.3 \label{eq:lzslide}\\
|c_{\perp}| < 0.2 \\
16 < r_{\rm cmod} < 19.6 \\
r_{\rm psf}-r_{\rm mod} > 0.3 
\end{eqnarray}
where here $i$ and $r$ indicate magnitudes and all magnitudes are corrected
for Galactic extinction (via the \citealt{SFD98} dust maps), the subscript ${\rm psf}$
denotes PSF magnitudes,
the subscript ${\rm mod}$ denotes `model' magnitudes \citep{Sto02},
the subscript ${\rm cmod}$ denotes `cmodel' magnitudes \citep{DR2}, and
\begin{equation}
c_{\parallel} = 0.7\,(g_{\rm mod}-r_{\rm mod})+
                1.2\,(r_{\rm mod}-i_{\rm mod}-0.18)
\end{equation}
and
\begin{equation}
c_{\perp} = r_{\rm mod}-i_{\rm mod}-(g_{\rm mod}-r_{\rm mod})/4.0 -0.18 .
\end{equation}
The resulting LOWZ galaxy sample has three times the spatial density
of the SDSS-II LRGs, as is shown in Fig.~\ref{fig:redshift}, with a
similar clustering amplitude to the CMASS sample \citep{Par13}.  

We define the effective redshift, $z_{\rm eff}$, as the mean redshift of
a sample weighted by the number of galaxy pairs with separations
$80 < s < 120\,h^{-1}$Mpc.
For the LOWZ sample $z_{\rm eff} =0.32$, slightly lower than that of the
SDSS-II LRGs as we place a redshift cut $z<0.43$ to ensure no overlap
with the CMASS sample, and hence independent measurements.
Further details can be found in \citet{Par13} and \citet{Tojlowz}.
Due to difficulties during the early phases of the project, the sky area
of the LOWZ sample lags that of the full survey by approximately
$1\,000\,{\rm deg}^2$, as can be seen in comparison of Tables
\ref{tab:basic_props_cmass} and \ref{tab:basic_props_lowzz}.

The CMASS sample is designed to be approximately stellar-mass-limited
above $z = 0.45$. These galaxies are selected from the SDSS DR8 imaging via
\begin{eqnarray}
 17.5 < i_{\rm cmod}  < 19.9\\
r_{\rm mod} - i_{\rm mod}  < 2 \\
d_{\perp} > 0.55 \label{eq:hcut}\\
i_{\rm fib2} < 21.5\\
i_{\rm cmod}  < 19.86 + 1.6(d_{\perp} - 0.8) \label{eq:slide}
\end{eqnarray}
where 
\begin{equation}
d_{\perp} = r_{\rm mod} - i_{\rm mod} - (g_{\rm mod} - r_{\rm mod})/8.0,
\label{eq:dp}
\end{equation}
and $i_{fib2}$ is the $i$-band magnitude within a $2^{\prime\prime}$ aperture radius.

\noindent For CMASS targets, stars are further separated from galaxies by only keeping objects with
\begin{eqnarray}
i_{\rm psf} - i_{\rm mod} > 0.2 + 0.2(20.0-i_{\rm mod})  \label{eq:sgsep1}\\
z_{\rm psf}-z_{\rm mod} > 9.125 -0.46\,z_{\rm mod} \label{eq:sgsep2},
\end{eqnarray}
unless the target also passes the LOWZ cuts (Eqs. 1-4) listed above. 

The CMASS selection yields a sample with a median redshift $z = 0.57$ and a
stellar mass that peaks at $\log_{10}(M/M_{\odot}) = 11.3$ \citep{Mar12} and
a (stellar) velocity dispersion that peaks at $240\,{\rm km}\,{\rm s}^{-1}$
\citep{Bolton12,Thomas13}.
Most CMASS targets are central galaxies residing in dark matter halos of
$\sim10^{13}\,h^{-1}M_\odot$, but a non-negligible fraction are satellites that
live primarily in halos about 10 times more massive \citep{Whi11,Nuza13}.
Further discussion can be found in \cite{Tojeiro12}.
Kinematics and emission line properties are described in \citet{Thomas13}.

\begin{table*}
\caption{Basic properties of the CMASS target class and corresponding mask as defined in the text.}
\begin{center}
\begin{tabular}{lrrrrrr}
\hline
\hline
 & \multicolumn{3}{c}{DR10} & \multicolumn{3}{c}{DR11}\\
 Property & NGC & SGC & total  & NGC & SGC & total\\ \hline
$N_{\rm targ}$ &479,625 & 137,079 & 616,704 & 630,749 & 212,651 & 843,400 \\
$N_{\rm gal}$ &420,696 & 119,451 & 540,147 & 556,896 & 186,907 & 743,803 \\
$N_{\rm known}$ &7,338 & 1,520 & 8,858 & 10,044 & 1,675 & 11,719 \\
$N_{\rm star}$ &11,524 & 3,912 & 15,436 & 13,506 & 6,348 & 19,854 \\
$N_{\rm fail}$ &7,150 & 2,726 & 9,876 & 9,059 & 4,493 & 13,552 \\
$N_{\rm cp}$ &25,551 & 6,552 & 32,103 & 33,157 & 9,427 & 42,584 \\
$N_{\rm missed}$ &7,366 & 2,918 & 10,284 & 8,087 & 3,801 & 11,888 \\
$N_{\rm used}$ &392,372 & 109,472 & 501,844 & 520,805 & 170,021 & 690,826 \\
$N_{\rm obs}$ &439,370 & 126,089 & 565,459 & 579,461 & 197,748 & 777,209 \\
Total area / deg$^2$ &5,185 & 1,432 & 6,617 & 6,769 & 2,207 & 8,976 \\
Veto area / deg$^2$ &293 & 58 & 351 & 378 & 100 & 478 \\
Used area / deg$^2$ &4,892 & 1,375 & 6,267 & 6,391 & 2,107 & 8,498 \\
Effective area / deg$^2$ &4,817 & 1,345 & 6,161 & 6,308 & 2,069 & 8,377 \\
\hline
\end{tabular}
\end{center}
\label{tab:basic_props_cmass}
\end{table*}

\begin{table*}
\caption{Basic properties of the LOWZ target class and corresponding mask as defined in the text.}
\label{tab:basic_props_lowzz}
\begin{center}
\begin{tabular}{lrrrrrr}
\hline
\hline
 & \multicolumn{3}{c}{DR10} & \multicolumn{3}{c}{DR11}\\
 Property & NGC & SGC & total  & NGC & SGC & total\\ \hline
$N_{\rm targ}$ &220,241 & 82,952 & 303,193 & 302,679 & 129,124 & 431,803 \\
$N_{\rm gal}$ &113,624 & 67,844 & 181,468 & 156,569 & 108,800 & 265,369 \\
$N_{\rm known}$ &89,989 & 8,959 & 98,948 & 124,533 & 11,639 & 136,172 \\
$N_{\rm star}$ &804 & 523 & 1,327 & 944 & 754 & 1,698 \\
$N_{\rm fail}$ &477 & 278 & 755 & 726 & 497 & 1,223 \\
$N_{\rm cp}$ &8,199 & 2,928 & 11,127 & 10,818 & 4,162 & 14,980 \\
$N_{\rm missed}$ &7,148 & 2,420 & 9,568 & 9,089 & 3,272 & 12,361 \\
$N_{\rm used}$ &157,869 & 61,036 & 218,905 & 219,336 & 94,444 & 313,780 \\
$N_{\rm obs}$ &114,905 & 68,645 & 183,550 & 158,239 & 110,051 & 268,290 \\
Total area / deg$^2$ &4,205 & 1,430 & 5,635 & 5,793 & 2,205 & 7,998 \\
Veto area / deg$^2$ &252 & 58 & 309 & 337 & 99 & 436 \\
Used area / deg$^2$ &3,954 & 1,372 & 5,326 & 5,456 & 2,106 & 7,562 \\
Effective area / deg$^2$ &3,824 & 1,332 & 5,156 & 5,291 & 2,051 & 7,341 \\
\hline
\end{tabular}
\end{center}
\end{table*}

Target lists are produced using these algorithms and are then
``tiled'' to produce lists of galaxies to be observed with a single
pointing of the Sloan telescope. Not all targets can be assigned
fibers, and not all that are result in a good redshift measurement. In
fact, there are three reasons why a targeted galaxy may not obtain a
BOSS spectrum:
\begin{enumerate}
\item SDSS-II already obtained a good redshift for the object; these are
denoted {\it known}.
\item A target of different type (e.g., a quasar) is within
  62$^{\prime\prime}$; these are denoted {\it missed}.
\item another target of the same type is within 62$^{\prime\prime}$;
  these are denoted {\it cp} for ``close pair''.
\end{enumerate}
The second and third conditions correspond to hardware constraints on the
closest that two fibers can be placed on a plate. In regions where plates overlap, 
observations of close pairs are achieved.
There are two reasons why a spectrum might not result in a good
redshift measurement: 
\begin{enumerate}
\item The spectrum reveals that the object is a star (i.e., it was not properly classified by the imaging data and targeted as a galaxy); denoted {\it star}.
\item The pipeline fails to obtain a good redshift determination from
  the spectrum. These are denoted {\it fail}.
\end{enumerate}
The numbers of targets over the sky-region used in our analyses that
fall into these categories are given in
Table~\ref{tab:basic_props_cmass} for CMASS and
Table~\ref{tab:basic_props_lowzz} for LOWZ.  
We also report $N_{\rm gal}$, the total number of galaxies with good BOSS
spectra, and $N_{\rm used}$, the subset of $N_{\rm gal} + N_{\rm known}$ that
pass our redshift cuts.
As in \cite{And12}, missed close pairs and redshift failures are accounted for by
up-weighting the nearest target of the same target class with a successful spectral identification/redshift (regardless of its
category). The LOWZ sample is then cut to $0.15 < z < 0.43$ and the
CMASS sample is cut to $0.43 < z < 0.7$ to avoid overlap, and to make
the samples independent. The regions of sky included for the DR10 and DR11
samples are described in the next section. In order to provide results that use the largest publicly available BOSS data sets, we analyse both the DR10 and DR11 samples throughout this paper.

\subsection{Masks}

We use the {\sc Mangle\/} software \citep{Mangle} to track the areas
covered by the BOSS survey and the angular completeness of each distinct region.
The mask is constructed of spherical polygons, which form the base unit
for the geometrical decomposition of the sky.
The angular mask of the survey is formed from the intersection of the imaging
boundaries (expressed as a set of polygons) and spectroscopic sectors
\citep[areas of the sky covered by a unique set of spectroscopic tiles,see][]{Tiling,Teg04,DR8}.
In each sector, we determine an overall completeness
\begin{equation}
  C_{BOSS} = \frac{N_{\rm obs}+N_{cp}}{N_{\rm targ}-N_{\rm known}}
\end{equation}
where $N$ is the number of objects in the sector, ${\rm obs}$ denotes observed
and ${\rm targ}$ denotes target.  We discard any sectors where $C_{BOSS}<0.7$.
We define the redshift completeness
\begin{equation}
  C_{\rm red} = \frac{N_{\rm gal}}{N_{\rm obs}-N_{\rm star}}
\end{equation}
and discard any sector with $C_{\rm red}<0.8$. Further details can be found in \cite{And12}, which defined
and applied these same two masking choices.

In addition to tracking the outline of the survey region and the
position of the spectroscopic plates, we apply several ``vetos'' in
constructing the catalogue. Regions were masked where the imaging was
unphotometric, the PSF modelling failed, the imaging reduction pipeline
timed out (usually due to too many blended objects in a single field),
or the image was identified as having critical problems in any of the
5 photometric bands. We mask the small regions around the centre posts of the plates, where fibres
cannot be placed due to physical limitations and also around bright stars in
the Tycho catalogue \citep{tycho2}, with area given by Equation 9 of \cite{And12}.
We also place a mask at the locations of objects with higher priority
(mostly high-$z$ quasars) than galaxies, as a galaxy cannot be observed
at a location within the fibre collision radius of these points.
In total we masked $\sim5$ per cent of the area covered by the
set of observed tiles due to our ``veto'' mask.

The sky coverage of the LOWZ and CMASS galaxies is shown in
Fig.~\ref{fig:footprints} for both the Northern Galactic Cap (NGC) 
and Southern Galactic Cap (SGC). The ratio of total edge length to total area has
decreased significantly with each release, and the effective area has
increased from 3,275 deg$^2$ for DR9, to 6,161 deg$^2$, to 8,377
deg$^2$ for the CMASS DR10 and DR11 samples respectively.
Tables~\ref{tab:basic_props_cmass} and \ref{tab:basic_props_lowzz} list the
total footprint area $A_{\rm total}$, the area removed by the veto
masks $A_{\rm veto}$, and the total area used
$A_{\rm used} = A_{\rm total} - A_{\rm veto}$.
The total effective area is the used area
weighted by $C_{BOSS}$.

The raw volume enclosed by the survey footprint and redshift cuts is 10 Gpc$^3$
for the DR11 CMASS sample and 3 Gpc$^3$ for the DR11 LOWZ sample,
for a total of 13 Gpc$^3$.
For these samples, we have also calculated the effective volume, summing
over 200 redshift shells
\begin{equation}
  V_{\rm eff} = \sum\limits_{i}
  \left(\frac{\bar{n}(z_i) P_0}{1+\bar{n}(z_i) P_{0}}\right)^2 \Delta V(z_i)\,,
\end{equation}
where $\Delta V(z_i)$ is the volume of the shell at $z_i$, and we
assume that $P_0=20\,000\,h^{-3}$Mpc$^3$, approximately matching the
power spectrum amplitude where the BAO information is strongest. The
``holes'' in the survey introduced by the veto mask are small, and are
better approximated by a reduction in the galaxy density than the
volume covered for the large-scale modes of interest. We therefore
estimate the galaxy density $\bar{n}(z_i)$ by dividing the number of
galaxies in each shell by the shell volume calculated using area
$A_{\rm total}$, and the volume of each shell is estimated using area
$A_{\rm total}$. For DR10, the LOWZ sample then has an effective volume
of 1.7\,Gpc${}^3$, and the CMASS sample 4.4\,Gpc${}^3$.
For DR11, these increase to 2.4\,Gpc${}^3$ for LOWZ and 6.0\,Gpc${}^3$
for CMASS.

\subsection{Weighting galaxies} \label{sec:systematics}

\begin{figure*}
  \centering
  \centerline{\includegraphics[width=0.9\textwidth]{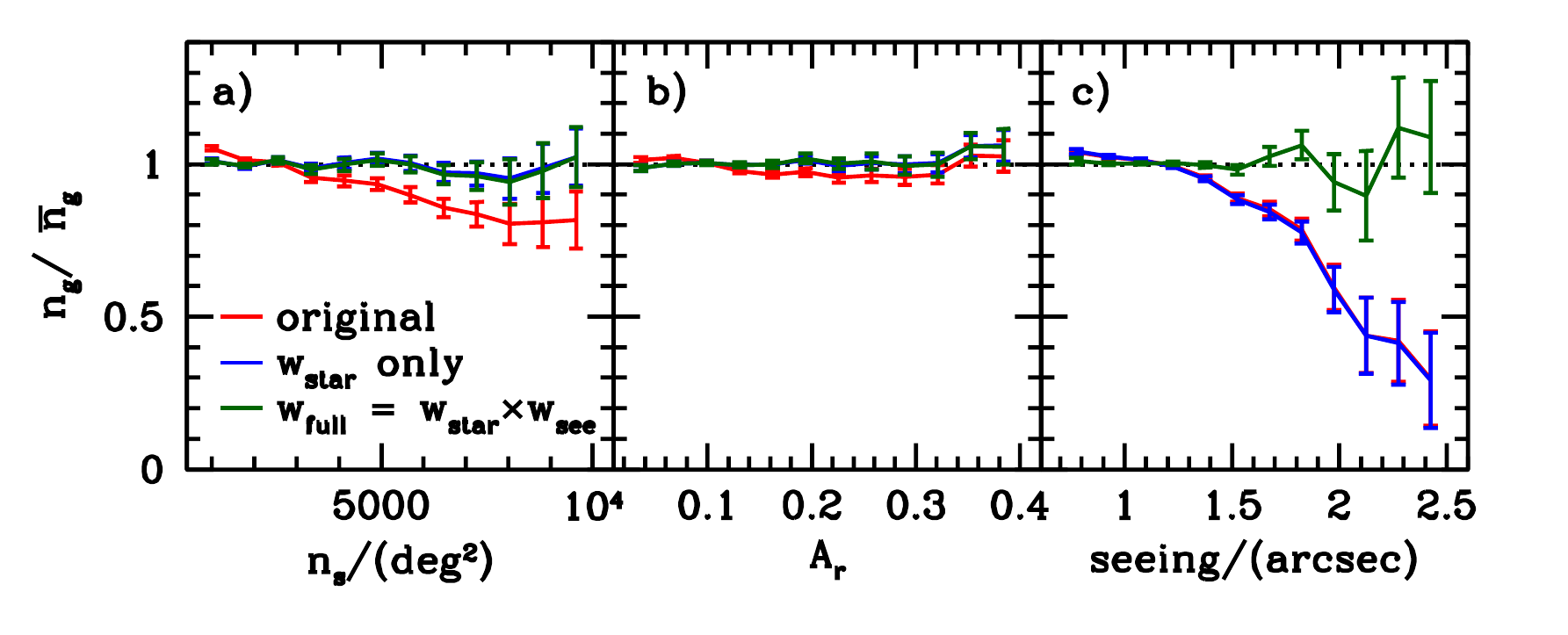}}
 \caption{
Dependence of the CMASS galaxy surface number density on the density of SDSS stars with $17.5 < i < 19.9$ (panel a), $r$-band Galactic extinction (b) and the $i$-band seeing of the imaging data (c). These lines deviate from $n_g=1$, indicating the presence of systematics
affecting the galaxy distribution. We correct for the systematic relationships using weights, with the relationships after applying weights shown in green. The relationship with seeing is dramatic, but only one per cent of the DR11 footprint has $i$-band seeing worse than 1$\farcs6$.
}
\label{fig:systematics}
\end{figure*}

To correct for the effects of redshift failures and fiber collisions, each
galaxy is given a set of weights.  A galaxy is upweighted if its nearest
neighbour (of the same target class) had a redshift failure ($w_{zf}$) or a redshift of that neighbour
was not obtained because it was in a close pair ($w_{cp}$).
For CMASS, we additionally apply weights to account for the systematic
relationships we find between the number density of observed galaxies
and stellar density and seeing (weights $w_{\rm star}$ and $w_{\rm see}$,
respectively).  Each galaxy is thus counted as
\begin{equation}
  w_{\rm tot} = (w_{cp}+w_{zf}-1)\,w_{\rm star}\,w_{\rm see},
\end{equation}
where $w_{\rm star}$ and $w_{\rm see}$ are equal to 1 for all LOWZ
galaxies. In this section, we justify the
application of these weights and describe how they are determined.

\citet{Ross11} created a photometric redshift catalog of the CMASS
sample over the full DR8 area, using early BOSS redshifts as a
training sample. Using this photometric redshift catalog,
\citet{Ross11} and \citet{Ho12} found that there exists a significant anti-correlation
between the surface number density of CMASS galaxies selected from the SDSS
DR8 imaging and stellar density. This relationship was found to impart
spurious large-scale clustering in angular distribution of CMASS
galaxies.

\citet{Ross11} and \citet{Ho12} also found a significant anti-correlation between the
number density of CMASS galaxies and seeing of the imaging data. It
was found that in areas with poorer seeing the star-galaxy separation
algorithm was more restrictive inducing the
observed anti-correlation. Using the same catalog, \citet{Ho12} derived
corrections based on measurements of the galaxy-seeing cross-power and
applied them to their angular power spectrum measurements, showing
that the seeing impacts the measured clustering. Over
the DR9 footprint, the impact of the systematic with seeing was found
to be insignificant \citep{Ross12}, as the pattern of seeing over the
DR9 area has negligible large-scale power. However, the effect on clustering
measured for any given footprint will scale with the pattern of seeing
in that particular footprint and any impact on the DR10 and DR11
clustering measurements must be re-tested.

\citet{Ross12} determined that weights applied to the DR9
CMASS galaxies as a function of stellar density and the $i_{\rm fib2}$
magnitude effectively removed any angular and redshift dependence of
the CMASS galaxy field on the number density of stars. They
found that, while a significant relationship existed between the
observed density of CMASS galaxies and seeing, the relationship did
not affect the measured clustering. Additional potential systematics
such as Galactic extinction, airmass, and sky background were tested
and the relationships were consistent with the expected angular
variation in galaxy number density. No significant systematic trends
were detected in the LOWZ sample.

For the DR10 and DR11 samples, we followed the same procedure as in
\citet{Ross12} to test and model the relation between the density of
spectroscopically identified galaxies and stellar density, seeing,
Galactic extinction, airmass and sky background.  To perform these
tests, we made HEALPix \citep{healpix} maps of the DR11 galaxies and
compared them to maps of the number of stars with $17.5 < i < 19.9$, where $i$
is the extinction-corrected $i$-band magnitude,
and to maps of the mean values of the potential systematic based on
data from the SDSS DR8 Catalog Archive Server (CAS), using various map
resolution parameters $N_{\rm side}$.

The solid red lines of Fig.~\ref{fig:systematics} show the
relationships between the surface number density of galaxies in the CMASS
sample, obtained after applying the completeness and close-pair
corrections described above, and the stellar density (panel a),
Galactic extinction (panel b) and $i$-band seeing (panel c). These lines
systematically deviate from $n_g/\bar{n}_g=1$, indicating the presence of
systematics affecting the galaxy distribution. The error bars in
these relations were obtained by applying the same test to the mock
catalogues described in Section \ref{sec:mocks}.  The systematic
effect associated with the surface density of stars, $n_{\rm s}$, is clearly visible in panel
(a), causing a decrease in the number of galaxies of as much as 20
per cent in regions with high stellar density. A weak relation between
the observed number of galaxies and the galactic extinction can be
seen in panel (b). This is due to the correlation between $A_r$ and
$n_{\rm s}$ and not to an independent systematic. Panel (c)
illustrates the strong impact of poor seeing conditions on the
observed galaxy number density: an $i$-band seeing of $S \simeq 2\arcsec$ leads to a
loss of approximately 50 per-cent of the galaxies. While this effect is 
dramatic, only 1 per cent of the survey footprint has $S>1\farcs6$. The systematic
relationship we find between the DR11 CMASS sample and the seeing in the imaging catalog
 is consistent with relationship found in the DR9 data \citep{Ross12}.

\begin{table}
\begin{tabular}{ccccc}
\hline
\hline
 & \multicolumn{2}{c}{DR10} & \multicolumn{2}{c}{DR11}\\
$i_{\rm fib2}$ range & $A_{\rm fib2}$ & $B_{\rm fib2}$ & $A_{\rm fib2}$ & $B_{\rm fib2}$\\
\hline
$< 20.3$ & 1.015 & -6.3$\times10^{-6}$ & 0.998 & 1.1$\times10^{-6}$\\
20.3,20.6 & 0.991 & 3.8$\times10^{-6}$ & 0.983 & 7.8$\times10^{-6}$\\
20.6,20.9 & 0.952 & 2.03$\times10^{-5}$ & 0.953 & 2.11$\times10^{-5}$\\
20.9,21.2 & 0.902 & 4.20$\times10^{-5}$ & 0.906 & 4.33$\times10^{-5}$\\
$> 21.2$ & 0.853 & 6.42$\times10^{-5}$ & 0.870 & 6.06$\times10^{-5}$\\
\hline
\end{tabular}
\caption{The coefficients we determine to apply weights for stellar
  density, as defined by Eq.~\ref{eq:wstar}. The stellar density
  weights are determined in bins of $i_{\rm fib2}$ magnitude.}
\label{tab:wstar}
\end{table}

We use the method to determine the corrective weight for stellar density, $w_{\rm star}$, defined in
\citet{Ross12}. This method weights galaxies as a function of the local stellar density and the
the surface brightness of the galaxy. We use the $i_{\rm fib2}$
as a measure of surface brightness and adopt a form for 
\begin{equation}
  w_{\rm star} (n_{\rm s}, i_{\rm fib2}) =
       A_{i_{\rm fib2}} + B_{i_{\rm fib2}} n_{\rm s},
  \label{eq:wstar}
\end{equation}
where $A_{i_{\rm fib2}}$ and $B_{i_{\rm fib2}}$ are coefficients to be fit empirically.
To construct these weights we divide the CMASS catalogue into five bins of 
$i_{\rm fib2}$, and fit the coefficients $A_{i_{\rm fib2}}$ and $B_{i_{\rm fib2}}$ in each bin
so as to give a flat relation between galaxy density and $n_{\rm s}$.
The stellar density map used for this task is based on a HEALPix grid with
$N_{\rm side} = 128$, which splits the sky into equal area pixels of 0.21 deg$^{2}$. 
This relatively coarse mask is enough to reproduce the large-scale variations of the
stellar density. The values of the $A_{i_{\rm fib2}}$ and $B_{i_{\rm fib2}}$ coefficients for DR10 and DR11 are given in Table~\ref{tab:wstar}.
The final weight $w_{\rm star}$ for a given galaxy is then computed according to the
local stellar density by interpolating the binned values of the coefficients
$A_{i_{\rm fib2}}$ and $B_{i_{\rm fib2}}$ to its observed $i_{\rm fib2}$.
The blue lines in Fig.~\ref{fig:systematics} illustrate the effect of applying these weights,
which correct for the systematic trend associated with $n_{\rm s}$ while leaving the 
relationship with the seeing unchanged, implying there is no significant correlation between the seeing and the stellar density.

Previous analyses of CMASS data (\citealt{Ross11,Ho12,Ross12}) found a systematic dependency with seeing consistent with the one we find for the DR11 CMASS data. In DR9, the relationship was not found to significantly impact the measured clustering and no weight was applied. For DR11, we now find a detectable impact of the relationship with seeing on the measured clustering. We therefore extend the DR9 analyses  include a weight, $w_{\rm see}$, for the $i$-band seeing, $S$, defined as
\begin{equation}
  w_{\rm see} (S) = A_{\rm see}\left[1-{\rm erf}\left(\frac{S-B_{\rm see}}{\sigma_{\rm see}}\right)\right]^{-1},
  \label{eq:wsee}
\end{equation}
which gives a good description of the observed relation.  Here the
coefficients $A_{\rm see}$, $B_{\rm see}$ and $\sigma_{\rm see}$ are fitted using the full sample,
as opposed to bins of $i_{\rm fib2}$. For this task we use a HEALPix map
with $N_{\rm side} = 1024$ (each pixel as a area 0.003 deg$^2$) as high resolution is required to sample
the intricate structure of the seeing in the footprint of the survey.
The green lines in Fig.~\ref{fig:systematics} show the effect of
applying the full weights $w_{\rm sys} = w_{\rm star}w_{\rm see}$,
which correct for all the observed systematic trends.  To avoid
applying large weights we set $w_{\rm sys}$ to a constant value for $S>2\farcs5$.
Introducing $w_{\rm see}$ is necessary, as we find the pattern
of seeing in the SGC has significant angular clustering and thus the
systematic induces spurious clustering into SGC measurements. The
$w_{\rm see}$ weights have negligible impact on measurements of the NGC
clustering (and, indeed, the DR9 SGC clustering); there is negligible large-scale power
in the pattern of the seeing in the NGC data. The best-fit
coefficients for the seeing weights we find and apply to the DR10
CMASS data are $A_{\rm see}=1.034$, $B_{\rm see}=2.086$ and
$\sigma_{\rm see}=0.731$ and for DR11
$A_{\rm see}=1.046$, $B_{\rm see}=2.055$ and $\sigma_{\rm see}=0.755$. We find no trend
in the relationship between galaxy density and seeing as a function of redshift. This implies that weighting
based on Eq. \ref{eq:wsee} removes from the CMASS density field any dependency on seeing in its full 3D space.

\section{Analysis changes common to isotropic and anisotropic clustering since DR9}
\label{sec:analysis}
We analyse the BAO feature and fit for distances using the 2-point
function in both configuration space (the correlation function, $\xi$)
and in Fourier space (the power spectrum, $P$). In Section \ref{sec:fit-2pt} 
we present the analysis techniques we use to obtain spherically averaged $P$
and $\xi$ and extract isotropic distance scale measurements.
In Section \ref{sec:aniso_fit}, we present the analysis techniques we use
measure the distance scale along and perpendicular to the line-of-sight using
Multipoles and Wedges in configuration space.
In this section, we detail the changes common to both the isotropic and
anisotropic clustering analysis since DR9.
These include changes in:
(i) density-field reconstruction,
(ii) mock catalogs, and
(iii) estimation of errors on these measurements by analyzing mock catalogues.

\subsection{Reconstruction}

The statistical sensitivity of the BAO measurement is limited by
non-linear structure formation.  Following \citet{Eis07a} we apply a
procedure to {\it reconstruct} the linear density field.  This procedure
attempts to partially reverse the effects of non-linear growth of structure and
large-scale peculiar velocities from the data. This is accomplished using the 
measured galaxy density field and Lagrangian theory relations between density 
and displacement. Reconstruction reduces the anisotropy in the clustering, reverses the
smoothing of the BAO feature due to second-order effects, and significantly reduces the
expected bias in the BAO distance scale that arises from these same second-order
effects. Reconstruction thus improves the precision of our BAO scale measurements 
while simplifying our analyses.

We apply reconstruction to both the LOWZ and CMASS samples. Briefly, we 
use the galaxy density field, applying an assumed bias for the galaxies, in order to
estimate the matter density field and solve for the displacement field. A correction is
applied to account for the effect of linear redshift space distortions. Full details 
of the reconstruction algorithm we apply can be found in \citet{Pad12} and \citet{And12}.
Compared to \citet{And12}, we have increased the number of points in the random 
catalogues used both when estimating the displacement field, and when sampling this 
field to give the shifted field \citep[see][for definitions]{Eis07a,Pad12,And12}.
Internal tests have shown that the results can be biased if the number of points
in the random catalogue is too small.  Given the large separation between the data 
in the NGC and SGC, we continue to run reconstruction on these two regions separately.

\subsection{Mock catalogs}
\label{sec:mocks}

To create mock galaxy catalogs for LOWZ and CMASS samples we use the the 
PTHalos methodology described in \citet{Man12} assuming the same fiducial 
cosmology as the data analysis.  The mocks reproduce the monopole and quadrupole 
correlation functions from the observed galaxies, and are randomly down-sampled 
to have the same mean $n(z)$ as a fitted 10-node spline to the sample $n(z)$.  
This achieves a smooth redshift distribution for the mean of the mocks.  We 
mask each mock to the area of the observed samples, simulate close-pair 
completeness (fiber collisions) and randomly downsample to the overall sky 
completeness based on regions defined by the specific tiling geometry of the 
data. 

To analyse the DR10 and DR11 CMASS samples, 600 mock CMASS galaxy catalogs were used 
with a slightly updated method as described in \citet{Man13}.  For the LOWZ sample, 
1000 mock LOWZ catalogs were created (again assuming the same fiducial cosmology) using 
a new incarnation of the PTHalos methodology \citep{Man13} that includes a redshift dependent
halo occupation distribution. The redshift dependence is fit to the data based jointly on the 
observed clustering and the observed $n(z)$. 

The analysis presented in this paper uses an earlier version of the mocks than the ones that will be publicly released in \citet{Man13}.
The differences are small and include an early estimate of the redshift distribution, a small difference in the way redshifts are assigned to random points, and lower intra-halo peculiar velocities. The mock catalogs are used to test our methodology and estimate covariance matrices. We expect these differences to have negligible statistical and systematic effects, especially when taking the approximate nature of the PTHalos methodology into account. Our systematic error budget is discussed further in Section 9.1.

\subsection{Covariance matrices}
\label{sec:cov_mat}

For each clustering metric we measure on the data, we also measure on 
the each mock galaxy catalog.  We use the distribution of values to 
estimate the sample covariance matrices that we use in the fitting. 
We use 600 mock catalogs for CMASS and 1000 for the LOWZ analysis. 
As the same underlying simulation was used to construct NGC and SGC 
versions of each mock catalog, we carefully combine a total measurement 
for each mock by using NGC and SGC measurements from different boxes.  
The full procedure we adopted is described in detail in \citet{Per13}, 
which focuses on understanding the error in the derived covariance matrix.
\citet{Per13} also includes how we propagate errors in the covariance matrix 
through to the parameter errors for all results presented in this paper.

\section{Measuring isotropic BAO positions} 
\label{sec:fit-2pt}
The BAO position in spherically averaged 2-point measurements is fixed
by the projection of the sound horizon at the drag epoch, $r_d$, and
provides a measure of
\begin{equation}
  D_V(z) \equiv \left[ cz (1+z)^2 D_A(z)^2 H^{-1}(z) \right]^{1/3},
\end{equation}
where $D_A(z)$ is the angular diameter distance and $H(z)$ is the Hubble parameter.
Matching our DR9 analysis \citep{And12} and previous work on SDSS-II LRGs
\citep{Per10}, we assume that the enhanced clustering amplitude along the
line-of-sight due to redshift-space distortions does not alter the relative
importance of radial and angular modes when calculating spherically averaged
statistics.  This approximation holds best for our results including
reconstruction, which are also our statistically most constraining measures.
If we measure the correlation function or power spectrum using a fiducial
cosmological model, denoted by a subscript ${\rm fid}$, to convert angles
and redshifts into distances, then to an excellent approximation the observed
BAO position depends simply on the scale dilation parameter
\begin{equation}
  \alpha \equiv \frac{D_V(z)r_{d,{\rm fid}}}{D^{\rm fid}_V(z) r_d} \,,
\end{equation}
which measures the relative position of the acoustic peak in the data versus
the model, thereby characterising any observed shift.
If $\alpha>1$, the acoustic peak is shifted towards smaller scales,
and $\alpha<1$ shifts the observed peak to larger scales.
We now outline the methodology we use to measure $\alpha$, tests made using
mock catalogues, and how we combine results from $\xi(s)$, and $P(k)$
measurements and from different binning schemes.

\subsection{Methodology}  \label{sec:method}

We have created separate pipelines to measure the average BAO position
in the BOSS data in configuration space using the correlation
function, $\xi(s)$, and in Fourier space using the power spectrum,
$P(k)$. 
The BAO position presents as a single peak in $\xi(s)$ and an oscillation
in $P(k)$. 

To calculate $\xi(s)$ we use the \citet{LanSza93} estimator, summing
pair-counts into bins of width $8\mpcoh$ \citep[as discussed further
in][]{Per13}. As a fiducial choice, the smallest $s$ bin is centred at
$6\mpcoh$, but we will also obtain results for the eight binning
choices shifted by increments of $1\mpcoh$.  For each binning, we
calculate $\xi(s)$ for bin centres in the range $29<s<200\mpcoh$ (22
bins, for our fiducial choice).

To calculate $P(k)$, we use the \citet{FKP94} estimator.
We use a Fourier grid of size $2048^3$, $4000\mpcoh$ along each side:
this comfortably encloses the survey including both the NGC and SGC
components; we use with sufficient zero-padding that aliasing is not
a problem which was confirmed by consistency between results from
other box sizes.
Compared to our DR9 analysis presented in \citet{And12}, we modify our
normalisation to properly account for the weights of galaxies introduced
to account for nearby close-pair or redshift failures.
We calculate $P(k)$ in Fourier modes averaged over bin widths of
$\Delta k=0.008\hompc$.
\cite{Per13} find this bin width minimises the combined error when
fluctuations in the covariance matrix are also included.
Our fiducial choice has the smallest $k$-bin centred at $k=0.004\hompc$.
We will also use the nine additional binning schemes that shift the bin
centres by increments of $0.0008\hompc$.
We calculate $P(k)$ for bin centres in the range $0.02<k<0.3\hompc$,
giving 35 bins for our fiducial choice.
These limits are imposed because the BAO have effectively died out for
smaller scales, and larger scales can be sensitive to observational systematics.

We fit the measured, spherically averaged, correlation function and
power spectrum separately and then combine results using the mocks to
quantify the correlation coefficient between measurements. Our fits
use polynomial terms to marginalise over the broad-band shape in
either 2-point measurement, while rescaling a model of the damped BAO
to fit the data. We use slightly different template BAO models for
$\xi(s)$ and $P(k)$ fits, as they enter the model functions in
different ways.

To produce a template model for the $P(k)$ fit, we first compute a
linear power spectrum $P^{\rm lin}$ produced by {\sc Camb}
\citep{lewis00}.  We then split into two components, 
one oscillatory $O^{\rm lin}$ and the other smooth $P^{\rm sm,lin}$,
that return the {\sc Camb} derived power spectrum when multiplied together. 
To perform the split, we fit $P^{\rm lin}$ using the same method 
that we use to fit to the data, but with
a BAO model calculated using the fitting formulae of \citet{Eis98}. The
resulting smooth model is taken to be $P^{\rm sm,lin}$, and 
$O^{\rm lin}$ is calculated by dividing $P^{\rm lin}$ by this. This follows
the procedure used in \citet{And12}.

The full model fitted to the data power spectrum is then
\begin{equation}  \label{eq:mod_pk}
  P^{\rm fit}(k)=P^{\rm sm}(k)
  \left[1+(O^{\rm lin}(k/\alpha)-1)e^{-\frac{1}{2}k^2
      \Sigma_{nl}^2}\right]\,,
\end{equation}
where 
\begin{equation}\label{eq:mod_pksm}
  P^{\rm sm}(k)= B_P^2P^{\rm sm,lin}(k)+A_1k+A_2+\frac{A_3}{k}
               + \frac{A_4}{k^2}+\frac{A_5}{k^3}\,.
\end{equation} 
There are therefore six ``nuisance'' parameters: a multiplicative constant
for an unknown large-scale bias $B_P$, and five polynomial parameters,
$A_1$, $A_2$, $A_3$, $A_4$, and $A_5$, which marginalise over broad-band
effects including redshift-space distortions, scale-dependent bias and any
errors made in our assumption of the model cosmology.
These effects may bias our measurement of the acoustic scale if not removed.

The damping was treated as a free parameter, with a Gaussian prior with
conservative width $\pm2\,h^{-1}$Mpc centered at the best-fit values
recovered from the mocks:
for the CMASS sample these are $\Sigma_{nl}=8.3\,h^{-1}$Mpc
pre-reconstruction, and $\Sigma_{nl}=4.6\,h^{-1}$Mpc post-reconstruction and
for LOWZ they are $\Sigma_{nl}=8.8\,h^{-1}$Mpc pre-reconstruction and
$\Sigma_{nl}=4.8\,h^{-1}$Mpc post-reconstruction.
This model, which differs from that used to fit the power spectrum in
\citet{And12}, is better matched to the now standard model for the
correlation function \citep[e.g.][]{And12} that we adopt.

To fit to the correlation function, we adopt the template model for the
linear correlation function given in \citet{EisSeoWhi07}, with damped
BAO
\begin{equation} \label{eqn:xu_xim}
  \xi^{\rm mod}(s) = \int \frac{k^2dk}{2\pi^2}P^{\rm mod}(k)j_0(ks) e^{-k^2a^2}\,,
\end{equation}
where the Gaussian term has been introduced to damp the oscillatory
transform kernel $j_0(ks)$ at high-$k$ to induce better numerical
convergence. The exact damping scale used in this term is not
important, and we set $a=1\,h^{-1}$Mpc, which is significantly below
the scales of interest. The power spectrum is given by
\begin{equation}  \label{eqn:template}
  P^{\rm mod}(k) = P^{\rm nw}(k)
    \left[1+\left(\frac{P^{\rm lin}(k)}{P^{\rm nw}(k)}-1\right)
    e^{-\frac{1}{2}k^2 \Sigma_{nl}^2}\right]\,,
\end{equation}
where $P^{\rm lin}(k)$ is the same model produced by {\sc Camb}, and
used to create the power spectrum fit template. $P^{\rm nw}(k)$ is
a model created using the no-wiggle fitting formulae of \citet{Eis98}, in 
which the BAO feature is erased. We refer to this template as the
``De-Wiggled'' template.

Using this template, our correlation function model is given by
\begin{equation} \label{eqn:fform}
  \xi^{\rm fit}(s) = B_\xi^2\xi^{\rm mod}(\alpha s)+A^\xi(s)\,.
\end{equation}
where $B_\xi$ is a multiplicative constant allowing for an unknown
large-scale bias, and the additive polynomial is
\begin{equation}\label{eq:xia}
  A^\xi(s) = \frac{a_1}{s^2} + \frac{a_2}{s} + a_3\,,
\end{equation}
where $a_1, a_2, a_3$ help marginalize over the broadband signal.

Unlike for the power spectrum, we do not allow the damping parameter to
vary and instead fix it at the average best-fit value recovered from
the mocks: the interplay between $B_\xi$ and the additive polynomial
$A_\xi$ in our fit to $\xi(s)$ means that the amplitude of the BAO peak
has more freedom already.

Apart from the differences in damping correction, the parallel between
$\xi(s)$ and $P(k)$ fitting methods is clear and follows from the
match between Eq.~(\ref{eq:mod_pk}) \&~(\ref{eqn:template}) and between
Eq.~(\ref{eq:mod_pksm}) and the combination of  Eqns.~(\ref{eqn:fform})
\&~(\ref{eq:xia}). There are three subtle differences: For the power
spectrum we only shift the BAO with the parameter $\alpha$, while for
$\xi(s)$ we shift the full model. As the nuisance parameters are
marginalising over the broadband, this should have no effect. For the
correlation function, the nuisance parameters are added to the final
model compared to the data $\xi^{\rm fit}(s)$; for the power
spectrum, they are added to the smooth model $P^{\rm sm}(k)$. This
slightly changes the meaning of the BAO damping term. We also
split the {\sc CAMB} power spectrum into BAO and smooth components in
different ways, utilising the \citet{Eis98} functions for the $\xi(s)$
template, whereas for the $P(k)$ fit we can applying the same fitting
method to the {\sc CAMB} power spectrum as used to fit the data. The
effect of this is expected to be small.

For fits to both the correlation function and power spectrum, we obtain
the best-fit value of $\alpha$ assuming that $\xi(s)$ and $\log P(k)$
were drawn from multi-variate Gaussian distributions, calculating $\chi^2$
at intervals of $\Delta\alpha=0.001$ in the range $0.8<\alpha<1.2$.
Our final error on $\alpha$ is determined by marginalising over the
likelihood surface and then correcting for the error in the covariance
matrix as described in \citet{Per13}.

\subsection{Testing on Mock Galaxy Catalogs}
\label{sec:isomocktest}

\begin{figure}
\resizebox{84mm}{!}{\includegraphics{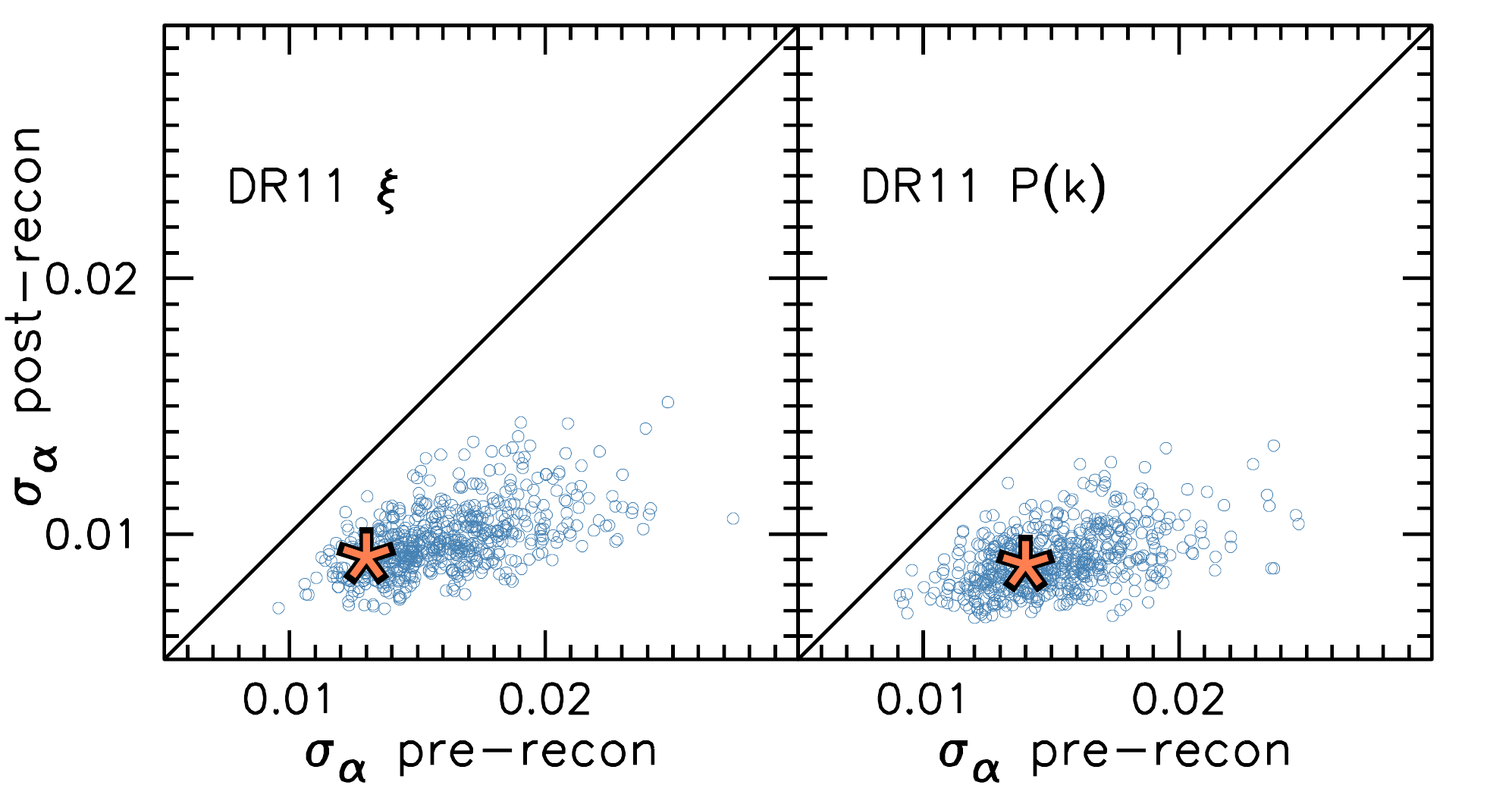}}
\resizebox{84mm}{!}{\includegraphics{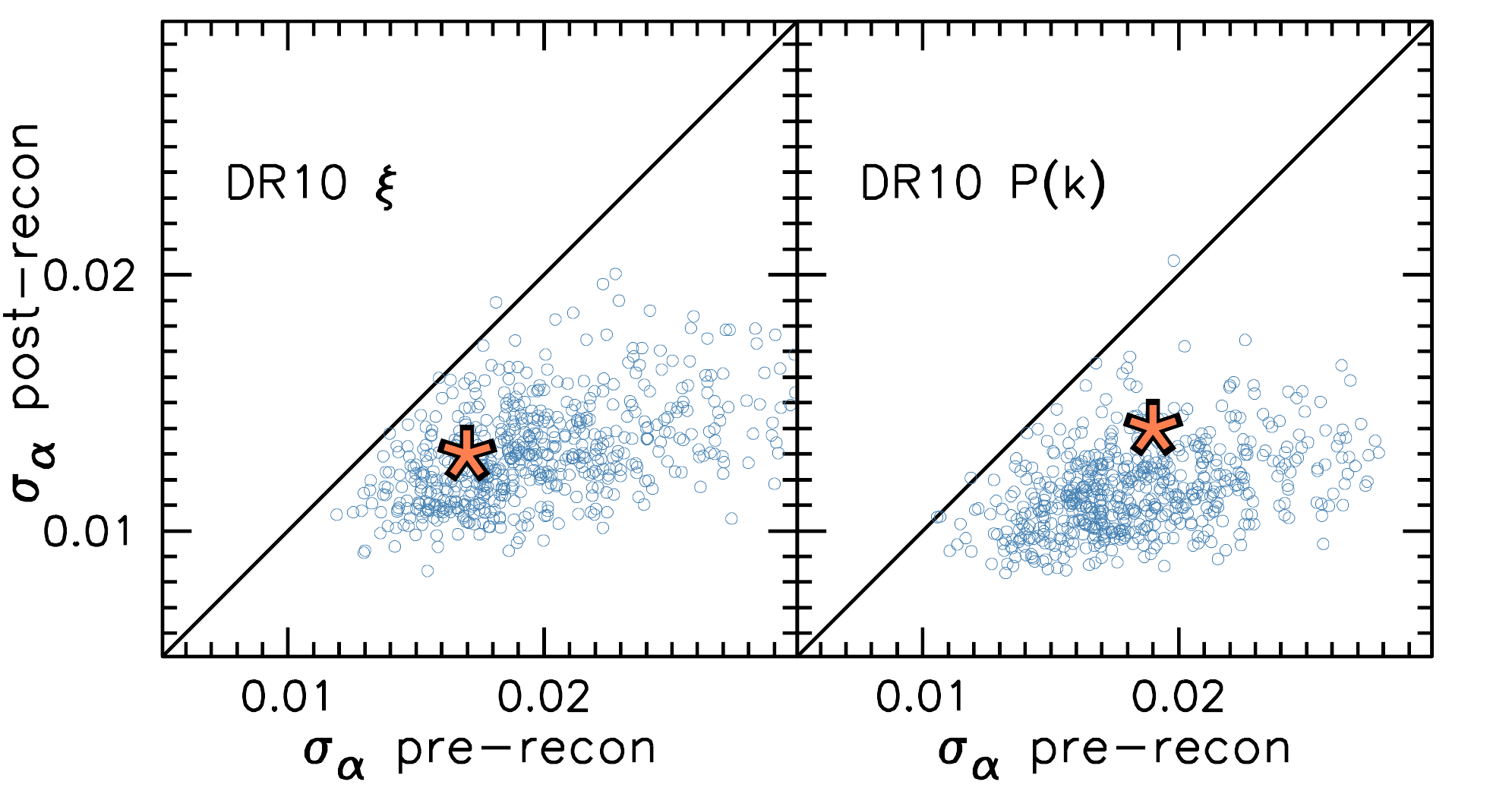}}
\caption{Scatter plots of $\sigma_\alpha$ pre- and post-reconstruction:
mocks (circles) + data (star) for $\xi$ and $P(k)$ CMASS DR10 and DR11. For the DR11 data, reconstruction improves the precision in each of the 600 mock realisations, for both $\xi(s)$ and $P(k)$. }
\label{fig:reconcomcmass}
\end{figure}

We test our $\xi(s)$ and $P(k)$ isotropic BAO fitting procedure on each of
our CMASS mock galaxy samples, both pre- and post-reconstruction.
The results are summarised in Table \ref{tab:mockbao}. \cite{Tojlowz} presents similar tests on the LOWZ mock galaxy samples.

\begin{table}
\caption{The statistics of isotropic BAO scale measurements recovered from the mock  galaxy samples. The parameter $\langle \alpha \rangle$ is the mean $\alpha$ value determined from 600 mock realisations of each sample, $S_{\alpha} = \sqrt{\langle(\alpha-\langle \alpha \rangle)^2\rangle}$ is the standard deviation of the $\alpha$ values, $\langle \sigma \rangle$ is the mean 1 $\sigma$ uncertainty on $\alpha$ recovered from the likelihood distribution of each realisation.  The ``combined'' results are post-reconstruction measurements optimally combined across a set of bin centre choices based on the correlation matrix determined from the mock realisations, as described in the text.}
\begin{tabular}{lccccc}
\hline
\hline
 Estimator  &  $\langle \alpha \rangle$ &$S_{\alpha}$&$\langle \sigma \rangle$&$\langle\chi^2\rangle$/dof\\
\hline
DR11  & & & \\
 {\bf Consensus $P(k)$+$\xi(s)$} & {\bf 1.0000} & {\bf 0.0090} & {\bf 0.0088}\\
combined $P(k)$ & 1.0001 & 0.0092 & 0.0089 \\
combined $\xi(s)$ & 0.9999 & 0.0091 & 0.0090 \\
post-recon $P(k)$ & 1.0001 & 0.0093  & 0.0090 & 28.6/27  \\
post-recon $\xi_0(s)$ & 0.9997 & 0.0095  &  0.0097 & 17.6/17 \\
pre-recon $P(k)$ & 1.0037 & 0.0163  & 0.0151 &  27.7/27 \\
pre-recon $\xi_0(s)$ & 1.0041 & 0.0157  & 0.0159  & 15.7/17\\
\hline
DR10   & & & & \\
 post-recon $P(k)$ & 1.0006 & 0.0117  & 0.0116 & 28.4/27  \\ 
 post-recon $\xi_0(s)$ & 1.0014 & 0.0122  & 0.0126 & 17.2/17 \\
pre-recon $P(k)$ & 1.0026 & 0.0187  & 0.0184 & 27.7/27  \\
pre-recon $\xi_0(s)$ & 1.0038 & 0.0188  & 0.0194 & 15.8/17  \\
\hline
\label{tab:mockbao}
\end{tabular}
\end{table}

Overall, we find a small, positive bias in the mean recovered
$\langle\alpha\rangle$ values pre-reconstruction, varying between
0.0026 (DR10 $P(k)$) and 0.0041 (DR11 $\xi(s)$).
This bias is significantly reduced post-reconstruction, as expected
\citep{EisSeoWhi07,PadWhi09,Noh09,Meh11}.
For the post-reconstruction DR11 samples, given that the uncertainty on one
realisation is 0.009, the statistical ($1\,\sigma$) uncertainty on
$\langle\alpha\rangle$ is 0.0004.
The $P(k)$ and $\xi(s)$ $\langle\alpha\rangle$ results are both
consistent with 1 (i.e.~unbiased).
This result is independent of bin size.

In general, the mean $1\,\sigma$ uncertainties recovered from the individual likelihood surfaces are close to the standard deviation in the recovered $\alpha$. All of these values include the appropriate factors to correct for the biases imparted by using a finite number of mocks, determined using the methods described in \cite{Per13}. The agreement between the recovered uncertainty and the standard deviation suggests that our recovered uncertainties are a fair estimation of the true uncertainty.

Applying reconstruction to the mock galaxy samples improves the uncertainty in BAO fits substantially. Fig. \ref{fig:reconcomcmass} displays scatter plots of $\sigma_\alpha$ before and after reconstruction for the DR11 (top) and DR10 (bottom) samples for $\xi(s)$ (left) and $P(k)$ (right). For DR11 reconstruction reduces the uncertainty in every case. The mean improvement, determined by comparing $\langle\sigma\rangle$ pre- and post-reconstruction, is more than a factor of 1.5 in every case and is even more for the DR11 $P(k)$ results. 

In summary, DR11 CMASS post-reconstruction $\xi(s)$ and $P(k)$ measurements are expected to yield estimates of the BAO scale, with statistical uncertainties that are less than 1 per cent, obtained from likelihood errors that agree with the standard deviation found in the measurements obtained from the mock samples. Furthermore, post-reconstruction, the systematic errors on the value of $\alpha$ measured from the mocks are consistent with zero for both correlation function and power spectrum fits, with an error on the measurement of $0.04$ per cent. Section~\ref{sec:sys_err} considers possible systematic errors on our measurements in more detail.

\subsection{Combining Results from Separate Estimators}\label{sec:combiso}

We have used $\xi(s)$ and $P(k)$ to measure the BAO scale for a number of different binning choices, with different values of bin centres and bin sizes in $s$ and $k$ respectively. These do not yield perfectly correlated BAO measurements because shot noise varies within each binning choice. Each estimate is un-biased, and we can therefore combine BAO measurements using different binning schemes and different estimators, provided we take the correlation into account, which we will do using the mocks. This results in more precise measurements of the BAO scale.

\begin{figure}
\resizebox{84mm}{!}{\includegraphics{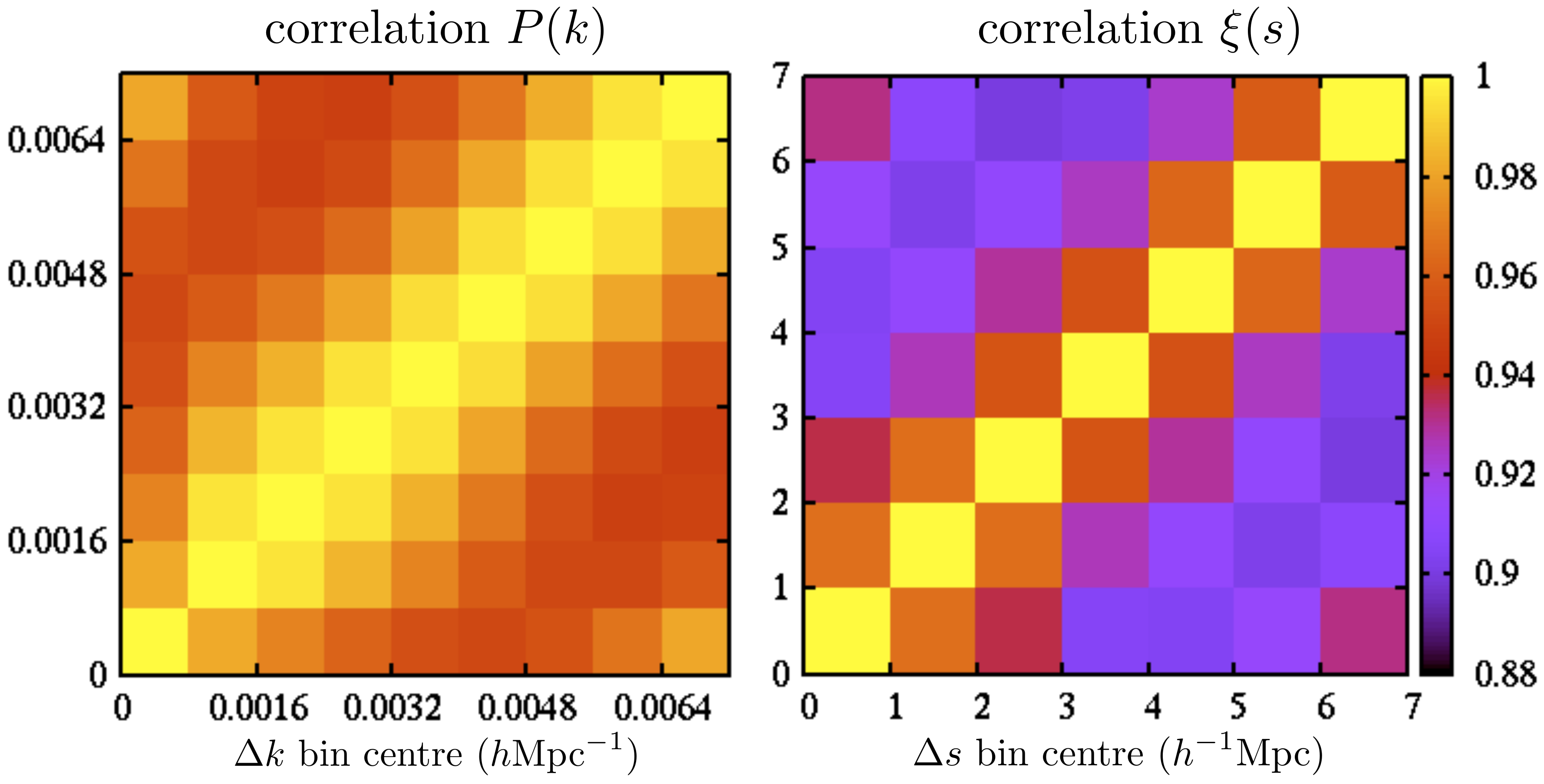}}
\caption{The correlation between recovered $\alpha$ values calculated using different bin centres for the DR11 CMASS reconstructed power spectrum ($P(k)$; left) and  correlation function ($\xi(s)$; right). The correlation between bins is of lower amplitude for $\xi(s)$ compared with $P(k)$, implying that combining results across $\xi(s)$ bin centres will improve the precision more than doing the same for $P(k)$.}
\label{fig:bincorr}
\end{figure}
 The dispersion in values of $\alpha$ that we recover from the mocks for a single choice of bin width but with different bin centres, is greater for $\xi(s)$ than for $P(k)$. There is therefore more to be gained by combining results from offset bins for our analysis of $\xi(s)$. The correlation matrices for $\alpha$ recovered from the eight $\xi(s)$ and the ten $P(k)$ bin centres tested (see Section~\ref{sec:method}) are displayed in Fig.~\ref{fig:bincorr}. For $\xi(s)$, the correlation is as low as 0.89. The $P(k)$ results are more correlated, as all of the correlations are greater than 0.94. 

\begin{figure}
\centering
\includegraphics[width=84mm]{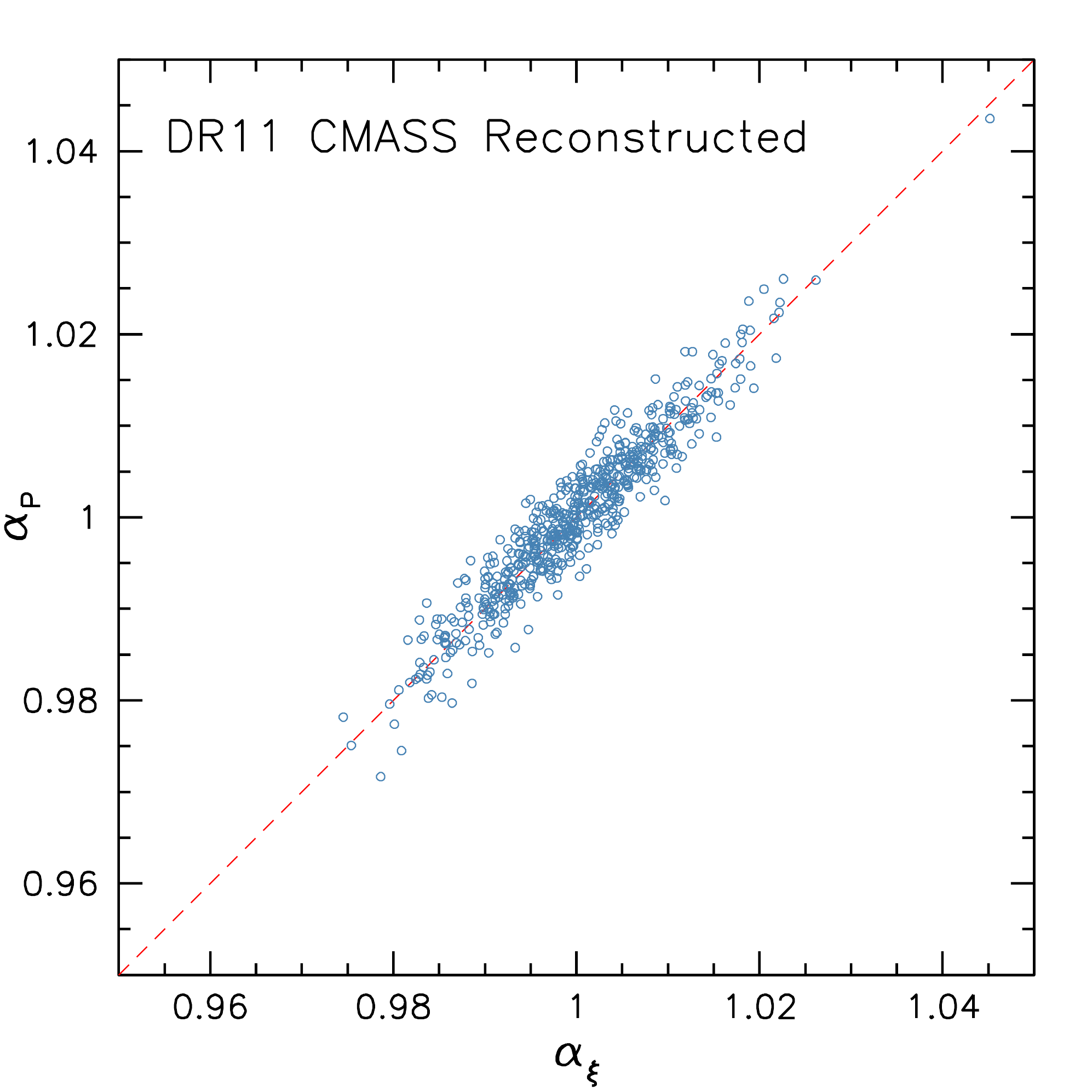}
  \caption{Scatter plot showing the measured BAO positions ($\alpha$) using DR11 CMASS reconstructed mock galaxy $P(k)$ versus those obtained from $\xi(s)$ (blue circles). The measurements are strongly correlated, with $C_{P,\xi}=0.95$ and an r.m.s. difference of 0.0027.}
\label{fig:Pkxicom}
\end{figure}
The fact that results using different bin centres are not perfectly correlated implies that an optimised $\alpha$ measurement can be made by calculating the weighted mean of $\alpha$ across all of the bin centre choices. The process we use is as follows: We find the weighted mean uncertainty, $\sigma_b$, using the correlation matrix, $D$
\begin{equation}
  \sigma_b = \frac{\sum_{i,j} \sigma_i D^{-1}_{i,j}}{\sum_{i,j}D^{-1}_{i,j}}.
\end{equation}
We then scale the elements of $D$ by $\sigma^2_b$ to obtain the covariance matrix, $C$, for the measurements at each bin centre. The BAO scale measurement, $\alpha_o$, and its uncertainty, $\sigma_{\alpha,o}$, obtained combining the results across bin centres are then given by
\begin{equation}
  \alpha_o = \frac{\sum_{i,j} \alpha_i C^{-1}_{i,j}}{\sum_{i,j}C^{-1}_{i,j}}
  \qquad , \qquad
  \sigma^2_{\alpha,o} = \frac{1}{\sum_{i,j}C^{-1}_{i,j}}.
\end{equation}
Applying this to $\xi(s)$ and $P(k)$ decreases the uncertainty and standard deviation such that they are nearly identical for $\xi(s)$ and $P(k)$, shown as the ``combined'' results in Table~\ref{tab:mockbao}. 

The method we apply to measure the BAO position from $P(k)$ has changed from the method applied in \cite{And12}; it is now more similar to the method applied to the $\xi(s)$ measurements (e.g., the smooth polynomials are similar). We combine information across bin centre choices for both fits. This results in BAO measurements that are more correlated between fits to $\xi(s)$ and $P(k)$. We use
\begin{equation}
C_{1,2}(X) = \frac{\sum_i^N(X_{1,i}-\langle X_1\rangle)(X_{2,i}-\langle X_1\rangle)}{(N-1)\sigma_1\sigma_2}
\label{eq:cov12}
\end{equation}
to quantify the correlation between two measures, where $\sigma$ in this case represents the standard deviation of sample $X$. For the DR11 CMASS reconstructed mock galaxy samples, we find $C_{P,\xi}(\alpha) = 0.95$.
Fig.~\ref{fig:Pkxicom} displays a scatter plot illustrating this tight
correlation. 

In order to combine BAO measurements from $P(k)$ and $\xi(s)$, we take
$\alpha_{\xi+P}$ as the mean of the two estimates and its uncertainty as
$\sigma_{\xi+P} = 0.987\bar{\sigma}$, where $\bar{\sigma}$ is the mean
uncertainty of the two $\alpha$ estimates.
This allows the uncertainty to vary for any given realisation, but assumes
that the uncertainty is equal and Gaussian distributed for $\alpha_P$ and
$\alpha_\xi$.
The 0.987 factor is the reduction in uncertainty obtained by averaging two
measurements with a 0.95 correlation factor that we identified from the 
mock measurements.

\section{Measuring Anisotropic BAO Positions}
\label{sec:aniso_fit}
Assuming an incorrect cosmology when calculating the galaxy
correlation function or power spectrum will differentially shift the
BAO feature in both the transverse and line-of-sight directions. These
shifts are typically parameterized by $\alpha_\perp$ and
$\alpha_{||}$, which are the natural extension of the isotropic scale
dilation factor $\alpha$ introduced in
Section~\ref{sec:fit-2pt}. Together, they allow us to measure the angular diameter distance (relative to the sound horizon at the drag epoch $r_d$)
$D_A(z)/r_d$, and  the Hubble parameter $H(z)$ via $cz/(H(z)r_d)$ separately through
\begin{equation}
\alpha_{\perp} = \frac{D_{A}(z)r_d^{\rm fid}}{D^{\rm fid}_{A}(z) r_d} \,,\,\,\,\,
\alpha_{\parallel} = \frac{H^{\rm fid}(z)r_d^{\rm fid}}{H(z) r_d} \,.
\end{equation}

Another parameterization that exists in the literature decomposes the
anisotropic shifts into $\alpha$ and an anisotropic warping factor
$\epsilon$, which can be defined in terms of $\alpha_\perp$ and
$\alpha_{\parallel}$ as
\begin{equation}
\alpha = \alpha_{\perp}^{2/3} \alpha_{\parallel}^{1/3} \,, \,\,\,\,
1 + \epsilon = \left( \frac{\alpha_{\parallel}}{\alpha_{\perp}} \right)^{1/3} \,.
\end{equation}
Note that in the fiducial cosmology, $\alpha = \alpha_\perp =
\alpha_\parallel = 1$ and $\epsilon = 0$.
In this paper, we concentrate on $\alpha_{\parallel}$ and $\alpha_{\perp}$, but there are discussions that use $\alpha$ and $\epsilon$ parameterization for the ease of explanation. In particular, we use $\alpha$-$\epsilon$ and $\alpha_{\parallel}$-$\alpha_{\perp}$ interchangeably for multipoles as we can convert one to another parameterization easily. Note that the $\alpha$ measured through anisotropic clustering is in theory the same as $\alpha$ measured using isotropic clustering. However, there can be a small amount of scatter between the two measured $\alpha$s. 

We have developed separate pipelines using either multipoles of the correlation
function, or top-hat windows in $\mu$ \citep[called wedges][]{Kaz12}, to
estimate $\alpha_{\parallel}$ and $\alpha_{\perp}$.
We now outline the methodology behind each pipeline and present the results
of tests on both using mock data.

\subsection{Methodology}\label{sec:aniso_method_5.1}

For the CMASS data we measure the average BAO position in configuration space
using moments of the correlation function, $\xi(s,\mu)$, where $\mu$ is the
cosine of the angle between a galaxy pair (we use the mid-point of the two
galaxy positions in redshift space) and the line-of-sight.
We use the CMASS galaxy catalog only and we don't do an anisotropic Fourier
space analysis in this paper
\citep[see][for a complementary analysis]{Beutler13}.
We measure $\xi(s,\mu)$ using the \citet{LanSza93} estimator, with radial
bins of width $8\mpcoh$ and angular bins of $\Delta\mu=0.01$
\citep[see][for the effect of bin-sizes on the measurement]{Per13,Vargas13}.
We then project the $\mu$-dependence to obtain both ``multipoles''
\begin{equation}
  \xi_{\ell}(s) = \frac{2 \ell +1}{2} \int_{-1}^{1} d\mu
    \ \xi(s, \mu) L_{\ell}(\mu) \,\,,
 \label{eqn:defn_multipoles}
\end{equation}
and ``wedges'', 
\begin{equation}
  \xi_{\Delta\mu}(s)=
    \frac{1}{\Delta\mu}\int_{\mu_{\rm min}}^{\mu_{\rm min}+\Delta\mu}
    d\mu\ \xi(s,\mu)\,.
\label{eqn:defn_wedges}
\end{equation}
Throughout we shall denote the Legendre polynomial of order $\ell$ as $L_\ell$,
since $P_\ell$ will be reserved for moments of the power spectrum.
As wedges and multipoles are alternative projections of $\xi(s,\mu)$, we
expect similar constraints from both.  We perform both analyses principally
as a test for systematic errors. 

For both cases we only measure and fit to two projections.
For the multipoles we use $\ell=0$ and 2.
In linear theory there is information in the $\ell= 4$ multipole as well,
and beyond linear theory there is information in all even multipoles,
but we do not include the higher multipoles as the increase in
signal-to-noise ratio is small compared to the increase in modeling complexity.
Furthermore, after reconstruction, the effect of redshift space distortions
is significantly reduced, decreasing the information in $\ell\ge 4$ further.
For the wedges, we choose $\Delta\mu=0.5$ such that we have a bin which is
primarily ``radial", $\xi_\parallel(s)\equiv\xi(s,\mu>0.5)$, and a bin which is
primarily ``transverse", $\xi_\perp(s)\equiv\xi(s,\mu<0.5)$.
This matches the methodology adopted for the anisotropic DR9 BAO measurements
presented in \citet{And13}.

We model the moments of the correlation function as the transform of
\begin{equation}
  P(k,\mu) = (1+\beta\mu^2)^2 F(k,\mu,\Sigma_s)P_{\rm pt}(k,\mu)\,,
  \label{eqn:tdp}
\end{equation}
where
\begin{equation}
  F(k,\mu,\Sigma_s) = \frac{1}{(1+k^2\mu^2\Sigma_s^2/2)^2}\,,
  \label{eqn:fog}
\end{equation}
is a streaming model for the Finger-of-God (FoG) effect
\citep{PeacockDodds1994} and $\Sigma_s$ is the streaming scale, which we set
to $3\mpcoh$.
This choice of the streaming scale has been tested in
\citet{Xu12b,And13,Vargas13}.
The $(1+\beta\mu^2)^2$ term is the linear theory prediction for redshift-space
distortions at large scales \citep{Kai87}.  In linear theory
$\beta=f/b\simeq \Omega_m^{0.55}/b$, where $f$ is the linear growth rate, but
we treat $\beta$ as a parameter which we vary in our fits.
This allows for modulation of the quadrupole amplitude, as $\beta$ is
degenerate with any quadrupole bias.
To exclude unphysical values of $\beta$ we a impose a prior.
This prior is discussed further in Section~\ref{sec:aniso_mocktest} and its
effects tested in Section~\ref{sec:fitting_data_aniso}.
We take $P_{\rm pt}$ to be:
\begin{equation}
  P_{\rm pt}(k)=P_{\rm lin}(k) e^{-k^2\sigma_v^2}
               + A_{\rm MC} P_{\rm MC} (k)\,,
\label{eqn:RPT}
\end{equation}
where the $P_{\rm MC}$ term includes some of the non-linearities to
second order, and  is given by \citep{GGRW,MakSasSut92,JaiBer94}: 
\begin{equation}
  P_{\rm MC}=2\int \frac{d^3q}{(2\pi)^3}
  \ |F_2(k-q,q)|^2 P_{\rm lin}(|k-q|)P_{\rm lin}(q)\,,
\end{equation}
with $F_2$ given by Eq.~(45) of the review of \citet{Bernardeau02} or
the references above.
The parameter $\sigma_v$ accounts for the damping of the baryonic acoustic
feature by non-linear evolution and $A_{MC}$ for the induced coupling between
Fourier modes.
We fit to the mocks with these parameters free and use the mean value of
the best-fits pre-reconstruction and post-reconstruction. 
In particular, $\sigma_v$ is  fixed to $4.85(1.9)\,h^{-1}$Mpc and
$A_{\rm MC}$ is fixed to $1.7(0.05)$ pre(post)-reconstruction. 

The template of Eq.~(\ref{eqn:RPT}) is different from the one used in
\citet{And13} and from the non-linear template used in
Section~\ref{sec:fit-2pt}.
The isotropic fitting in both configuration and Fourier space used the
``De-Wiggled'' template (Eq.~\ref{eqn:template}), while we use $P_{\rm pt}$,
inspired by renormalized perturbation theory.
This template was previously used by \citet{Kaz13} in the analysis of the
CMASS DR9 multipoles and clustering wedges and is described in more detail
in \citet{Sanchez13a}. 

We then decompose the full 2D power-spectrum into its Legendre moments:
\begin{equation}
P_{\ell}(k)=\frac{2\ell+1}{2}\int_{-1}^{1}P(k, \mu)L_{\ell}(\mu)d\mu
\end{equation}
using $P(k,\mu)$ from Eq.~(\ref{eqn:tdp}), 
which can then be transformed to configuration space using
\begin{equation}
\label{eqn:finalcorr} 
\xi_{\ell}(s) = i^{\ell} \int \frac{dk}{k}\ \frac{k^3 P_\ell(k)}{2 \pi^2}
  \ j_{\ell}(ks)
\end{equation}
where, $j_{\ell}(ks)$ is the $\ell$-th spherical Bessel function.

Similar to the isotropic BAO fitting procedure (Section~\ref{sec:fit-2pt}),
we use polynomial terms to marginalize over the broad-band shape for both
multipoles and wedges.
The model multipoles, $\xi_{0,2}^{\rm m}(s)$, and projections,
$\xi_{\perp,\parallel}^{\rm m}(s)$, are defined by our template
evaluated for the fiducial cosmology.
The model fit to the observed multipoles is then
\begin{align}
  \xi^{\rm fit}_0(s) &= B_{\xi,0}^2\xi_0^{\rm m}(\alpha,\epsilon,s)
  +A_0^\xi(s) \,, \nonumber \\ 
  \xi^{\rm fit}_2(s) &= \hphantom{B_{\xi,0}^2}\xi_2^{\rm m}(\alpha,\epsilon,s)
  + A_2^\xi(s)\,,
\end{align}
and to the observed wedges is
\begin{align}
  \xi^{\rm fit}_\perp(s)    &= \hphantom{r^2}B_{\xi,\perp}^2\xi_\perp^{\rm m}(\alpha_\perp,\alpha_\parallel, s)+A_\perp^\xi(s)\,, \nonumber \\ 
  \xi^{\rm fit}_\parallel(s) &= r^2B_{\xi,\perp}^2\xi_\parallel^{\rm m}(\alpha_\perp,\alpha_\parallel, s)+A_\parallel^\xi(s)\,,
\end{align}
where \cite{Xu12b} describe how to include $\alpha$ and $\epsilon$ in the
template $\xi_{0,2}^{\rm m}$ and \cite{Kaz13} describe the equivalent
methodology for $\xi_{\parallel,\perp}^{\rm m}$.  
The parameters $B_{\xi}$ are  bias factors that rescale the amplitude of
the input models, while $r$ regulates the amplitude ratio of the two wedges.
The polynomial terms
\begin{equation}
  A_{\ell}(s) = \frac{a_{\ell,1}}{s^2} + \frac{a_{\ell,2}}{s} + a_{\ell,3}
  \quad ;\quad \ell=0,2,\parallel,\perp\,.
  \label{eqn:fida}
\end{equation}
are used to marginalize out broadband (shape) information that contributes to
$\xi_{\ell}(s)$ due to, e.g., scale-dependent bias or redshift-space distortions.

In order to find the best-fit values of $\alpha_\parallel$ and $\alpha_\perp$,
we assume that the correlation function moments are drawn from a multi-variate
Gaussian distribution with a covariance matrix derived from our mocks
\citep{Man13}, corrected as summarized in Section \ref{sec:cov_mat}.
We fit to 40 points over the range $45<s<200\mpcoh$, including both the
monopole and the quadrupole or the two wedges.
Since there are 10 parameters in our fitting model, this gives 30 degrees
of freedom in the fit.

In our analysis of the wedges, we use a Markov chain Monte Carlo (MCMC)
to explore the parameter space
\begin{equation}
 \theta=(\alpha_{\perp},\alpha_{\parallel},B_{\xi,\perp},r,
         a_{i,\perp},a_{i,\parallel}) \quad .
\label{eqn:par_wedges}
\end{equation}
We impose flat priors in all these parameters and obtain our constraints on
$\alpha_{\perp}$ and $\alpha_{\parallel}$ by marginalizing over all the
remaining parameters. 

In our analysis of the multipoles, we explore the parameter space by
calculating the likelihood surface over a large grid of $\alpha$ and
$\epsilon$ with $\Delta \alpha=0.003$, and $\Delta \epsilon=0.006$
\footnote{We have tested the effect of grid size on $\sigma_\alpha$
  and $\sigma_\epsilon$ and have verified that finer grids results in  no
  difference to the errors recovered \citep{Vargas13}.}.
Before performing the fit, we normalize the model to the data at $s=50\mpcoh$
and hence $B_\xi^2\sim1$.
As mentioned previously, we allow $\beta$ to vary in our fits but apply two
priors:
\begin{itemize} 
\item Gaussian prior on $\log(B_\xi^2)$ centered on $1$, with standard deviation of $0.4$.
\item Gaussian prior on $\beta$ with a standard deviation of 0.2. The central
  value is set to $f/b\sim\Omega_m ^{0.55}(z)/b=0.4$ pre-reconstruction,
  and zero post-reconstruction \citep{Xu12b}.
\end{itemize}
For each grid point, ($\alpha$,$\epsilon$), we fit the remaining parameters
to minimize the $\chi^2$.
Assuming the likelihood surface is Gaussian allows us to estimate the
uncertainties of $\alpha$ and $\epsilon$ as the standard deviations of the
marginalized 1D likelihoods (for more details see \citealt{Xu12b} and
\citealt{Vargas13}).
The deviations are computed by integrating the likelihood surface over
$\alpha=[ 0.8,1.2]$ and $\epsilon=[-0.2,0.2]$.
We do however use an expanded likelihood surface covering a wider range of
$\alpha$ and $\epsilon$ as input for measuring cosmological parameters,
so the chosen integration intervals do not have any effect on the down-stream
cosmological analysis.  We test the effect of each of these priors in
Section~\ref{sec:aniso_mocktest}.
We can then easily convert any ($\alpha$,$\epsilon$) to
($\alpha_{\parallel}$,$\alpha_{\perp}$). 

\subsection{Testing on Mock Galaxy Catalogues} 
\label{sec:aniso_mocktest}

We test our anisotropic BAO fitting procedure with both multipoles and wedges,
pre- and post-reconstruction using mock catalogs.
The results are summarized in Table~\ref{tab:sigmamethodr11rec}.
We list the median values of the recovered $\alpha_{||}$, $\alpha_{\perp}$,
$\sigma_{||}$ and $\sigma_{\perp}$ from all the mock galaxy samples. 
Pre-reconstruction, we find that there is a small positive bias (0.006) in the
median $\alpha_{\parallel}$ using multipoles and a small negative bias (-0.004)
when using wedges.
The signs of biases are reversed for $\alpha_{\perp}$, as
(again pre-reconstruction) there is a small negative bias (-0.003) for
multipoles and a small positive bias (0.001) for wedges.
Reconstruction reduces the bias.
Post-reconstruction, the largest bias is 0.003 for the median multipole
$\alpha_{\parallel}$.  The others are all $\leq 0.001$.
Finally, we note that both the standard deviation of the $\alpha$s and the
median of their errors are very consistent.
The uncertainties are also significantly larger than the biases on $\alpha$
(the bias is at most 11\% of the uncertainty on $\alpha$s) for both methods.

\citet{And13} and \citet{Kaz13} describe detailed tests applied to the
``Wedges'' technique. Given the high degree of correlation between
wedge and multipole based measurements and fitting methodology of multipoles has changed slightly since
\citet{And13}, here we focus on tests based
on multipoles.

We tested the robustness of our fits to a number of parameter choices,
including the following:
\begin{itemize} 
\item Changing fitting ranges
\item Changing the number of nuisance parameters, $A_\ell(r)$
\item Changing the priors on $B_0$ and $\beta$. 
\end{itemize} 

The results of these and further tests are extensively detailed in
\citet{Vargas13}. Here we only highlight the specific findings that
are pertinent to this analysis (see Table~\ref{tab:sigmadr11rec}). 
None of the tests  resulted in
significantly biased values for the best fit parameters or their associated errors. 
In particular, the best fit values of $\alpha$ do not vary by more than
$0.2$ per cent for all cases, and most of the best fit values of $\epsilon$
do not vary by more than $0.3$ per cent.  It is particularly interesting to
note that the median errors of both $\alpha$ and $\epsilon$ do not
change at all for all of the different fitting parameter choices. Note
that this is not true if we extend the range of $\alpha$ and
$\epsilon$ over which we integrate to make these measurements. By
design, the priors act to exclude unphysical models, which otherwise
can affect the measured errors. However, the likelihood close to the
best-fit solution is not affected by these priors, and hence the
best-fit values and errors are not affected.

\begin{table*}
  \caption{ 
    Measurements of $\alpha_{||}$ and $\alpha_{\perp}$ and their
    1$\sigma$ errors for CMASS mock galaxy catalogs when we use different anisotropic
    clustering estimates (multipoles and wedges). We choose to show
    median values which are less affected by the range
    of parameters over which we integrate to determine best-fit values and their associated errors.  The columns are
    the median values of $\alpha, \epsilon, \alpha_{||,\perp}$ ($\widetilde{
      \alpha}, \widetilde{ \epsilon},\widetilde{ \alpha_{||,\perp}}$),
    and the standard deviations of $\alpha, \epsilon, \alpha_{||,\perp}$
    ($S_{\alpha, \epsilon,\alpha_{||,\perp}}$). Further details can be
    found in \protect\cite{Vargas13}.  The ``consensus'' results
    combine the likelihoods determined from multipoles and wedges, as
    described in the text. 
\label{tab:sigmamethodr11rec}}

\begin{tabular}{@{}lcccccccccccc}

\hline\hline
Method&
$\widetilde{\alpha}$&
$S_\alpha$&
$\widetilde{\epsilon}$&
$S_{\epsilon}$&
$\widetilde{\alpha_{\parallel}}$&
$S_{\alpha_{\parallel}}$&
$\widetilde{\sigma_{\alpha_{\parallel}}}$&
$S_{\sigma_{\alpha_{\parallel}}}$&
$\widetilde{\alpha_{\perp}}$&
$S_{\alpha_{\perp}}$&
$\widetilde{\sigma_{\alpha_{\perp}}}$&
$S_{\sigma_{\alpha_{\perp}}}$\\
\hline
\\[-1.5ex]
Post-Rec DR11 \\
{\bf Consensus} &-&-&-&-&1.0009&0.0252 &0.0270&0.0045 &0.9984&0.0143 &0.0149&0.0018\\
Multipoles &
1.0002&0.0092&
0.0011&0.0122&
1.0032&0.0266&
0.0248&0.0072&
0.9999&0.0149&
0.0137&0.0018\\
Wedges&
1.0003&0.0090&
0.0005&0.0124&
1.0006&0.0264&
0.0296&0.0052&
0.9993&0.0153&
0.0161&0.0026\\
\hline
Pre- Rec DR11 \\
Multipoles &
0.9995&0.0155&
0.0022&0.0189&
1.0058&0.0443&
0.0384&0.0150&
0.9965&0.0210&
0.0205&0.0033\\
Wedges&
0.9991&0.0152&
-0.0011&0.0207&
0.9965&0.0475&
0.0466&0.0137&
1.0007&0.0222&
0.0230&0.0086\\
\hline
\end{tabular}
\end{table*}

\begin{table*}
\caption{Variations in measured parameters and errors from the DR11 CMASS
  mock galaxy catalogs post-reconstruction for different changes to
  the fiducial fitting methodology.  The variation is defined as
  $\Delta v=v^{i}-v^{fid}$, where $v$ is the parameter or error of
  interest. These results confirm the robustness of the fitting
  methodology. The largest variation observed on the fitted parameters is
  in epsilon $\Delta\epsilon=0.003$ while the largest variation in
  alpha is only $\Delta\alpha=0.001$. Median variations $\Delta v$, and percentiles are multiplied by 100.}

 \label{tab:sigmadr11rec}

\begin{tabular}{@{}lrrrrrrrr}

\hline\hline

Model&
$100 \widetilde{\Delta \alpha}$&
$100 \widetilde{\Delta \sigma_{\alpha}}$&
$100 \widetilde{\Delta \epsilon}$&
$100 \widetilde{\Delta \sigma_{\epsilon}}$&
$100 \widetilde{\Delta \alpha_{\parallel}}$&
$100 \widetilde{\Delta \sigma_{\alpha_{\parallel}}}$&
$100 \widetilde{\Delta \alpha_{\perp}}$&
$100 \widetilde{\Delta \sigma_{\alpha_{\perp}}}$\\
\hline

$30<r<200$&
$0.05^{+0.13}_{-0.12}$&
$-0.03^{+0.02}_{-0.03}$&
$0.10^{+0.15}_{-0.11}$&
$-0.01^{+0.03}_{-0.03}$&
$0.25^{+0.41}_{-0.31}$&
$-0.01^{+0.10}_{-0.12}$&
$-0.06^{+0.09}_{-0.09}$&
$-0.04^{+0.03}_{-0.03}$\\
\\[-1.5ex]

$\mathrm{2-term} \; A_l(r)$&
$0.03^{+0.07}_{-0.06}$&
$0.02^{+0.02}_{-0.02}$&
$0.27^{+0.15}_{-0.12}$&
$-0.02^{+0.03}_{-0.04}$&
$0.58^{+0.32}_{-0.25}$&
$0.02^{+0.07}_{-0.07}$&
$-0.24^{+0.13}_{-0.18}$&
$-0.04^{+0.02}_{-0.03}$\\
\\[-1.5ex]

$\mathrm{4-term} \; A_l(r)$&
$-0.05^{+0.06}_{-0.08}$&
$-0.01^{+0.02}_{-0.02}$&
$-0.15^{+0.11}_{-0.12}$&
$0.00^{+0.03}_{-0.03}$&
$-0.35^{+0.27}_{-0.31}$&
$-0.01^{+0.07}_{-0.07}$&
$0.09^{+0.08}_{-0.07}$&
$0.01^{+0.01}_{-0.01}$\\
\\[-1.5ex]

%
%
%
\\[-1.5ex]

$\mathrm{Fixed \;}\beta=0.0$&
$-0.00^{+0.01}_{-0.03}$&
$-0.00^{+0.00}_{-0.01}$&
$0.02^{+0.10}_{-0.10}$&
$-0.01^{+0.02}_{-0.03}$&
$0.03^{+0.17}_{-0.21}$&
$-0.02^{+0.09}_{-0.12}$&
$-0.02^{+0.11}_{-0.12}$&
$-0.02^{+0.03}_{-0.03}$\\
\\[-1.5ex]

%
%
$\mathrm{No \; priors \;(RL)}$&
$0.00^{+0.04}_{-0.02}$&
$0.02^{+0.06}_{-0.01}$&
$-0.03^{+0.12}_{-0.12}$&
$0.05^{+0.12}_{-0.03}$&
$-0.05^{+0.24}_{-0.21}$&
$0.08^{+0.36}_{-0.10}$&
$0.02^{+0.15}_{-0.11}$&
$0.05^{+0.08}_{-0.03}$\\
\\[-1.5ex]

$\mathrm{Only} \;B_0\; \mathrm{prior}\;(RL)$&
$0.00^{+0.04}_{-0.01}$&
$0.02^{+0.04}_{-0.01}$&
$-0.02^{+0.11}_{-0.11}$&
$0.04^{+0.08}_{-0.02}$&
$-0.04^{+0.22}_{-0.20}$&
$0.06^{+0.28}_{-0.09}$&
$0.02^{+0.14}_{-0.11}$&
$0.04^{+0.06}_{-0.03}$\\
\\[-1.5ex]

$\mathrm{Only} \; \beta \; \mathrm{prior}\;(RL)$&
$-0.00^{+0.01}_{-0.01}$&
$0.00^{+0.01}_{-0.00}$&
$-0.00^{+0.01}_{-0.02}$&
$0.01^{+0.02}_{-0.01}$&
$-0.00^{+0.03}_{-0.05}$&
$0.02^{+0.04}_{-0.02}$&
$0.00^{+0.01}_{-0.01}$&
$0.01^{+0.01}_{-0.00}$\\
\\[-1.5ex]

\hline
\end{tabular}

\end{table*}

\begin{figure}
\resizebox{84mm}{!}{\includegraphics{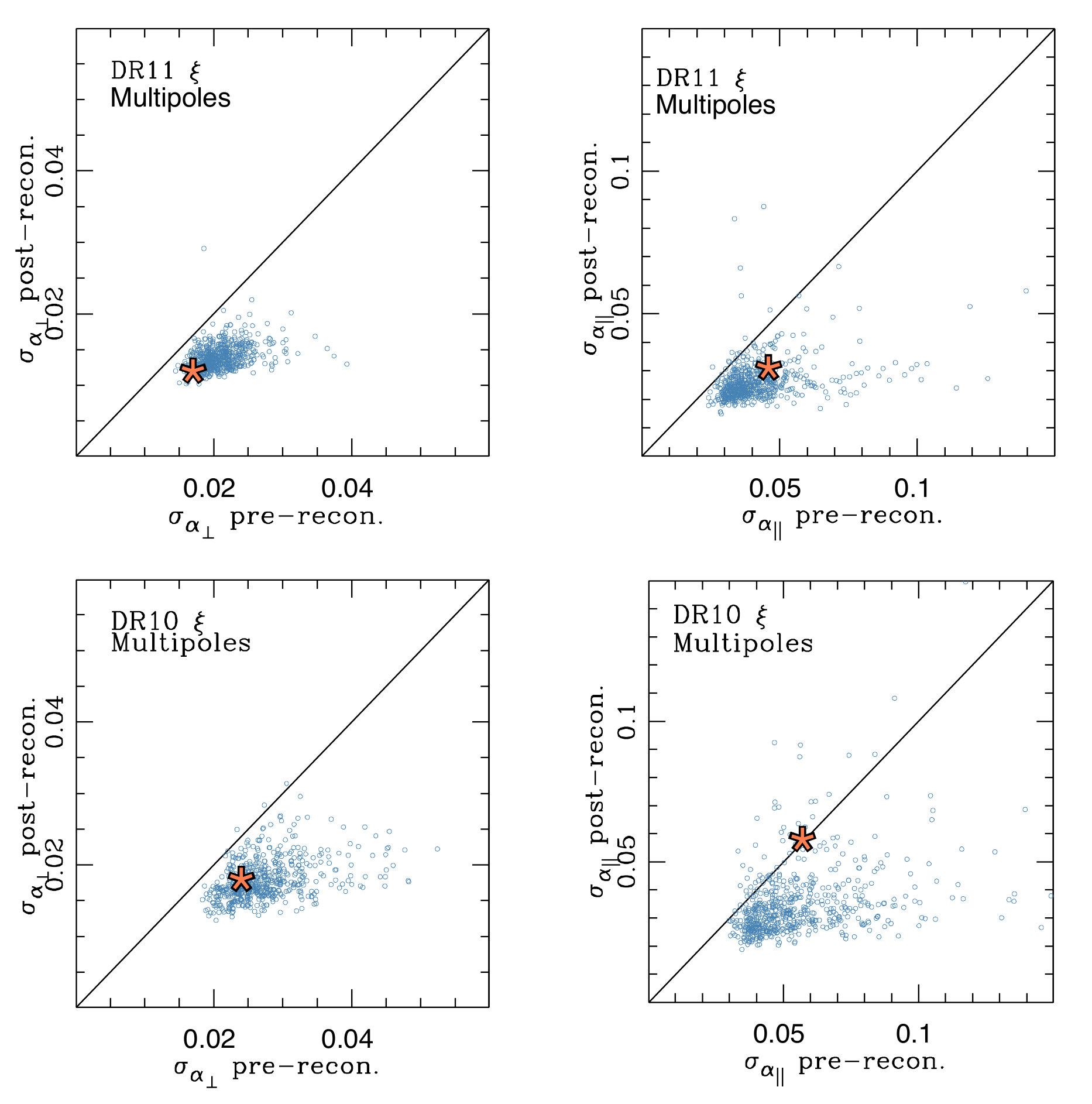}}
\caption{These are scatter plots of $\sigma_\alpha$ post-reconstruction:
mocks (circles) + data (star) for $\sigma_{\alpha_{\perp}}$ and $\sigma_{\alpha_{||}}$ for  CMASS DR10 and DR11. The reconstruction significantly improves the precision in nearly all  of the 600 mock galaxy samples for both DR10 and DR11. Note that we converted to one parameterization ($\alpha_{\parallel}$,$\alpha_{\perp}$) for ease of comparison between multipoles and wedges. }
\label{fig:sigdr11_multip}
\end{figure}

Finally, we further test our Multipoles method by looking at the error
on both $\alpha_{||}$ and $\alpha_{\perp}$ for all of our mock galaxy
samples (blue points), and compare it to our data in DR10 and DR11
(orange stars).  We show in Figure~\ref{fig:sigdr11_multip} that
reconstruction decreases the uncertainty on $\alpha_{\perp}$ and
$\alpha_{\parallel}$ in the vast majority of the 600 mock galaxy
samples. This is especially true for $\alpha_{\perp}$.  The DR10
footprint is less contiguous than the DR11 one and there are thus more
outliers in DR10 than in DR11 where reconstruction does not improve
the uncertainty.
The constraints obtained using the pre- and post-reconstruction wedges show a similar behavior.

\subsection{Comparing and Combining Methodologies}
\label{sec:com_multwedge}

Table \ref{tab:sigmamethodr11rec} compares the fitting results of our DR11 mock galaxy catalogs using
the multipoles and clustering wedges.  
There are slight differences in both the median and dispersion between methods in pre-reconstruction, but both are 
unbiased and give similar errors in both $\alpha_\parallel$ and $\alpha_\perp$. 
 For instance, the median and the  68\% confidence level of the variation between the two methods,
$\Delta\alpha_{||}=\alpha_{||,{\rm multipoles}}-\alpha_{||,{\rm wedges}}$, 
is
$\widetilde{\Delta \alpha_{||}}=+0.005_{-0.028}^{+0.025}$ while that for
$\alpha_\perp$ is
$\widetilde{\Delta \alpha_{\perp}}=-0.004_{-0.011}^{+0.014}$.
These are small differences, especially when compared to the standard deviations  ($S_{\alpha_{\parallel}}$ and $S_{\alpha_{\perp}}$) within the mocks, which are on the order of $\sim 0.046$ and $\sim 0.021$ respectively. 

\begin{figure}
\resizebox{84mm}{!} {\includegraphics{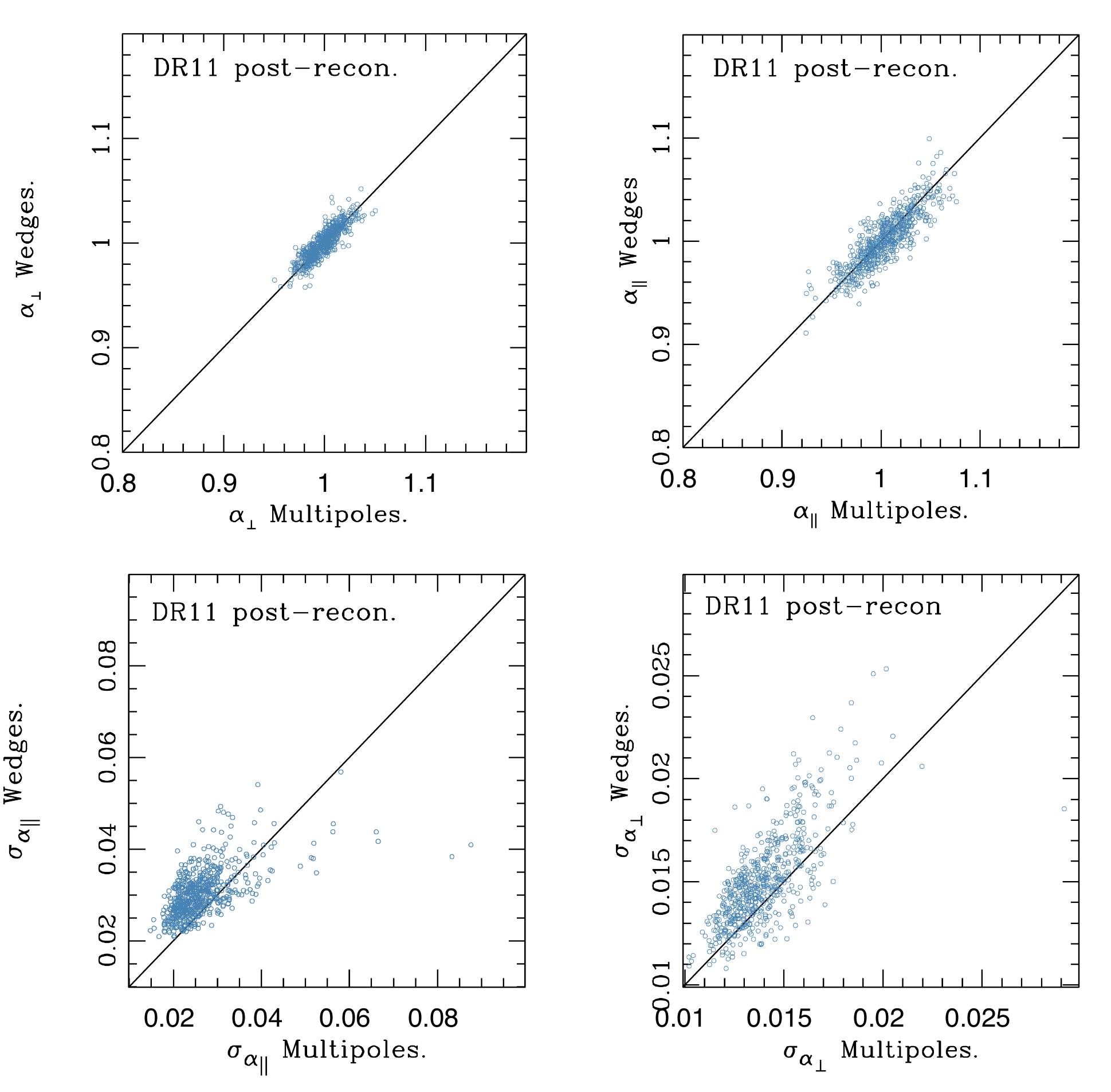}}
  \caption{ Top panels compare the $\alpha_{\parallel}$ and $\alpha_{\perp}$ recovered with the multipoles methodology with the values recovered from wedges for the  DR11 CMASS mock galaxy samples. 
Bottom panel compares the $\sigma_{\alpha_{||}}$ and $\sigma_{\alpha_{\perp}}$ recovered with the multipoles with the values recovered from wedges using the same mock galaxy samples as in the top panel.
  }
\label{fig:anisotropic_alphas_comp}
\end{figure}

As we can see from Table~\ref{tab:sigmamethodr11rec},  the fitting results of post-reconstructed mock catalogs from both methods are extremely similar. 
After reconstruction, the median BAO measurements become even more similar between the two methods and the scatter, relative to the standard deviation, decreases slightly: we find $\widetilde{\Delta \alpha_{||}}=
+0.001_{-0.016}^{+0.016}$ and $\widetilde{\Delta \alpha_{\perp}}=-0.001_{-0.007}^{+0.008}$. 
The top panels of Fig.\ref{fig:anisotropic_alphas_comp} show scatter plots between the BAO measurements for multipoles and those of wedges, post-reconstruction, determined from the 600 mock samples. 
The two measurements are clearly correlated.

We find that the multipole results are slightly more precise, on average. We obtain tight constraints 
on both $\alpha_{||}$ and $\alpha_\perp$.  
In particular, $\widetilde{\Delta \sigma_{\alpha_{||}}}=-0.008_{+0.007}^{-0.008}$ and $\widetilde{\Delta \sigma_{\alpha_{\perp}}}=-0.003_{-0.003}^{+0.003}$ pre-reconstruction and post reconstruction, while the median difference in best-fit values are $\widetilde{\Delta \alpha_{||}}=-0.005_{-0.004}^{+0.005}$ and $\widetilde{\Delta \alpha_{\perp}}=-0.002_{-0.002}^{+0.001}$. 
As measurements from the two fitting methodologies are clearly correlated, it is not surprising that the obtained precisions on the $\alpha$s are similar.

\begin{figure}
   \resizebox{84mm}{!} {\includegraphics{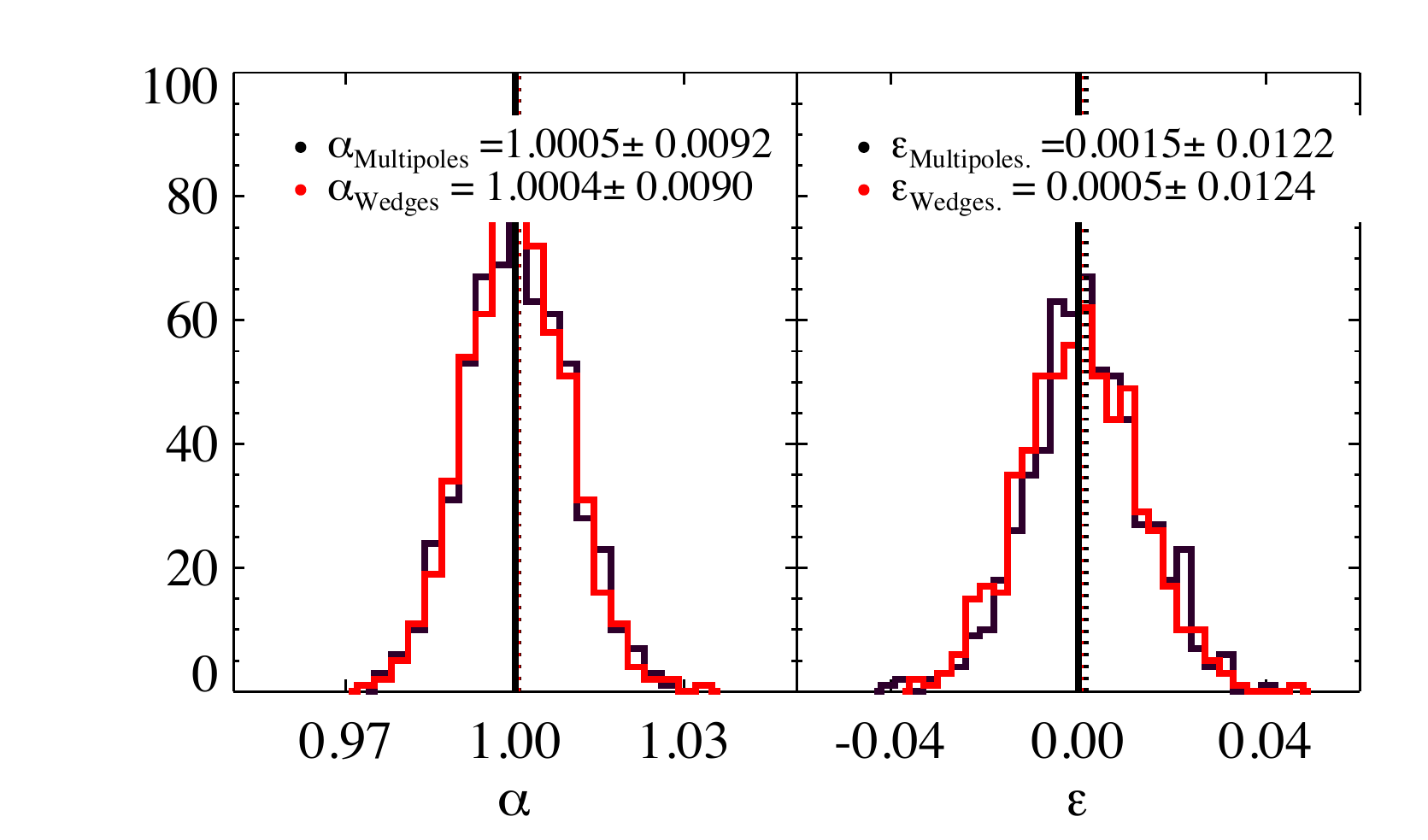}}
  \caption{The distributions of $\alpha$ (left) and $\epsilon$ (right) recovered with multipoles methodology compared to the values recovered from wedges for DR11 mock galaxy samples. 
The legend indicates the mean and r.m.s. of the distribution.
  }
\label{fig:alphaepsilondist}
\end{figure}

For reasons that will become apparent later in Sec~\ref{sec:results_aniso}, we also look at the differences in fitting results between wedges and multipoles when we measure $\alpha$ and $\epsilon$, in addition to $\alpha_{\perp}$ and $\alpha_{\parallel}$.
In Figure~\ref{fig:alphaepsilondist} , we show the histogram of the  fitted $\alpha$ and $\epsilon$ from the two different methodologies used in anisotropic clustering. The median values of $\alpha$s are almost identical in the two methodologies with close to zero median shift.  The $\epsilon$ distributions show small median shift  of 0.2 and -0.1 per cent   which point in different directions for multipoles and wedges approaches respectively. The standard deviations in both $\alpha$ and $\epsilon$ from both methods are also comparable. 

In general, the statistics indicate a good agreement between the distributions of the fitted parameters and errors obtained from multipoles and wedges. We do not find any indication that favours one technique over the other. Pre-reconstruction we find differences of 0.2$\sigma$ in the median values of $\alpha_{||}$ and $\alpha_{\perp}$ recovered by the two methods, but post-reconstruction these differences become negligibly small, less than 0.08$\sigma$. 
We therefore believe that the two methods are equally un-biased. 
The scatter in the results recovered by the Multipoles and Wedges methods in individual realizations come from 
shot-noise and differences in methodology, as explored further in \cite{Vargas13}.

Given that the multipoles and wedges results are both unbiased but are
not perfectly correlated, our results are improved by combining the
two results. We do this following the
procedure adopted in \cite{And13}. Briefly, we take the mean of the
log-likelihood surfaces obtained using each method and use this
averaged likelihood surface to obtain consensus results. We have
applied this procedure to the results from each mock realisation and
the statistics are listed as the ``consensus'' values in Table
\ref{tab:sigmamethodr11rec}. The standard deviation in both the BAO
measurements and their uncertainties have decreased, showing that a
small improvement is afforded by combining the two measurements.

\section{BAO Measurements from Isotropic Clustering Estimates}
\label{sec:results_iso}
\subsection{Clustering Estimates}

\begin{figure*}
\centering
  \resizebox{0.8\textwidth}{!}{\includegraphics{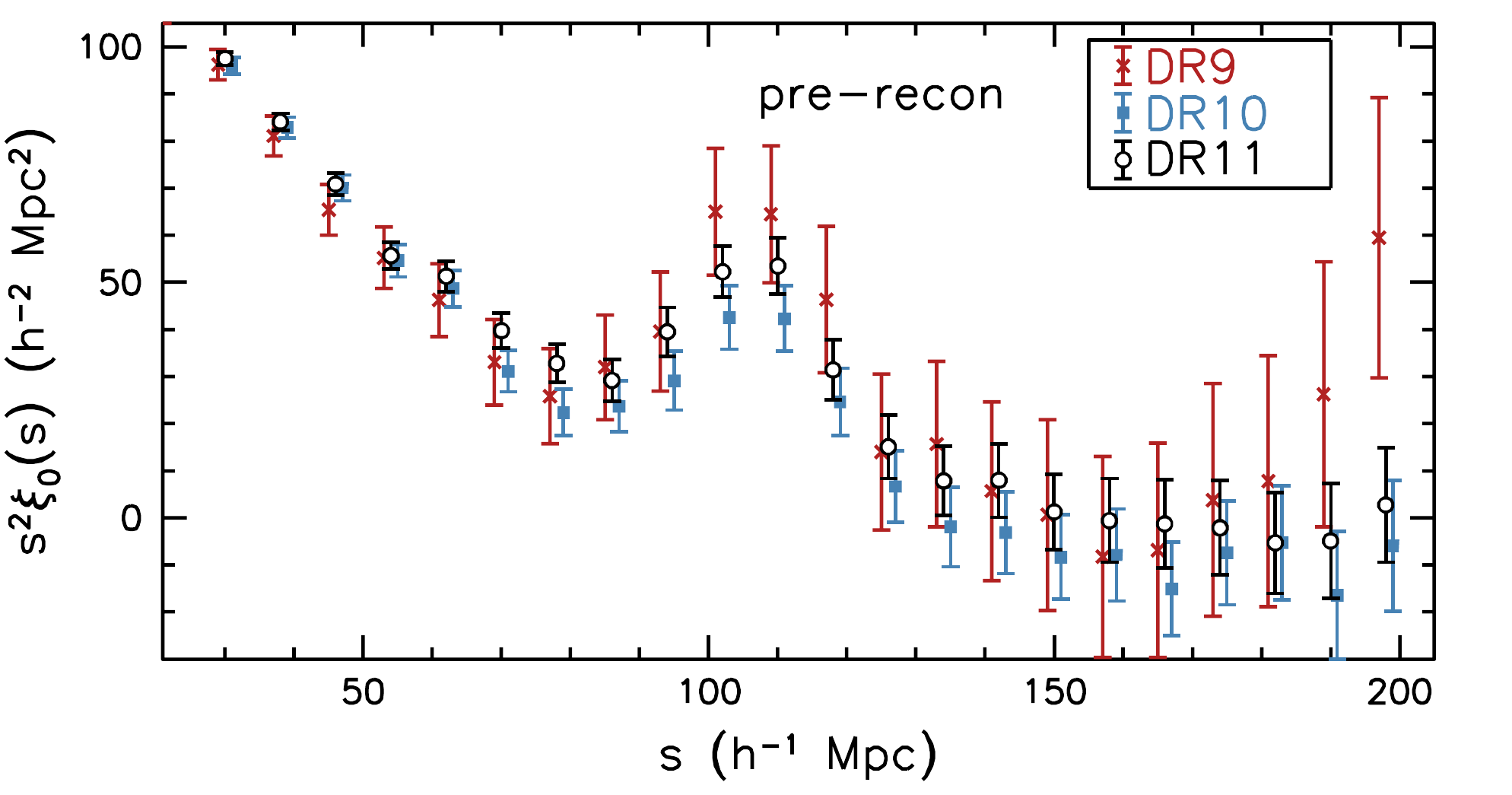}}
  \resizebox{0.8\textwidth}{!}{\includegraphics{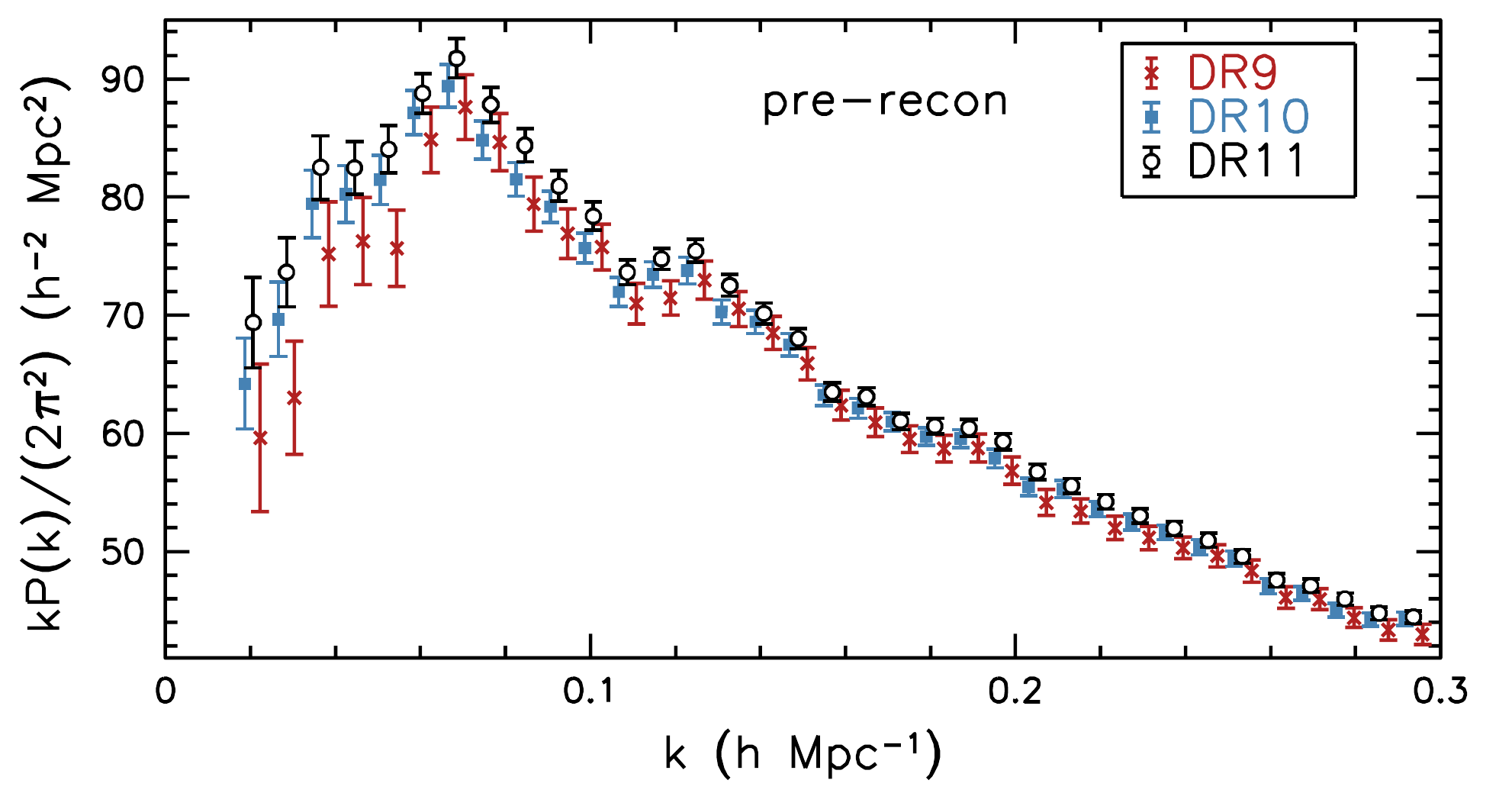}}
  \caption{Top panel: The measured monopole of the CMASS galaxy
    correlation function, multiplied by the square of the scale, $s$,
    for each of the BOSS data releases. These figures are shown pre-reconstruction. For clarity, the DR10 data have
    been shifted horizontally by $+1\mpcoh$ and the DR9 data by $-1\mpcoh$. Bottom
    panel: The measured spherically averaged CMASS galaxy power
    spectrum, multiplied by the frequency scale, $k$, for each of the
    BOSS data releases. For clarity, the DR9 data have been shifted by
    $+0.002\hompc$ and the DR10 data by $-0.002\hompc$. All of the error-bars shown in both panels represent the diagonal elements of the covariance matrix determined from the mocks. One can
    observe broadly consistent clustering, especially in the overall shape of each
    curve.}
\label{fig:xipkrel}
\end{figure*}

In the previous sections, we detailed our analysis techniques and demonstrated
they recover un-biased estimates of the BAO scale. We now apply our results to 
the BOSS data. We present our isotropic measurements in this section and our
anisotropic results in the following section.

The configuration space and Fourier space clustering measurements made
from the DR10 and DR11 CMASS samples are presented in
Fig.~\ref{fig:xipkrel} for $\xi(s)$ and $P(k)$, using our fiducial
binning choice. These points are compared against the DR9 clustering
results\footnote{We recalculate the DR9 $P(k)$ using the new method
  presented in Section~\ref{sec:method} for consistency.} presented in
\citet{And12}. For both $P(k)$ and $\xi(s)$, there are variations in the power observed in the different data sets, 
but the shapes of each are clearly consistent, suggesting
that we should expect to recover consistent results for the BAO scale. Measurements of the clustering in the LOWZ sample are presented in \cite{Tojlowz}.

The power is observed
to increase with each data release, and similar behaviour is observed in the correlation function for $s < 70 h^{-1}$Mpc. 
The difference in clustering amplitude can be explained by the tiling of the survey. In order to obtain the most complete sample,
 dense regions are observed using overlapping plates. Thus, as the survey progresses, a larger
percentage of observations using overlapping plates are completed and the mean density of the
survey increases. This increase in density occurs almost exclusively by adding over-dense regions and thus
increases the clustering amplitude. The measured increase in clustering amplitude is roughly the square of 
increase in density (4 per cent between DR9 and DR11 and 2 per cent between DR10 and DR11). As the survey
nears completion, the issue naturally becomes less important. For DR11, it represents, at worst, a 1 per cent
underestimate of the bias of the CMASS galaxies. Consistent trends are found in the LOWZ sample \citep{Tojlowz}.


\begin{table}
\centering
\caption{Isotropic BAO scale measurements recovered from BOSS
  data. The ``combined'' results are the optimally combined post-reconstruction
  $\alpha$ measurements across multiple bin centre choices, based on
  the correlation matrix obtained from the mock samples. The
  $P(k)$+$\xi(s)$ measurements are the mean of these combined results,
  with an uncertainty calculated as described in the text. The quoted errors are statistical only, except for
  the `Consensus'' measurements, where a systematic uncertainty has been included. This estimated systematic error is discussed in Section~\ref{sec:sys_err}.}
\begin{tabular}{llcc}
\hline
\hline
Estimator  &   ~~~~~~~~~~~~~~~$\alpha$ & $\chi^2$/dof \\
\hline
DR11 CMASS & & &  \\
{\bf Consensus} ${\bf z=0.57}$ & ${\bf 1.0144\pm0.0098}$~{\bf (stat+sys)}& \\
$P(k)$+$\xi(s)$ & $1.0144\pm0.0089$~(stat)& \\
combined $P(k)$ & $1.0110\pm0.0093$ & \\
combined $\xi(s)$ & $1.0178\pm0.0089$ & \\
post-recon $P(k)$ & $1.0114\pm0.0093$ &18/27   \\
post-recon $\xi_0(s)$ & $1.0209\pm0.0091$ & 16/17  \\
pre-recon $P(k)$ & $1.025\pm0.015$ & 33/27   \\
pre-recon  $\xi_0(s)$ & $1.031\pm0.013$ & 14/17  \\
\hline
DR10 CMASS & & &  \\
Consensus & $1.014\pm0.014$~(stat+sys) \\
post-recon $P(k)$ & $1.007\pm0.013$ & 23/28    \\ 
post-recon $\xi_0(s)$ & $1.022\pm0.013$ & 14/17   \\
pre-recon $P(k)$ & $1.023\pm0.019$ & 35/28   \\
pre-recon $\xi_0(s)$ & $1.022\pm0.017$ & 16/17  \\
\hline
DR9 CMASS & & &  \\
Consensus & $1.033\pm0.017$ \\ 
\hline
DR11 LOWZ & & &  \\
{\bf Consensus} ${\bf z=0.32}$ & ${\bf 1.018\pm0.021}$~{\bf (stat+sys)} & \\  
$P(k)$+$\xi(s)$ & $1.018\pm0.020$~(stat)& \\
\hline
DR10 LOWZ \\
Consensus & $1.027\pm0.029$~(stat+sys)\\
\hline

\label{tab:isobaoresults}
\end{tabular}
\end{table}

\begin{figure*}
\centering
  \resizebox{0.8\textwidth}{!}{\includegraphics{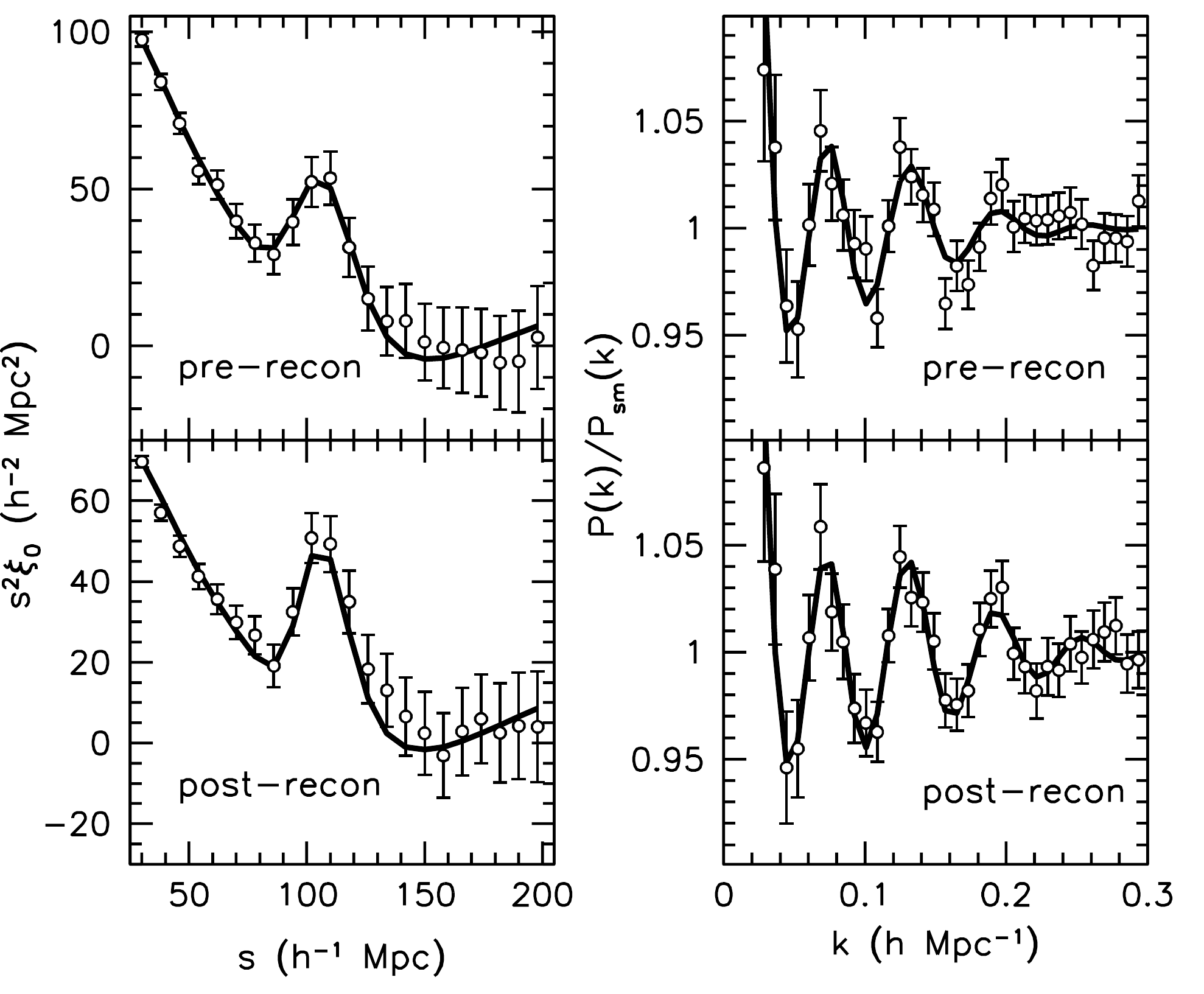}}
  \caption{ DR11 CMASS clustering measurements (black circles) with
    $\xi(s)$ shown in the left panels and $P(k)$ in the right
    panels. The top panels show the measurements prior to
    reconstruction and the bottom panels show the measurements after
    reconstruction. The solid lines show the best-fit BAO model in each case. One
    can see that reconstruction has sharpened the acoustic feature considerably
    for both $\xi(s)$ and $P(k)$. }
\label{fig:CMASSpkxi}
\end{figure*}

Fig.~\ref{fig:CMASSpkxi} displays the best-fit BAO model (solid
curves) compared to the data for $\xi(s)$ (left panels) and $P(k)$
(right panels) for DR11 only. The pre-reconstruction measurements  are
displayed in the top panels, and the post-reconstruction ones in the
bottom panels. The measurements are presented for our fiducial binning
width and centring, and show a clear BAO feature in both $P(k)$ and
$\xi(s)$, with the best-fit models providing a good fit. The effect of
reconstruction is clear for both the correlation function and power
spectrum, with the BAO signature becoming more pronounced relative to
the smooth shape of the measurements. Indeed, all of the BAO
measurements, listed in Table \ref{tab:isobaoresults}, have improved
post-reconstruction, in contrast to our DR9 results
\citep{And12}. This behaviour is expected given the results of
Section~\ref{sec:isomocktest}, which showed that, given the precision
afforded by the DR11 volume coverage, reconstruction improved the
results from all of our mock catalogues. Reconstruction is
particularly striking in the power spectrum plot, showing a clear 
 third peak in the post-reconstruction $P(k)$.

\subsection{DR11 Acoustic Scale Measurements}  \label{sec:dr11_measurement}

\begin{figure*}
  \centering
  \resizebox{0.8\textwidth}{!}{\includegraphics{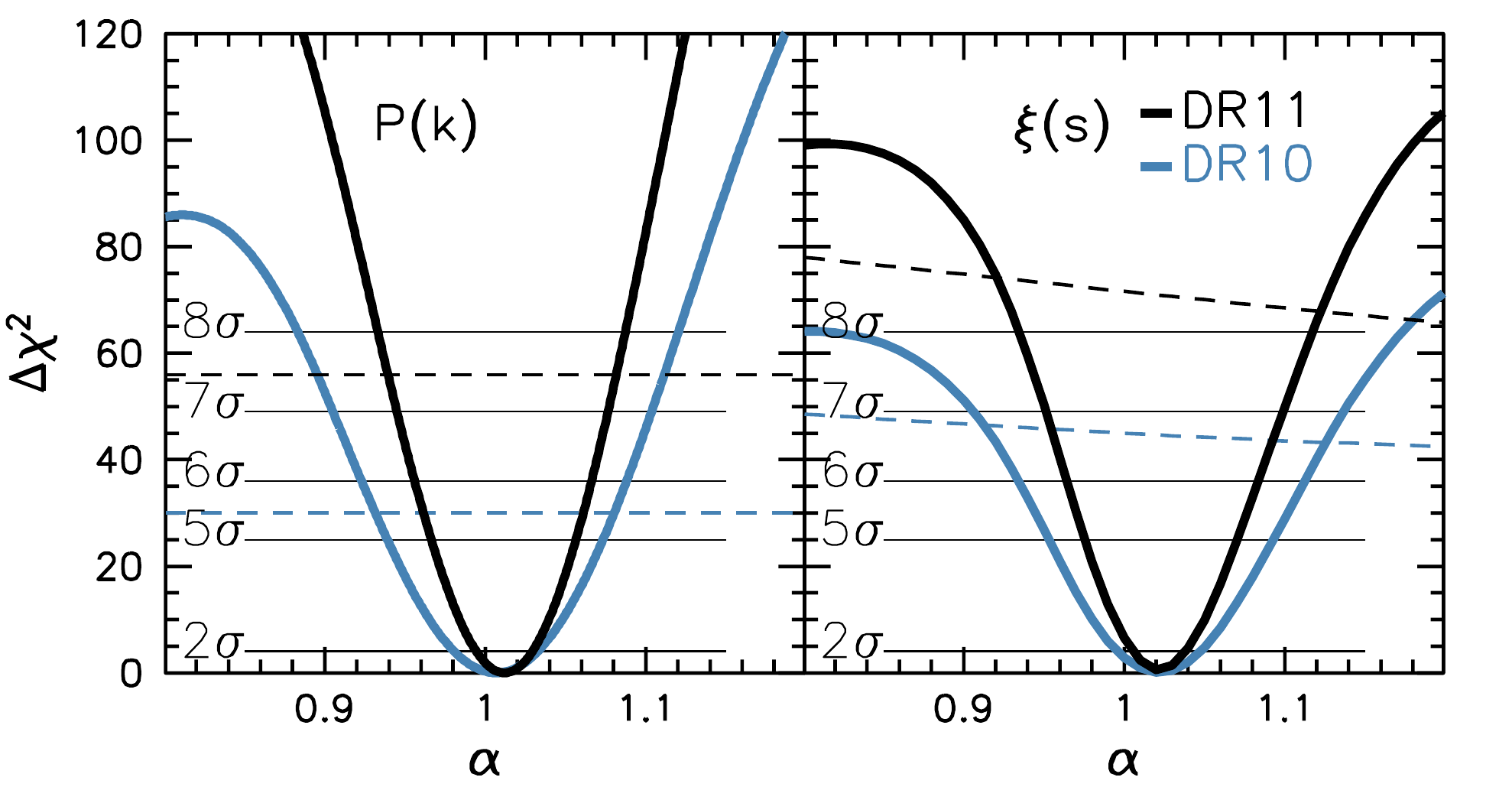}}
  \caption{Plot of $\chi^2$ vs.~$\alpha$, for reconstructed data from DR10 (blue), and DR11 (black) data,
    for $P(k)$ (left) and $\xi(s)$ (right). The dashed lines display
    the $\chi^2$ for a model without BAO, which we compute by setting
    $\Sigma_{NL}\rightarrow \infty$ in 
    Eqs.~(\ref{eq:mod_pk}) and (\ref{eqn:template}).
    In the $\xi(s)$ case, this limiting template still depends on $\alpha$,
    so the $\chi^2(\alpha)$ is not constant.  Our $P(k)$ model has no
    dependence on $\alpha$ in this limit.
    The DR11 detection significance
    is greater than 7$\sigma$ for $P(k)$ and 8$\sigma$ for $\xi(s)$.}
\label{fig:baosig}
\end{figure*}

 Our BAO measurements are listed in Table~\ref{tab:isobaoresults}. The
mocks for DR10 and DR11 show significant improvement with
reconstruction in most realisations, and we therefore adopt the reconstruction results
as our default measurements. Our consensus value for the CMASS BAO measurement,
$\alpha=1.0144\pm0.0089$, is determined from a combination of $P(k)$ and $\xi(s)$
measurements, and in what follows we describe the process of obtaining this value, and tests that validate it. 

Post-reconstruction, the significance of the
BAO detection in both the correlation function and the power spectrum are
greater than 7$\sigma$ for the reconstructed DR11 CMASS BAO
measurements. The significance of detection is shown in Fig.~\ref{fig:baosig}, where we
also see a difference in the detection significance between results from $\xi(s)$ and $P(k)$. This
variation is caused by the differential ability of the models for the broad-band
component to match the offset between the data and the no-baryon
model. The broad-band model for the power spectrum has more free
parameters than that for the correlation function, so it is perhaps
not surprising that the no-baryon model is a slightly better fit.

Table~\ref{tab:isobaoresults} also lists $\chi^2$/dof for the best-fit
models, showing that they are close to unity for DR10 and DR11 fits
using both the correlation function and power spectrum. The most unusual
is the $\chi^2/$dof $=18/27$ for the post-reconstruction DR11 $P(k)$
measurement. Such a low $\chi^2$ is expected in 10 per cent of
cases, thus we conclude that our best-fit models provide adequate
descriptions of the data.

The precision of the BAO measurements are typical of those achieved in
the mock samples. This consistency in shown in the top panels of
Fig.~\ref{fig:reconcomcmass}, where the orange stars show the
uncertainty in the data post-reconstruction versus the uncertainty 
pre-reconstruction. All of the CMASS data points lie within well the distribution of the
mock points. The most discrepant result is for the DR10 $P(k)$
measurement post reconstruction; it has an uncertainty of 0.014, while
the mean uncertainty from the mock realisations is 0.011, but one can see that many
mock realisations recover an uncertainty larger than 0.014.

\begin{table}
\caption{BAO scale measurements for DR11 reconstructed data using
  different bin centres. These results are combined using their
  correlation matrix to obtain optimised BAO measurements.}
\centering
\begin{tabular}{llcc}
\hline
\hline
{ ~~~~~~~~~~Shift}  &   { ~~~~~~~~~~~$\alpha$} & $\chi^2$/dof\\
\hline
$P(k)$ & \\
$\Delta k_i=0$ & $1.0115\pm0.0093$ & 18/27\\
$\Delta k_i=0.0008\hompc$ & $1.0113\pm0.0094$ & 19/27\\
$\Delta k_i=0.0016\hompc$ & $1.0101\pm0.0096$ & 21/27\\
$\Delta k_i=0.0024\hompc$ & $1.0097\pm0.0097$ & 21/27\\
$\Delta k_i=0.0032\hompc$ & $1.0103\pm0.0095$ & 20/27\\
$\Delta k_i=0.004\hompc$ & $1.0111\pm 0.0094$ & 19/27\\
$\Delta k_i=0.0048\hompc$& $1.0115\pm0.0094$ & 18/27\\
$\Delta k_i=0.0056\hompc$ & $1.0119\pm 0.0093$ & 16/27\\
$\Delta k_i=0.0064\hompc$ & $1.0125\pm0.0092$ & 16/27\\
$\Delta k_i=0.0072\hompc$ & $1.0122\pm 0.0092$ & 17/27\\
\hline
 $\xi(s)$ &  \\
  $\Delta s_i=-2\mpcoh$ & $1.0188\pm0.0104$ & 12/17\\
$\Delta s_i=-1\mpcoh$ & $1.0154\pm0.0094$ & 8/17\\
$\Delta s_i=0$  &  $1.0209\pm0.0091$ & 16/17  \\
$\Delta s_i =+1\mpcoh$ & $1.0186\pm0.0086$ & 14/17\\
 $\Delta s_i=+2\mpcoh$ & $1.0201\pm0.0087$ & 16/17\\
$\Delta s_i=+3\mpcoh$ & $1.0164\pm0.0087$& 19/17\\
$\Delta s_i=+4\mpcoh$ & $1.0153\pm0.0092$ & 17/17\\
$\Delta s_i=+5\mpcoh$ & $1.0191\pm0.0100$ & 13/17\\
\hline
\label{tab:binshift}
\end{tabular}
\end{table}

We combine the DR11 CMASS $\xi(s)$ BAO measurements using eight bin centres and
the $P(k)$ results using ten bin centres in the same manner as applied
to the mocks, as described in Section~\ref{sec:combiso}. The
individual fits determined for different bin centres are shown in
Table~\ref{tab:binshift}. For $\xi(s)$, our fiducial
choice recovered the largest $\alpha$ of any of the bin centres. Thus,
when combining the results across all of the bin centre choices,
$\alpha$ decreases to $1.0178\pm0.0089$. The uncertainty has decreased
by only 2 per cent (compared to the mean of 7 per cent found for the
mocks) in part because the estimated uncertainty of the fiducial bin
choice (0.0091) is less than the weighted mean uncertainty across all
of the bin choices (0.0092). For $P(k)$, the result changes little when we
combine across the results of the 10 bin centre choices; it changes from
1.0114$\pm$0.0093 to 1.0110$\pm$0.0093.

We obtain a BAO measurement with an expected error measured from the
likelihood surface that is less than 1 per cent for both the
reconstructed $\xi(s)$ and $P(k)$. The difference between the two values of $\alpha$ is
0.0068. While small in magnitude, this difference is unexpectedly
large in the context of the mock results, for which we found a
correlation factor of 0.95 between the $P(k)$ and $\xi(s)$ results
combined across all of the bin choices. Accounting for this
correlation factor, the expected 1$\sigma$ dispersion in the $P(k)$
and $\xi(s)$ measurements is
$(\sigma^2_{\alpha,P}+\sigma^2_{\alpha,\xi}-2C_{P,\xi}\sigma_{\alpha,P}\sigma_{\alpha,\xi})^{\frac{1}{2}}
= 0.0028$. The discrepancy in the data is thus 2.4$\sigma$.  Comparing
$|\alpha_{P}-\alpha_{\xi}|/(\sigma^2_{\alpha,P}+\sigma^2_{\alpha,\xi})^{\frac{1}{2}}$
to the results from the mocks, we find 7 (1.2 per cent) that have a
larger deviation, consistent with our estimation of a 2.4$\sigma$ discrepancy. Both estimates of
$\alpha$ are stable to a variety of robustness tests, as we will show in Section \ref{sec:isorobust}, and our
tests on mock samples demonstrate that each estimator is unbiased. We therefore conclude that, despite being
unusual, the difference between the two measurements is not indicative of an existence of a bias in
either measurement.

Our tests on mocks suggest no systematic effects for either the
$P(k)$ or $\xi(s)$ results when they are obtained by combining results
across bin centres. Our methodology applied to mock samples recovers
unbiased estimates of the BAO position for both $\xi(s)$ and $P(k)$
with nearly identical uncertainty. We therefore obtain the consensus
BAO scale measurement by assuming the mean uncertainty of the $\xi(s)$
and $P(k)$ measurements for each and using the 0.95 correlation
factor. The correct treatment of the data, assuming Gaussian statistics and no systematic
uncertainty is to take the mean of $P(k)$ and $\xi(s)$ measurements, reducing the uncertainty
based on their correlation factor. Thus, our consensus value for the CMASS BAO measurement is
$\alpha=1.0144\pm0.0089$, where this uncertainty is purely statistical. Our systematic error
budget is discussed in Section \ref{sec:sys_err}.

We obtain our consensus DR11 LOWZ isotropic BAO measurement, at an effective
redshift $z=0.32$ by applying the same process as applied to CMASS. The details 
can be found in \cite{Tojlowz}. The difference in the recovered BAO scale from LOWZ $P(k)$
and $\xi(s)$ is within $1\sigma$ of the expected difference and is opposite in sign to
the difference we find for CMASS. The consensus DR11 LOWZ measurement is
$\alpha = 1.018\pm0.020$, considering only the statistical uncertainty.

\subsection{DR10 BAO measurements}

For completeness, we also include DR10 BAO measurements in
Table~\ref{tab:isobaoresults}. Post-reconstruction, these data produce a 1.4
per cent BAO scale measurement that is consistent with the DR11
measurements discussed in the previous section. For pre-reconstruction
measurements the error on DR11 the result is 30 per cent lower than for DR10.
For the post-reconstruction results, the improvement increases to 40 per cent. The
 reconstruction is more efficient for DR11, which almost
certainly results from the more contiguous nature of the DR11 survey
mask.

As shown in Fig.~\ref{fig:baosig}, the detections for DR10 are both
greater than 5$\sigma$, with the significance for the $\xi(s)$
measurement being higher than that of the $P(k)$ measurement. As
discussed in Section~\ref{sec:dr11_measurement} the improved detection
observed in $\xi(s)$ is because the
$P(k)$ broad-band model is better able to model the full $P(k)$
when no BAO are included, compared with the broad-band $\xi(s)$ model.

The most obvious issue for the DR10 results in
Table~\ref{tab:isobaoresults} is that, for the DR10 $P(k)$, the
measurement of $\alpha$ shifts by $-0.020$ post-reconstruction, compared
to a mean shift of $-0.004\pm0.015$ observed in the mocks (here the
uncertainty is the standard deviation of the mock values). The size of
this shift is thus only just greater than 1$\sigma$ and is
consequently not a significant concern.

\subsection{DR11 Robustness Checks}
\label{sec:isorobust}

\begin{table}
\caption{Robustness checks on isotropic BAO scale measurements
  recovered from DR11 reconstructed data. }
\centering
\begin{tabular}{llcc}
\hline
\hline
Estimator & Change  &   $\alpha$ & $\chi^2$/dof\\
\hline

$P(k)$ & fiducial & $1.0114\pm0.0093$ & 18/27\\
& NGC only & $1.0007\pm0.0113$ & 16/27\\
& SGC only & $1.0367\pm0.0167$ & 15/27 \smallskip \\ 
& $0.02<k<0.25\hompc$ & $1.0082\pm0.0094$ & 14/21\\
& $0.02<k<0.2\hompc$ & $1.0121\pm0.0113$ & 11/15\\
& $0.05<k<0.3\hompc$ & $1.0120\pm0.0091$ & 15/23 \smallskip\\
& $\Sigma_{nl} = 3.6\pm0.0h^{-1}$Mpc & $1.0111\pm0.0085$ & 19/28\\  
& $\Sigma_{nl} = 4.6\pm0.0h^{-1}$Mpc & $1.0119\pm0.0089$ & 19/28\\  
& $\Sigma_{nl} = 5.6\pm0.0h^{-1}$Mpc & $1.0116\pm0.0097$ & 18/28\\  
& $A_1, A_2 = 0$ & $1.0136\pm0.0095$ & 40/29\\
& Spline fit & $1.0109\pm0.0094$ & 17/24\smallskip\\
& $\Delta k = 0.0032\hompc$ & $1.0122\pm0.0097$ & 71/79\\
 & $\Delta k = 0.004\hompc$ & $1.0082\pm0.0094$ & 55/62\\
 & $\Delta k = 0.006\hompc$ & $1.0091\pm0.0096$ & 33/39\\
 & $\Delta k = 0.01\hompc$ & $1.0120\pm0.0097$ & 16/20\\
 & $\Delta k = 0.012\hompc$ & $1.0133\pm0.0091$ & 9/15\\
 & $\Delta k = 0.016\hompc$ & $1.0100\pm0.0099$ & 5/9\\
 & $\Delta k = 0.02\hompc$ & $1.0186\pm0.0105$ & 5/6\\
 \hline
$\xi(s)$ & fiducial & $1.0209\pm0.0091$ & 16/17\\
& NGC only & $1.0132\pm0.0105$ & 12/17\\
& SGC only & $1.0441\pm0.0190$ & 15/17\smallskip\\
 &  $50 < s < 150\mpcoh$ & $1.0208\pm0.0094$ & 6/7\\
 & $a_1, a_2, a_3 = 0$ & $1.0210\pm0.0097$ & 24/20\\
 & $a_1, a_2 = 0$ & $1.0232\pm0.0098$ & 19/19\\
 & $a_1 = 0$ & $1.0231\pm0.0099$ & 19/18\\
 & $a_2 = 0$ & $1.0218\pm0.0097$ & 18/18\\
 & $B_\xi$ free & $1.0209\pm0.0091$ & 15/17\\
& $\Sigma_{nl} = 3.6\mpcoh$ & $1.0212\pm0.0089$ & 15/17\\ 
& $\Sigma_{nl} = 5.6\mpcoh$ & $1.0206\pm0.0095$ & 17/17\smallskip\\
&  recon $\beta = 0.318$ & $1.0195\pm0.0090$ & 11/17\\
& recon $\beta = 0.478$ & $1.0206\pm0.0094$ & 18/17\\
& recon $b = 1.50$ & $1.0224\pm0.0100$ & 23/17 \\
& recon $b = 2.24$ & $1.0183\pm0.0086$ & 14/17 \smallskip\\
 &  $\Delta s = 4\mpcoh$ & $1.0197\pm0.0090$ & 42/38\\
&  $\Delta s = 5\mpcoh$ & $1.0156\pm0.0093$ & 31/29\\
&  $\Delta s = 6\mpcoh$ & $1.0189\pm0.0093$ & 19/23\\
&  $\Delta s = 7\mpcoh$ & $1.0165\pm0.0088$ & 20/19\\
&  $\Delta s = 9\mpcoh$ & $1.0188\pm0.0089$ & 10/14\\
&  $\Delta s = 10\mpcoh$ & $1.0175\pm0.0099$ & 9/12 \\
\hline
\label{tab:isorobust}
\end{tabular}
\end{table}

In order to ensure that our measurements on the CMASS data are robust
to our methodological and binning choices, we re-measure the BAO scale
using the reconstructed DR11 power spectrum and correlation function,
changing the fitting methods, binning and fitting to the NGC and SGC
separately. Table~\ref{tab:isorobust} lists the results of these tests.

The absolute difference in the $\alpha$ values recovered from the NGC
and SGC regions has decreased considerably from \cite{And12}. For the
correlations function fits, the decrease if from $0.055$ to
$0.031$. Given the decrease in the uncertainty thanks to the larger
area coverage in both regions, the significance of the discrepancy is
similar to that found for DR9, 1.4$\sigma$. We find 79 out of the 600
mock samples (13 per cent) have a larger discrepancy, consistent with
the estimation of a 1.4$\sigma$ discrepancy.  We find a similar
picture when we fit to the $P(k)$ measurements from NGC and SGC
although, in this case, the discrepancy is slightly larger, at
1.8$\sigma$. Less significant differences, with opposite sign, are
found in the DR11 LOWZ sample \citep{Tojlowz}.

Table~\ref{tab:isorobust} also presents results fitting to the power
spectrum for different ranges in $k$, removing the largest and
smallest-scale data in turn. The recovered errors on
$\alpha$ do not change significantly if we remove data at
$k<0.05\hompc$ or at $k>0.25\hompc$. This is not surprising, given there is little BAO
signal on these scales. Only fitting to $0.02<k<0.25\hompc$ reduces
the best-fit value of $\alpha$ by $0.0039$, but cutting further in $k$
to $0.02<k<0.2\hompc$ returns the best fit back to the fiducial
value, suggesting that there is no wavelength-dependent systematic trend
present. Fixing the BAO damping at the best-fit value from the mocks
$\Sigma_{nl}=4.6h^{-1}$Mpc does not alter the best-fit value of $\alpha$, but
does decrease the size of the error, but we consider this action to be too
aggressive given that the true value of the damping is
unknown. Changing $\Sigma_{nl}$ by $\pm1h^{-1}$Mpc does not have a
large effect, although overdamping the BAO in the model
does increase the error on $\alpha$, as it removes the signal we wish
to match to the data.

Results from applying two alternatives to the model for the broad-band
power spectrum shape are also shown: Cutting the polynomial model back
to a 4-parameter model by setting $A_1=0$ and $A_2=0$ in
Eq. \ref{eq:mod_pksm} only slightly affects $\alpha$ and the recovered
error, but does significantly increase the best-fit value of $\chi^2$,
showing that this model inadequately describes the shape of the
power. Changing to the bicubic spline broad-band model used previously
\citep{And12} does not significantly affect either the best-fit value
of $\alpha$ or the recovered error.

Table~\ref{tab:isorobust} also presents results reducing the range of
scales fitted in the correlations function from $28<s<200\mpcoh$ to
$50<s<150\mpcoh$: we find a negligible change in the best-fit value of
$\alpha$, and a revised error that only increases by a small amount,
demonstrating that this reduced range of scales contains all of the
BAO signal as expected. We also present results from possible changes
to the model used to fit the broad-band correlation function, where we
remove various polynomial terms, or remove the prior on $B_\xi$
(Eq.~\ref{eqn:fform}). The greatest change is an increase in the
recovered $\alpha$ value of 0.0023 when only the constant $a_3$ term
is used to fit $\xi(s)$ (Eq.~\ref{eq:xia}). Indeed, for $\xi(s)$, the
preference for the inclusion of a polynomial is not strong; $\Delta
\chi^2= 8$ for three additional parameters. While the correlation
function adds terms to a full linear model (Eq.~\ref{eqn:fform}), the
power spectrum only includes the BAO component (Eq.~\ref{eq:mod_pk}),
which is why the polynomial term is less important for $\xi(s)$. As we
did for the power spectrum, we vary the non-linear BAO damping,
finding consistent results.

The reconstruction algorithm requires an assumed amplitude
for the real-space clustering of the galaxy field ($b$) and its
associated velocity field ($\beta$). In the fiducial case, we assume
$b=1.87$ and $\beta= 0.398$, which are measured from the
mocks. However, our results are not sensitive to these assumptions: if
we change each by $\pm20$ per cent and re-calculate the reconstructed
field for the DR11 data and re-determine $\xi(s)$, the resulting
measurements of $\alpha$ show negligible change. 

For both the power spectrum and correlation function,
Table~\ref{tab:isorobust} also presents results where we change the bin
size, revealing significant scatter. The equivalent comparison for the
mock catalogues was presented in \citet{Per13}. For both $P(k)$ and
$\xi(s)$ measured from the data, a dispersion of 0.002 is found in the
best-fit $\alpha$ values. The weighted mean across bin sizes
(accounting for the covariance between bins) is 1.0180$\pm$0.0089 for
$\xi(s)$ and 1.0117$\pm$0.0091 for $P(k)$. These measurements are 
similar to the results obtained when combining across bin centres,
suggesting the combined bin centre results largely capture the same
information as changing the bin size. The $\Delta k = 0.02\hompc$ bin
size recovers $\alpha = 1.0186\pm0.0105$. While this value is significantly
larger than any of the other bin sizes, this bin size has a relatively
small correlation factor, 0.8, with the weighted mean of the other bin
sizes. It is thus only 1.2$\sigma$ from the BAO fit to $P(k)$
averaged into narrower bins.

For all of the tests presented in this section, we find no evidence
for changes in the best-fit value of $\alpha$ that are sufficiently
outside of the statistical expectation to indicate the presence of systematic effects. The most significant discrepancy we have
observed is the different values of $\alpha$ recovered from $\xi(s)$
and $P(k)$, but the robustness checks presented in this section have
not pointed to any origin for this difference, other than simply it
being a $2.4\sigma$ statistical fluctuation.

\subsection{Displaying the BAO Feature}
\label{sec:dis_BAO}

\begin{figure}
\centerline{
\hspace*{0pt}\resizebox{0.62\columnwidth}{!}{\includegraphics{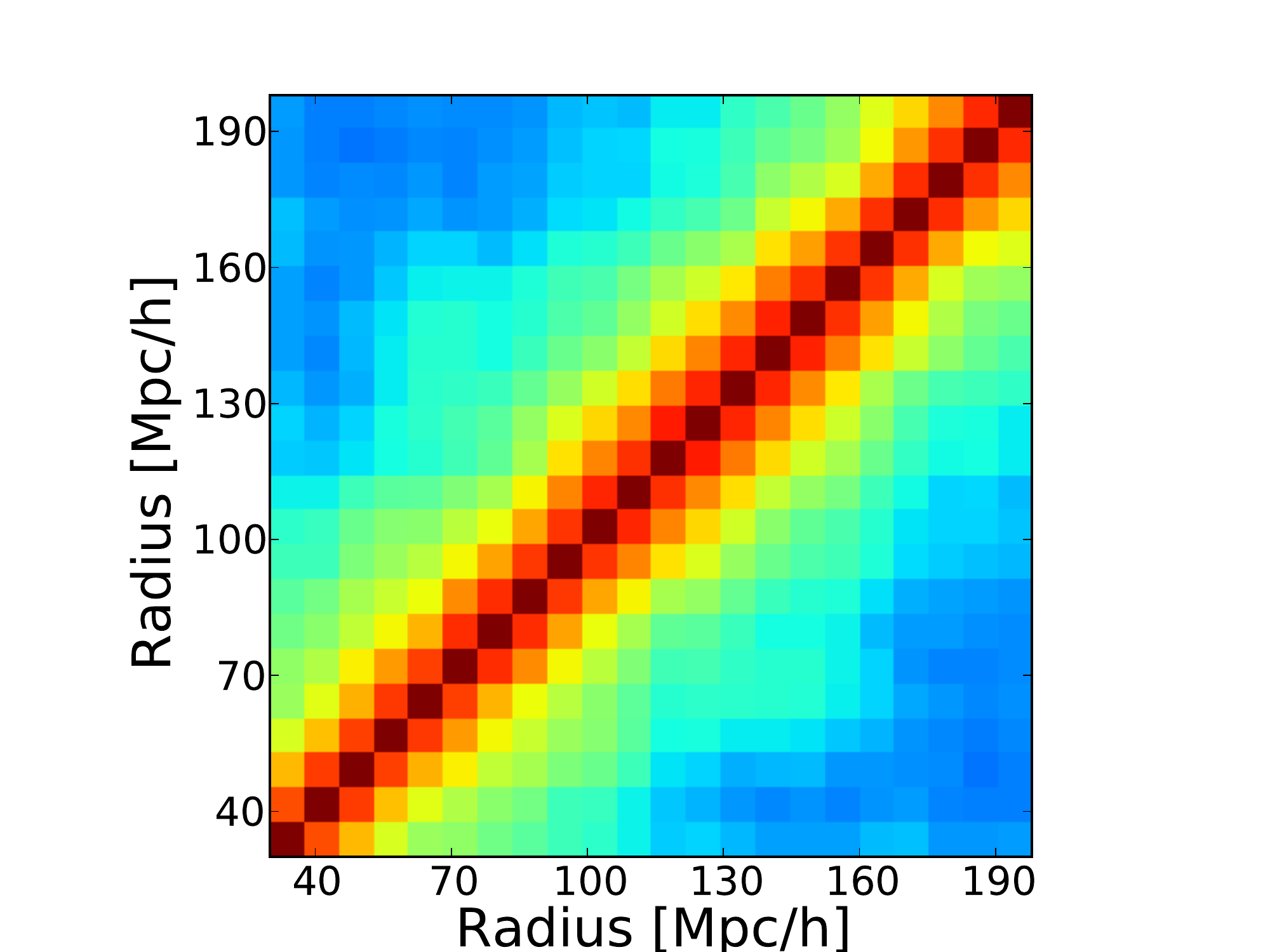}}
\hspace*{-30pt}\resizebox{0.62\columnwidth}{!}{\includegraphics{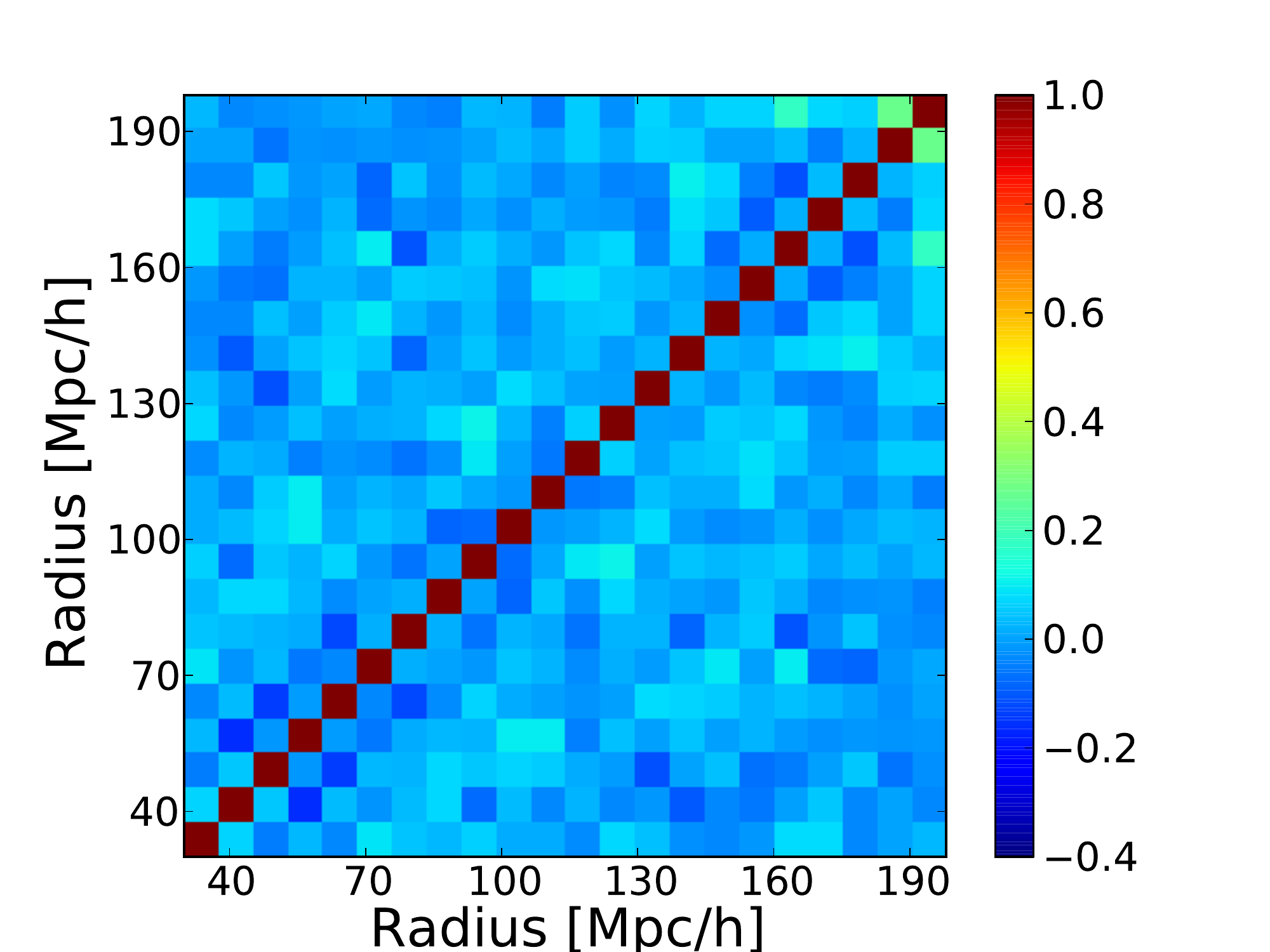}}
}
\caption{\label{fig:Xcov}The reduced covariance matrix of $\xi(r)$ ({left}) and 
$X(r)$ ({right}), for the analysis of the DR11 CMASS sample post-reconstruction.
One can see that the substantial correlations between separations in $\xi(r)$
have been largely cured in $X(r)$, save in the first two and last two bins
where the pentadiagonal transformation must be modified.
}
\end{figure}

\begin{figure}
\centering
\resizebox{0.99\columnwidth}{!}{\includegraphics{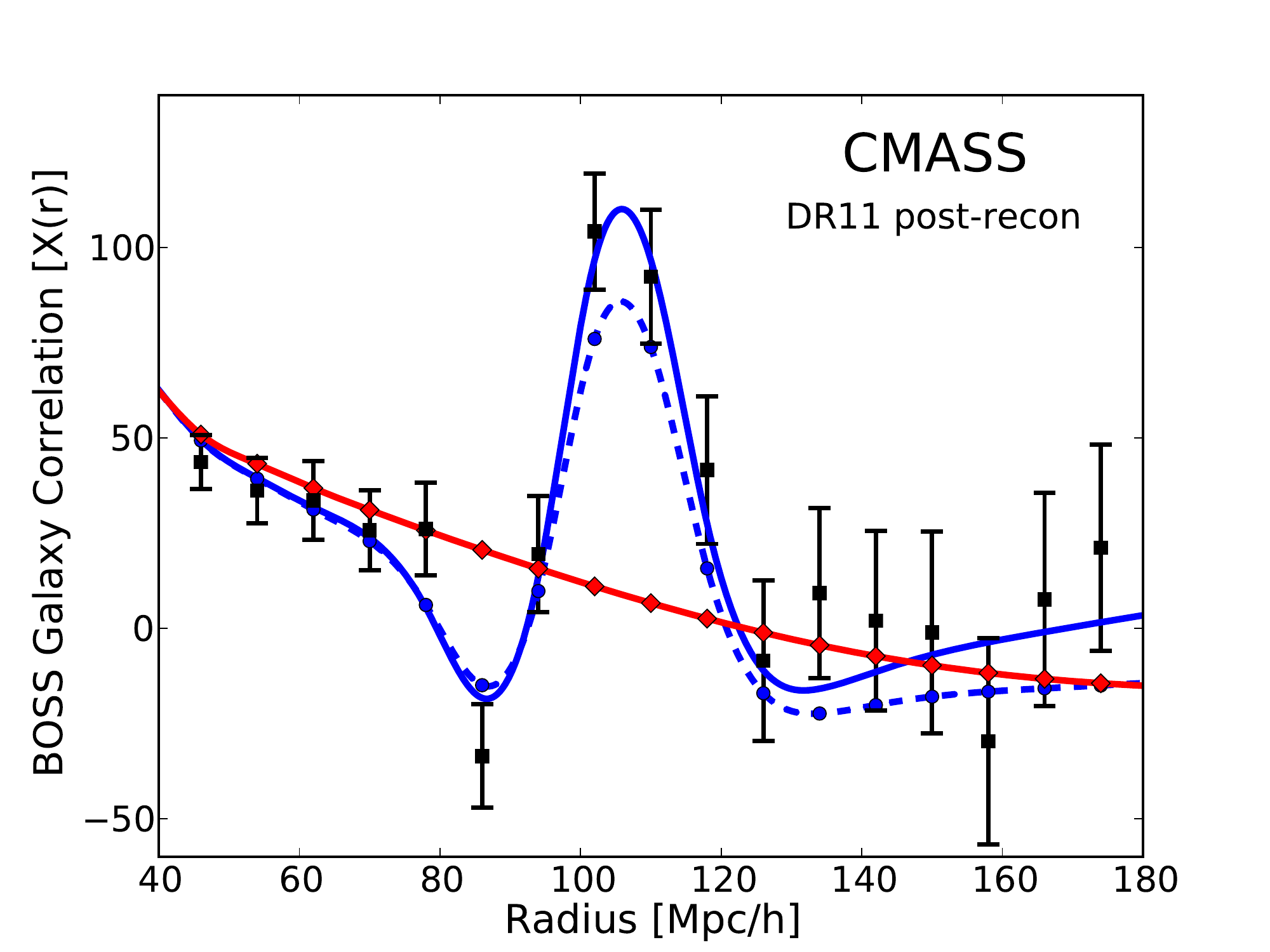}}
\caption{\label{fig:Xcmass}%
The DR11 CMASS correlation function, transformed as defined by 
Eq.~\ref{eq:X} with $a=0.30$ and $b=0.10$.  Unlike the usual correlation function, these error bars
are nearly independent.  The off-diagonal elements of the reduced
covariance matrix deviate from zero only by an rms of 5 per cent,
compared to 80 per cent covariance between neighboring bins of the original 
correlation function.  The blue solid line is the best-fit BAO model
with no marginalization of broadband terms; the dashed line marginalizes
over our standard quadratic polynomial. 
The red solid line is the best-fit
non-BAO model without marginalization; this model is rejected by 
$\Delta\chi^2\approx70$.  We note that since the transformation is defined
on the binned estimators, the models are formally not curves but 
simply predictions for the discrete estimators.  We plot those predictions
as the small dots; the curve is a spline connecting those dots.
}
\end{figure}

\begin{figure}
\centering
\resizebox{0.99\columnwidth}{!}{\includegraphics{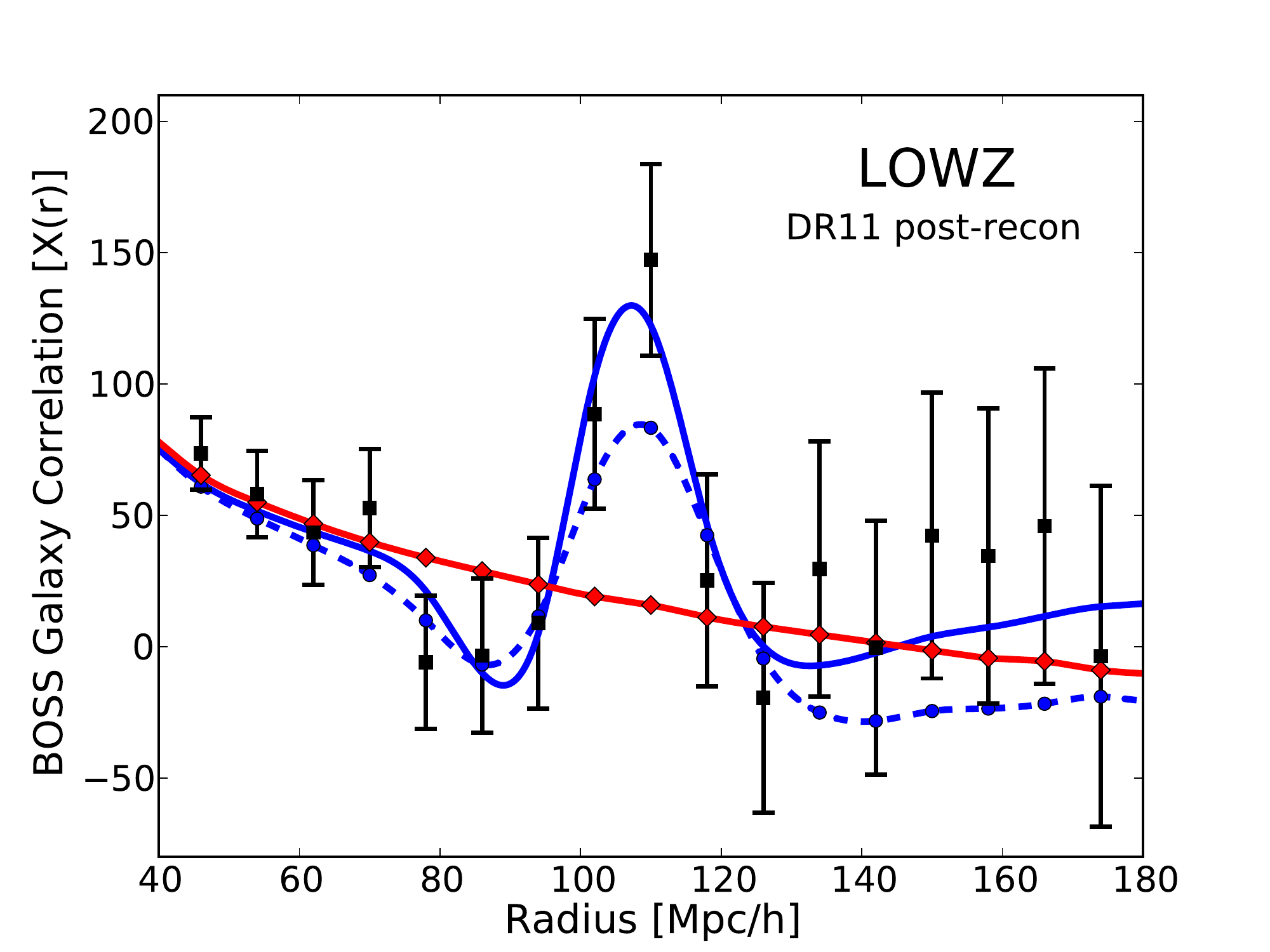}}
\caption{\label{fig:Xlowz}%
As Figure \ref{fig:Xlowz}, but for the DR11 LOWZ correlation function transformed as defined by 
Eq.~\ref{eq:X} with $a=0.39$ and $b=0.04$. As before, these error bars 
are nearly independent, with a worst case of 12 per cent and an r.m.s.~of
3.4 per cent in the off-diagonal elements of the reduced covariance matrix.
}
\end{figure}

\begin{figure}
\centering
\resizebox{0.99\columnwidth}{!}{\includegraphics{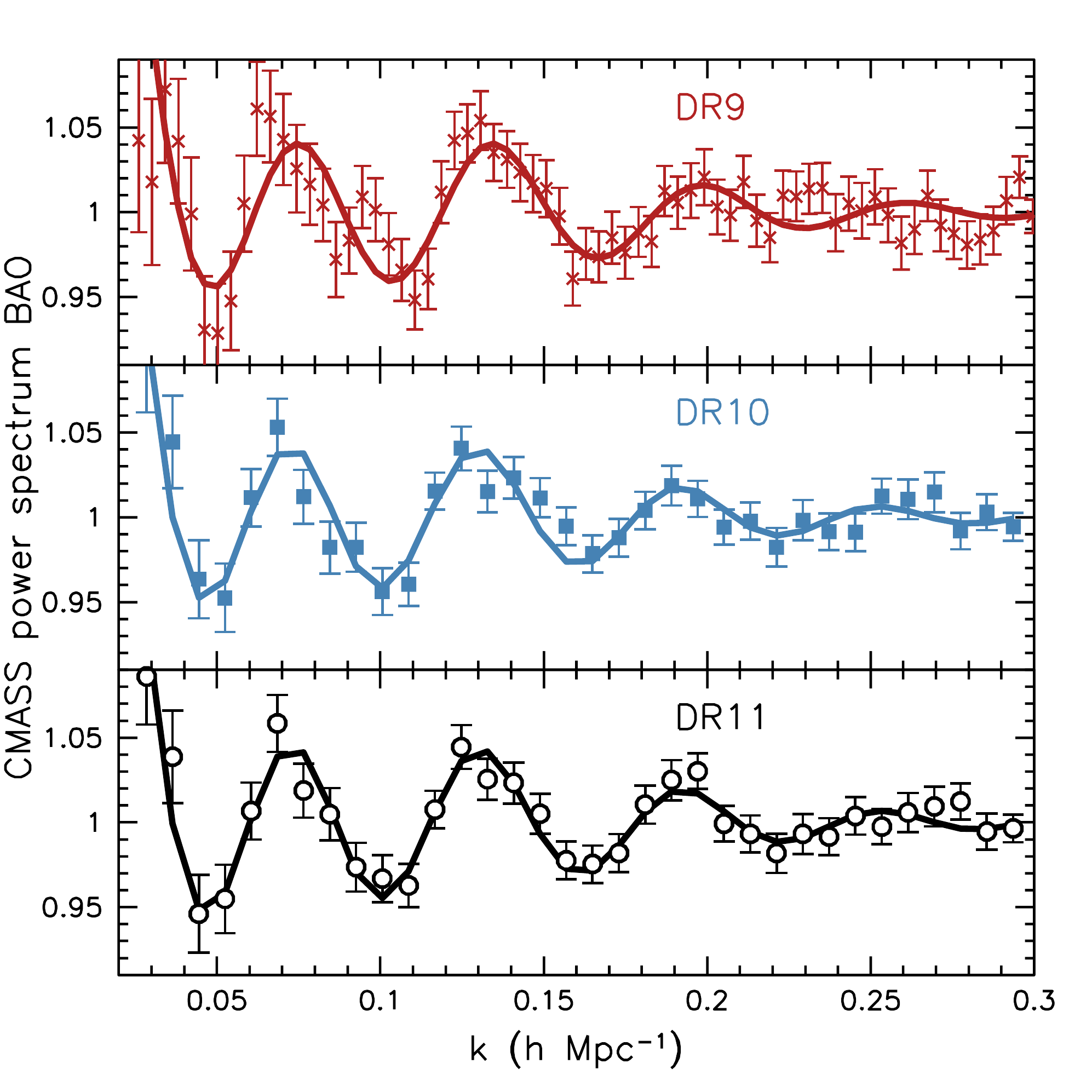}}
\caption{The CMASS BAO feature in the measured reconstructed power spectrum of each of the BOSS data releases, DR9, DR10, and DR11. The data are displayed with points and error-bars and the best-fit model is displayed with the curves. Both are divided by the best-fit smooth model.  We note that a finer binning was 
used in the DR9 analysis.}
\label{fig:PkBAOev}
\end{figure}

\begin{figure}
\centering
\resizebox{0.99\columnwidth}{!}{\includegraphics{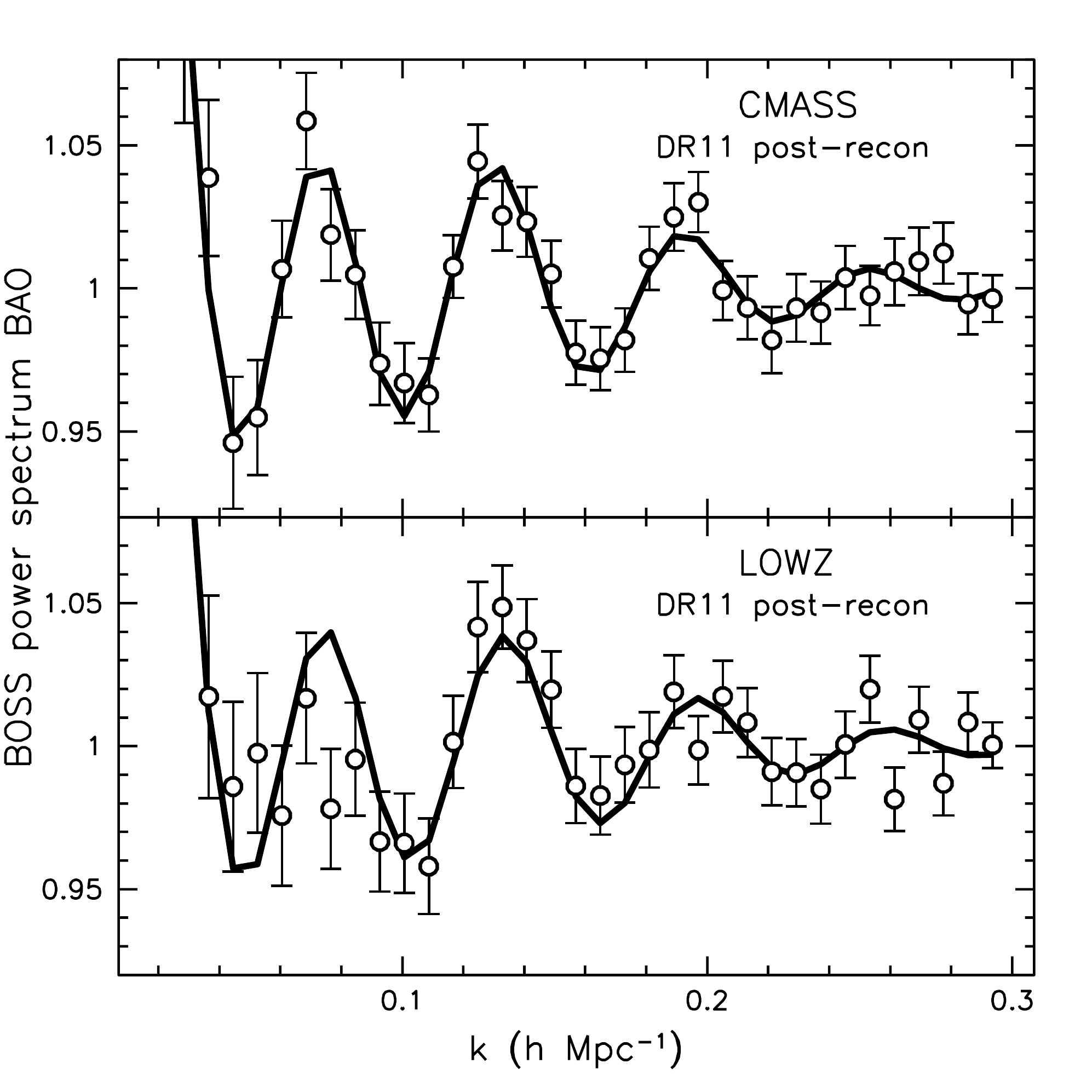}}
\caption{The BAO feature in the measured power spectrum of the DR11 reconstructed CMASS (top) and LOWZ (bottom) data. The data are displayed with black circles and the best-fit model is displayed with the curve. Both are divided by the best-fit smooth model.}
\label{fig:PkBAO}
\end{figure}

Plots of the two-point clustering statistics can be difficult to 
interpret because of the correlations between the data points.  This effect
is particularly severe for the correlation function: as the density of
the data set increases, different scales become heavily correlated.
For example, fluctuations in the amplitude in poorly constrained modes of very low 
wavenumbers cause the entire correlation function to shift up and down.
This means that the diagonal of the covariance matrix is a poor representation
of the actual uncertainties.  In the case of the acoustic peak, this 
leads to the data being {\it more} constraining than it appears!  
This effect is of no consequence for the formal analysis---one simply uses
the covariance matrix when fitting models---but it is a challenge for pedagogy.

The correlations of estimators can be avoided by adopting a new
basis, i.e., choosing new estimators that are linear combinations
of the original correlation function bins.  Such transformations are
extensively discussed in \citet{Ham00}.  There are an infinite
number of choices of bases that will produce diagonal covariance
matrices.  The pedagogical challenge is that the new estimators now
represent a mixture of all scales and hence it is not clear how to
plot the measurements.

Here, we present a hybrid approach in which one adopts a
simply-defined estimator with compact support as a function of scale,
but chooses the estimator so that the covariances are significantly 
suppressed.  In particular, \citet{Ham00} noted that transformations based 
on the symmetric square root of the Fisher matrix had surprisingly
compact support for their power spectrum analysis.  When we formed
this matrix for the DR11 CMASS correlation function, we found that 
the first and second off-diagonal terms are nearly constant and that
subsequent off-diagonals are small.  This suggests that a basis transform 
of the pentadiagonal form
\begin{equation}
X(s_i) = \frac{x_i-a\left(x_{i-1}+x_{i+1}\right)-b\left(x_{i-2}+x_{i+2}\right)}
	      {1-2a-2b} 
\label{eq:X}
\end{equation}
will approach a diagonal form.
Here, $x_i = s_i^2\xi_0(s_i)$ and $s_i$ is the bin center of measurement
bin $i$.
We introduce the $1-2a-2b$ factor so as to normalize $X$ such
that it returns $X=x$ for constant $x$.  For the first two and last two
bins, the terms beyond the end of the range are omitted and the normalization
adjusted accordingly. 

We find that for DR11 CMASS after reconstruction, values of $a=0.3$ and $b=0.1$
sharply reduce the covariances between the bins.  The reduced covariance
matrices for $\xi(r)$ and $X(r)$ are shown in Figure \ref{fig:Xcov}.
The bins near the edge 
of the range retain some covariances, but the off-diagonal terms of the
central $10\times10$ submatrix of the reduced covariance matrix have a
mean and r.m.s. of $0.008\pm0.044$, with a worst value of 0.11.  For display
purposes, this is a good approximation to a diagonal covariance matrix,
yet the definition of $X(s)$ is well localized and easy to state.
For comparison, the reduced covariance matrix of $s^2 \xi_0$ has typical 
first off-diagonals values of 0.8 and second off-diagonals values of 0.6.

We display this function in Figure \ref{fig:Xcmass}.  One must also transform
the theory to the new estimator: we show the best-fit BAO models with and
without broadband marginalization, as well as the best-fit non-BAO model
without broadband marginalization.
The presence of the BAO is clear, but now the error bars are 
representative.  For example, the significance of the detection as
measured by the $\Delta\chi^2$ of the best-fit BAO model to the best-fit
non-BAO model is 69.5 using only the diagonal of the covariance matrix
of $X$, as opposed to 74 with the full covariance matrix.  We do not 
use this transformation when fitting models, but we offer it as a pedagogical view.

The same result is shown for DR11 LOWZ post-reconstruction in Figure \ref{fig:Xlowz}.  Here we use $a=0.39$ and $b=0.04$.  The level of the off-diagonal 
terms is similarly reduced, with an r.m.s. of 3.4 per cent and a worst value of 12 per cent.

It is expected that the best values of $a$ and $b$ will depend on the data
set, since data with more shot noise will have covariance matrices of the
correlation function that are more diagonally dominant.  Similarly,
the choice of a pentadiagonal form may depend on the binning of the 
correlation function, as it likely reflects a physical scale of the 
covariances between bins.  However, some of the simplicity likely results
from the fact that the covariances between nearby bins 
are dominated by small-scale correlations in the density field that become 
independent of separation at large separation.  This property gives the matrix
a regularity: bins at 90 and $100\,$Mpc will be correlated to each 
other similarly to 
bins at 110 and $120\,$Mpc.  Tridiagonal matrices
have inverses with exponentially decreasing off-diagonal terms 
\citep{Ryb95}.  Apparently, treating the off-diagonal
covariances as exponentially decreasing with only weak dependences on
separation provides a good approximation.

For $P(k)$, the measurements in $k$-bins are already fairly
independent, as one would expect for a near-Gaussian random field. 
Correlations between bins can occur because of the finite survey
volume and because of non-Gaussianity in the density field.  For CMASS,
we find the
mean first off-diagonal term of the reduced covariance matrix is 0.28 (with a
standard deviation of 0.06). When the $P(k)$ measurements are divided
by the best-fit smooth model, $P^{\rm sm}(k)$, they are, generally, even
less correlated. We determine $P(k)/P^{\rm sm}(k)$ for each mock sample
and construct a revised ``BAO'' covariance matrix from this. We
do not use this covariance matrix to perform any fits---our fits are
to the full $P(k)$ and use the original covariance matrix. For the
revised covariance matrix, the mean first off-diagonal term of the
correlation matrix is reduced to 0.03 (with a standard deviation of
0.15). The diagonal elements within this covariance matrix are also
reduced in amplitude, reflecting the smaller variance available once a
smooth fit has been removed. The errors derived from this matrix thus
better represent the errors on the measured BAO; the data when
presented as $P(k)/P_{\rm sm}(k)$ are more independent and provide a more
accurate visualisation of the measurements.

Fig. \ref{fig:PkBAOev} displays the measured post-reconstruction
values of $P(k)/P^{\rm sm}(k)$, for the BOSS CMASS sample in DR9, DR10,
and DR11 (from top to bottom), showing the evolution in the
signal-to-noise ratio of the BAO as BOSS has increased its observed
footprint. In the DR11 sample, the third peak is clearly visible. In
Fig. \ref{fig:PkBAO}, we display the DR11 post-reconstruction
$P(k)/P^{\rm sm}(k)$ for the two BOSS samples; the CMASS sample at
$z_{\rm eff}=0.57$ is presented in the top panel and the LOWZ sample at
$z_{\rm eff} = 0.32$ is shown in the bottom panel. The LOWZ sample possesses a
clear BAO feature, but the signal-to-noise ratio is considerably lower than
that of the CMASS sample.

\section{BAO Measurements from Anisotropic Clustering Estimates} 
\label{sec:results_aniso} 

\subsection{Anisotropic Clustering Estimates}

\begin{figure}
\centering
\centerline{\includegraphics[width=3.0in]{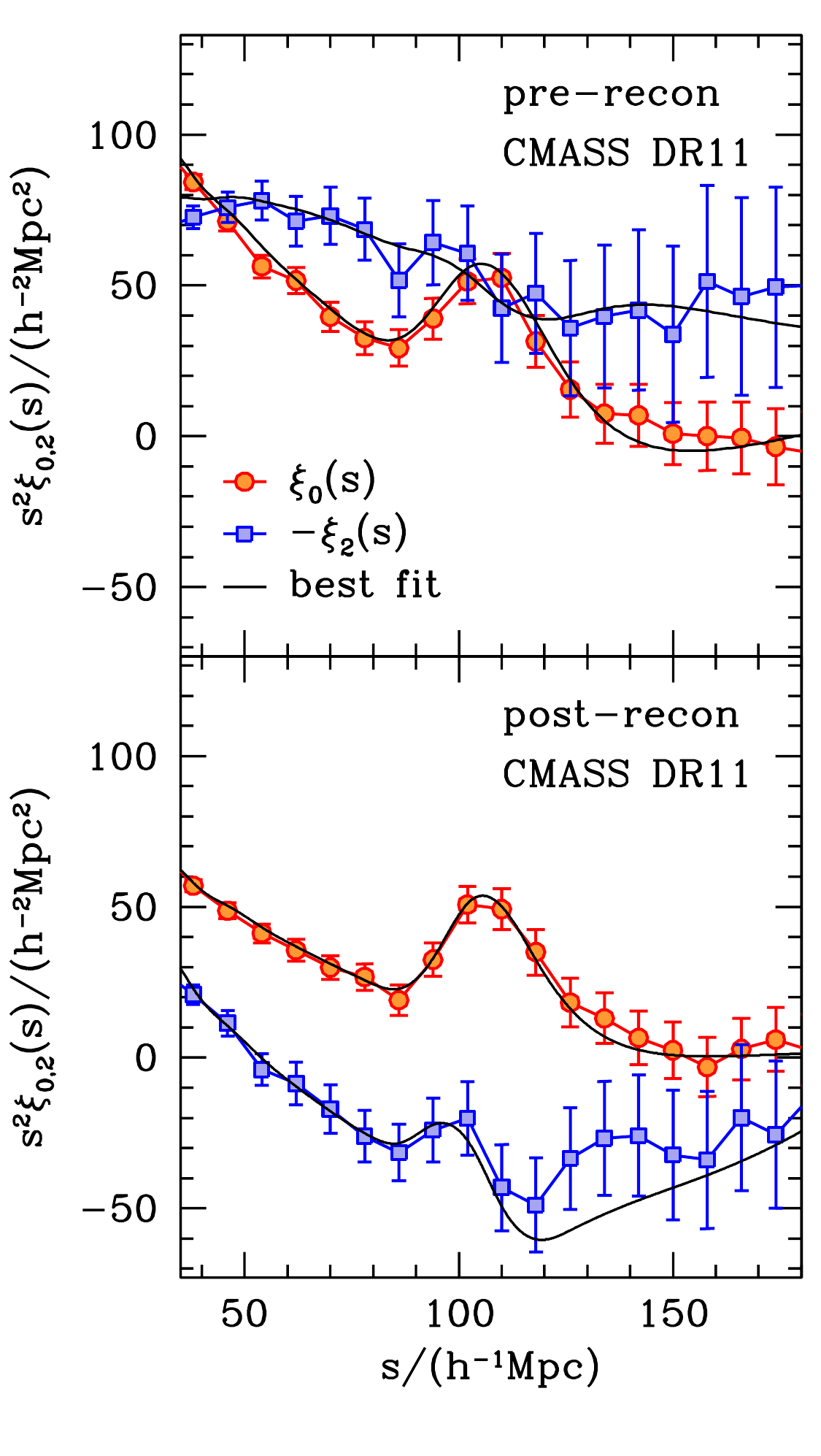}}
\caption{The DR11 multipole measurements along with their fits using the
method described in Sec~\ref{sec:aniso_fit}.  The top panel is
pre-reconstruction while the bottom one is post-reconstruction.}
\label{fig:multi_dr11_bao}
\end{figure}

\begin{figure}
\centering
\centerline{\includegraphics[width=3.0in]{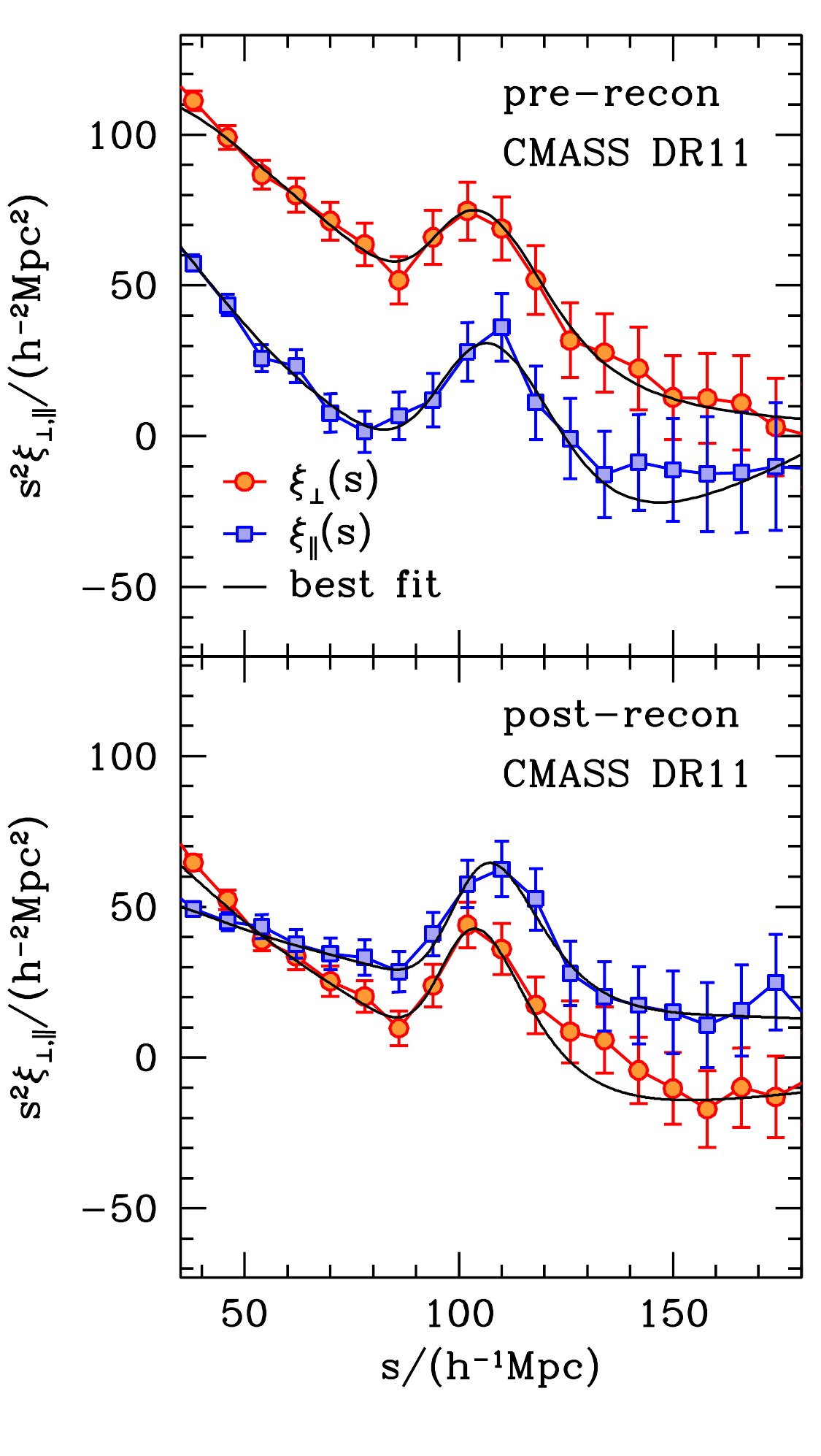}}
\caption{The DR11 wedge measurements along with their fits. The top panel is
pre-reconstruction while the bottom one is post-reconstruction.}
\label{fig:wedges_dr11_bao}
\end{figure}

In Section~\ref{sec:aniso_fit}, we detailed our analysis techniques (multipoles and wedges statistics), and 
demonstrated they recover un-biased estimates of the BAO scales both along and perpendicular to line-of-sight with similar uncertainties. 
We now apply these two techniques to BOSS CMASS sample (at $z=0.57$). 
Fig.~\ref{fig:multi_dr11_bao} displays the multipoles, $\xi_{0,2}$, of the
DR11 CMASS sample correlation function pre- and post-reconstruction, using
our fiducial binning choice, for the range of scales fitted ($45<s<200\mpcoh$).
For the quadrupole ($\xi_2$), we see a dramatic change from the pre- to
post-reconstruction results, as the reconstruction algorithm has removed 
almost all of the redshift space distortion contribution.
Further, an apparent dip is now seen in the data on scales slightly larger than the peak in
the monopole.
The strength of this feature is related to the deviation in $\epsilon$ from
0 (or the deviation in $\alpha_\perp$ from 1).

Fig.~\ref{fig:wedges_dr11_bao} displays the correlation function divided into
two wedges ($\xi_{||,\perp}$), once again with the pre-reconstruction
measurements displayed in the top panel and the post-reconstruction
measurements in the bottom panel.
Reconstruction has made the BAO peak sharper for both
$\xi_{||}$ and $\xi_{\perp}$.
Further, reconstruction has decreased the difference in their amplitudes as the 
redshift space distortion signal has been reduced.

\subsection{DR11 Acoustic Scale Measurement from Anisotropic Clustering}

As for our isotropic analysis, the results of our anisotropic BAO fits to the DR10 and DR11 mocks show significant improvement on average with reconstruction (see Table~\ref{tab:sigmamethodr11rec}), and therefore we adopt post-reconstruction results as our default. Our consensus value for the CMASS anisotropic BAO measurement, $\alpha_{||} = 0.968\pm0.032$, $\alpha_{\perp} = 1.044\pm0.013$, is determined from a combination of the measurements using the multipoles and the wedges methodologies, and we describe the individual measurements and the process of arriving at our consensus measurement in what follows. 

The curves in Figs.~\ref{fig:multi_dr11_bao} and~\ref{fig:wedges_dr11_bao} show the best-fit BAO models\footnote{The best fits to both $\xi_{\ell}(r)$ where $\ell = 0,2$ and $\ell=\parallel, \perp$ respectively.} to the pre- and post-reconstruction data using the multipoles and wedges methodology. The fits, with characteristics listed in Table~\ref{tab:data_res_clustering}, provide a good description of the data for 30 dof: the largest $\chi^2$ is 35 (a larger $\chi^2$ would be expected 24 per cent of the time) and the smallest is 21 (a smaller $\chi^2$ would be expected 11 per cent of the time). 

\begin{figure*}
{\includegraphics[trim=0cm 0cm 0cm 7cm, clip=true, width=3.in]{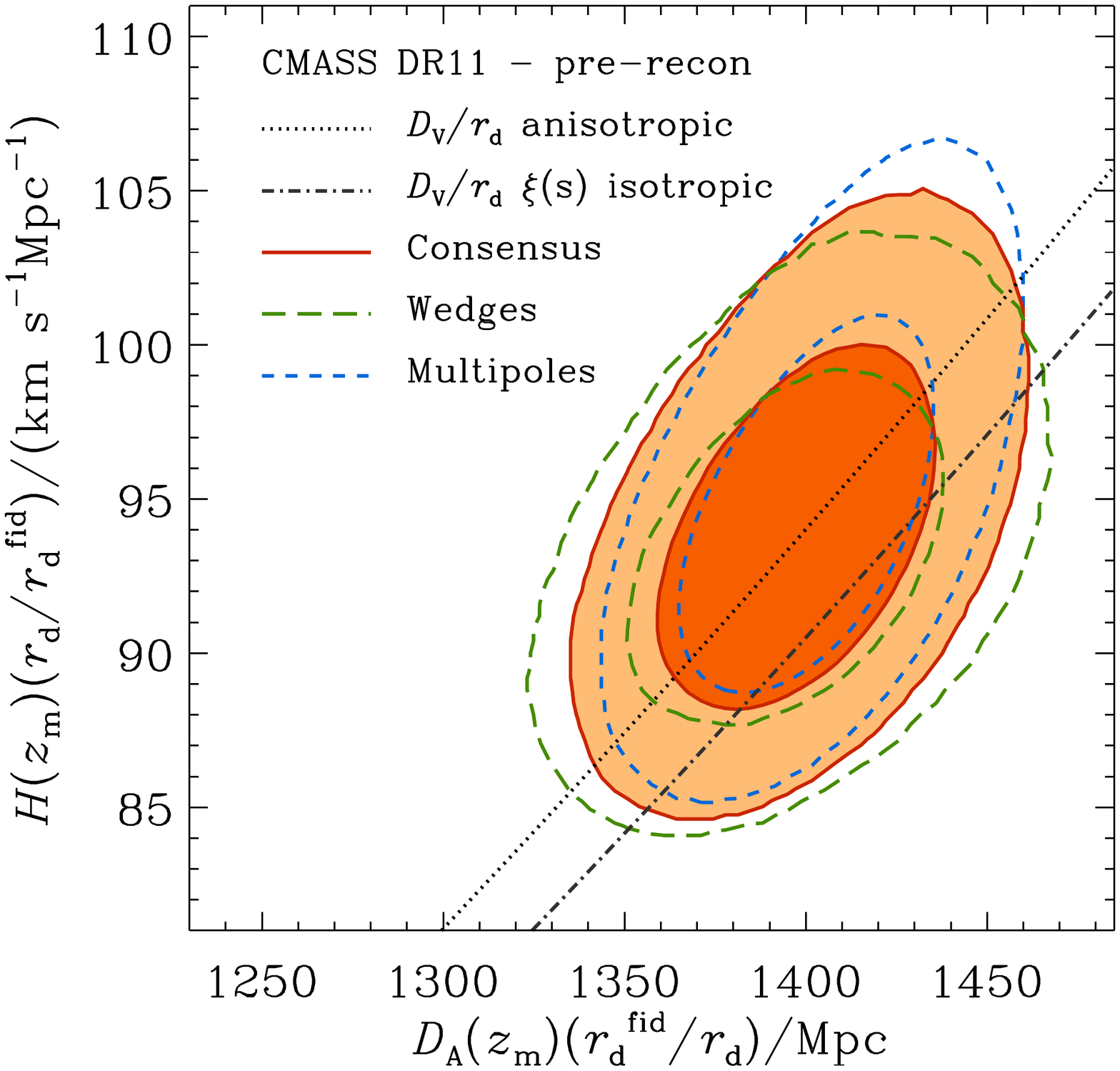}}
{\includegraphics[trim=0cm 0cm 0cm 0cm, clip=true, width=3.in]{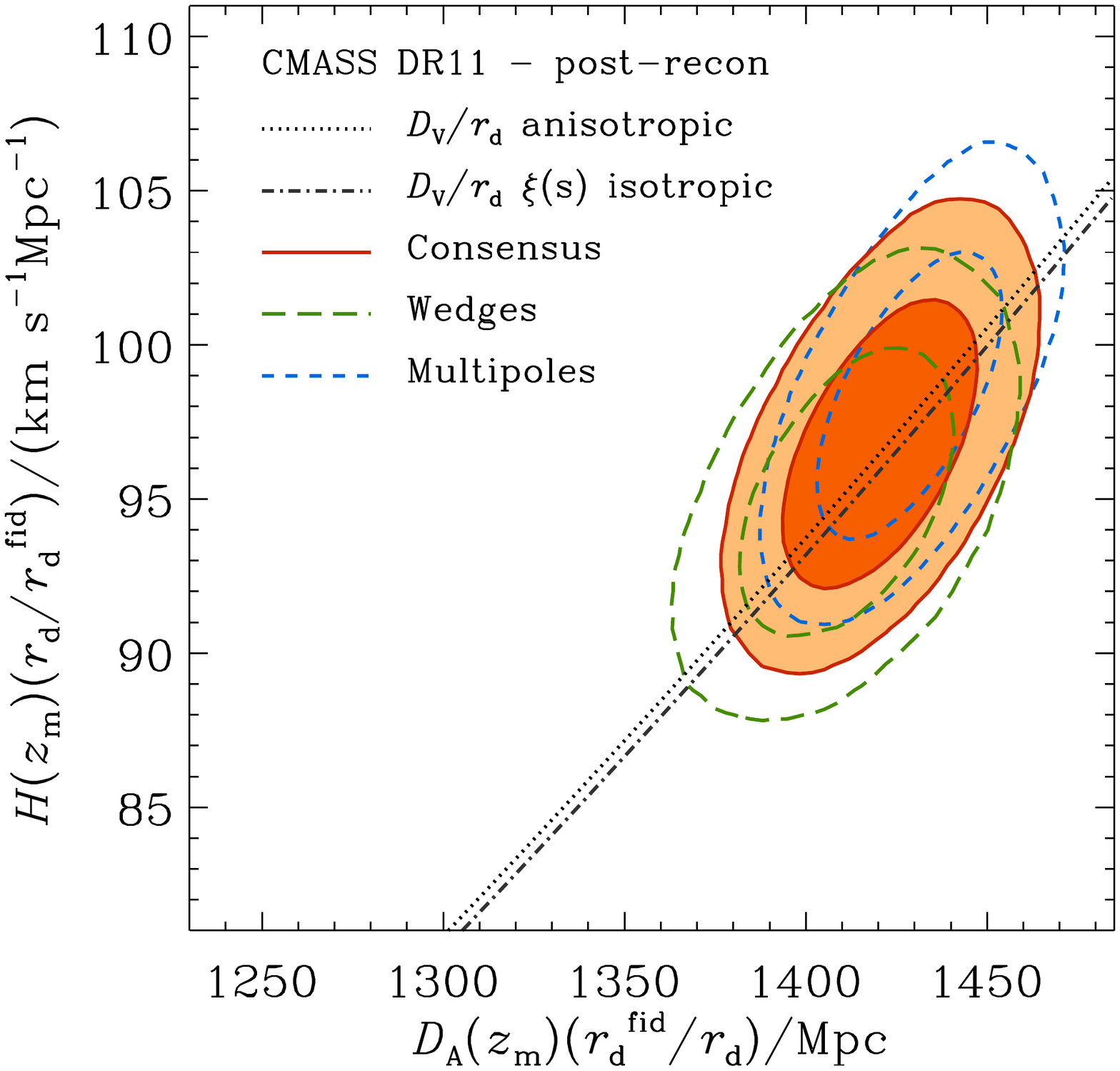}}
  \caption{
Comparison of the 68 per cent and 95 per cent constraints in the $D_A(z)$-$H(z)$ plane scaled by ($r_d^{\rm fid}/r_d$) obtained from multipoles gridded analysis (blue short-dashed line) and wedges (green long-dashed line), for DR11 pre-reconstruction (left)  and post reconstruction (right). The solid contours are the consensus values issues from combining the $\mathrm{log}(\chi^2)$ from both approaches. The multipoles provide slightly tighter constraints, the consensus contours follows a more elongated form  aligned with the axis of constant $\alpha$.
We also show the central values from fits of $D_V/r_d$ from isotropic and anisotropic clustering.  We note the slight difference
between the isotropic and anisotropic constraints on $D_V/r_d$ and the slight shift in direction of the contours of $D_A(z)-H(z)$ compared to $D_V(z)$.
We note that anisotropic clustering measurements provide stronger cosmological constraints than isotropic clustering measurements.
We thus adopt CMASS anisotropic values as our best cosmological dataset (as discussed in both Sec~\ref{sec:ani_iso_compare} and Sec~\ref{sec:datasets}) and will label it as ``CMASS'' in subsequent sections. 
}
\label{fig:DaH}
\end{figure*}

The uncertainties on the anisotropic BAO measurements are typical of those we find in the mock samples. For the multipoles result, this is illustrated in Fig. \ref{fig:sigdr11_multip}, which shows that the uncertainties recovered from the data (orange stars) are within the range of those recovered from mock samples (blue points). The uncertainty on the BAO measurements using the wedges methodology are similar to the multipoles results, with a small increase for $\alpha_{\perp}$, that exactly matches that seen fitting mock catalogues. We further illustrate the constraints obtained from each method in Fig.~\ref{fig:DaH} where one can see the comparison 
of the 60 per cent and 95 per cent constraints in the $D_A$ and $H(z)$ plane scaled by $r^{\rm fid}/r_d$ using the two methods. The size of the contours from both methods agree very well, with a slightly more elongated contour from multipoles, showing that the multipoles and wedges contain slightly different information.

The precision of the DR11 results are improved by reconstruction, as expected. This is illustrated in Fig.~\ref{fig:DaH}, where the post-reconstruction $D_A(z)$, $H(z)$ contours in the right-hand plot show a dramatic decrease compared to the pre-reconstruction results displayed in the left-hand panel.  
Based on our testing of 600 mock CMASS samples, we found 
(as shown in Figure~\ref{fig:sigdr11_multip}) that reconstruction is expected to improve the precision of the multipoles method measurement of $\alpha_{\perp}$ by $\sim 40$ per cent (the median uncertainty decreases from 0.021 to 0.015) and of $\alpha_{\parallel}$ by 63 per cent (the median uncertainty decreases 0.044 to 0.027), with very similar results using the wedges methodology. 
We find that for the DR11 data, the results are similar to our expectation, as the improvements in the precision of the results gained by reconstruction are all between 39 and 50 per cent. The improvement in $\alpha_{\perp}$ (50 per cent for multipoles and 42 per cent for wedges) is slightly better than expected and the improvement in $\alpha_{||}$ (39 per cent for multipoles and 48 per cent for wedges) is slightly worse, but Fig. \ref{fig:sigdr11_multip} shows that the results (orange stars) are well within the range of the results determined from mock samples (blue points). 

Table~\ref{tab:data_res_clustering} shows that the DR11 post-reconstruction multipoles and wedges results disagree by close to 1-$\sigma$: $\alpha_{\parallel,{\rm Mult}}= 0.952 \pm 0.031$, $\alpha_{\parallel,{\rm Wedges}}= 0.982 \pm 0.031$; $\alpha_{\perp,{\rm Mult}}= 1.051 \pm 0.012$, $\alpha_{\perp,{\rm Wedges}}= 1.038 \pm 0.012$ . 
The difference in $\alpha_\parallel$ is $0.030$. We then turn to the galaxy mock catalogs to see whether this behavior is common. 
We find that 39  out of 600 mocks show the same or larger differences between the two methods. 
The mean difference is $0.005$ with a RMS of $0.016$ suggesting that this difference in the data is a 1.9 $\sigma$ event. 
The difference in $\alpha_\perp$ is  $0.013$, we also found 45 out of 600  cases that show the same or larger differences between the two methods. The mean difference found in the mocks is $0.001$ and the r.m.s. is $0.008$, this suggests that the difference in the data is a 1.6$\sigma$ event. 
This is mostly driven by differences in the fitted results of $\epsilon$, Table~\ref{tab:data_res_clustering} shows us that the fitted values of 
$\alpha$ from both methodologies only differ by $0.2$ per cent, while $\epsilon$ is different by $1.5$ per cent, which is comparable to the 1$\sigma$ error on $\epsilon$. 
We thus turn to a discussion using $\alpha$-$\epsilon$ parametrization in the following discussion. 

Pre-reconstruction, the multipoles and wedges measurements in $\alpha$ and $\epsilon$ differ by less than 0.25$\sigma$ as shown in Table~\ref{tab:data_res_clustering}. 
Fig.~\ref{fig:DaH} shows that, as reconstruction tightens the constraints from both methods, the central values shift slightly along the axis of constant $\alpha$ by  $1.5$ per cent  in $\epsilon$. 
When we look at this comparison in our mocks,
we find a r.m.s. difference in $\epsilon$ fits of $0.007$, indicating that the data is a $2\sigma$ outlier.  $27$ of 600 mocks have differences more extreme than $\pm 0.015$.  The other three cases (DR10 and DR11 pre-reconstruction) show smaller variations.  We conclude that this event is consistent with normal scatter of the two estimators.

Our tests on our fitting methodology, presented for the mock samples in Section \ref{sec:aniso_mocktest} and on the DR11 data in Section \ref{sec:fitting_data_aniso} suggest no systematic issue causing the observed difference between the results of the two methods. 
Thus, we combine the likelihood distributions recovered from the multipoles and wedges measurements, using the method described in Section \ref{sec:com_multwedge}, to recover our consensus anisotropic BAO measurement,  $\alpha_{\perp}= 1.045 \pm 0.015$ and $\alpha_{\parallel}= 0.968 \pm 0.033$.  
We quote the statistical and systematic error\footnote{The systematic errors are described in Sec~\ref{sec:fitting_data_aniso} and Sec~\ref{sec:sys_err}. The addition of the two types of error is described in Sec~\ref{sec:sys_err}. } here for consensus values, while the remaining values in Table~\ref{tab:data_res_clustering} consider only the statistical errors. 
 
\begin{table*}
\label{tab:aniso_bao}
\caption{Fits to anisotropic clustering measurements recovered from BOSS DR10
  and DR11 pre- and post-reconstruction. We also show the fit to isotropic
  correlation function $\xi_0(s)$ for comparison (extracted from
  Table~\ref{tab:isobaoresults}). The isotropic results we extracted refers to the one that we find closest in fitting methodology to the anisotropic fits. Therefore, the isotropic results here are fits to correlation functions only and without combining the bins across different bin center choices.  We include here anisotropic fits made
  using the ``De-Wiggled'' template \citep[see][]{And13} since this
  matches the fit to the isotropic clustering measurements.  It is not 
  surprising that the $\alpha$ values fit from anisotropic clustering
  using this template are in even better agreement with the isotropic 
  clustering measurement. \label{tab:data_res_clustering}}

\begin{tabular}{@{}lccccccccc}

\hline
\hline
Model&
$\alpha$&
$\epsilon$&
$\rho_{\alpha,\epsilon}$&
$\alpha_{\parallel}$&
$\alpha_{\perp}$&
$\rho_{\alpha_{||},\alpha_{\perp}}$&
$\chi^2$ \\

\hline
DR11\\[1.5ex]
%
{\bf Consensus}&
{\bf 1.019}$\pm${\bf 0.010}&
{\bf -0.025}$\pm${\bf 0.014}&
{\bf 0.390}&
{\bf 0.968}$\pm${\bf0.033}&
{\bf 1.045}$\pm${\bf 0.015}&
{\bf -0.523}&-&\\

Post-Recons. Multipoles&
$1.017\pm0.009$&
$-0.033\pm0.013$&
0.505&
$0.952\pm0.031$&
$1.051\pm0.012$&
-0.311&
$21/30$&
\\

Post-Recon. Wedges&
$1.019\pm 0.010$&
$-0.018\pm 0.013$&
0.389&
$0.982\pm0.031$&
$1.038\pm0.014$&
-0.501&
21./30&
\\

 Post-Recon.  De-Wiggled&
 $1.017 \pm 0.009$&
 $-0.032\pm0.013$ &
 $0.512$ &
 $0.952\pm0.032$&
 $1.051\pm0.012$&
 -0.304  & 
 $21/30$&

 \\

Post-Recon. Isotropic&
$1.021\pm0.009$  &
-- &
--&
--&
--&
-&
$16/17$&
\\
 
%

Pre-Recon. Multipoles&
$1.017\pm0.013$&
$-0.012\pm0.019$&
0.495&
$0.992\pm0.046$&
$1.030\pm0.017$&
-0.246&
$35/30$&
\\

Pre-Recon. Wedges&
$1.018\pm0.015$&
$-0.008\pm0.018$&
0.236&
$1.001\pm0.043$&
$1.027\pm0.021$&
-0.453&
33/30&
\\

 Pre-Recon. De-Wiggled&
 $1.025\pm 0.014$&
 $-0.010 \pm 1.035$&
 0.572&
 $1.004\pm 0.049$&
 $1.035 \pm 0.017$ &
 -0.149&
 $33/30$&

 \\

Pre-Recon. Isotropic&
$1.031\pm0.013$  &
- &
-&
-&
-&
-&
$14/17$&
\\
 
\hline

DR10\\[1.5ex]

{\bf Consensus}&
{\bf 1.019}$\pm${\bf 0.015}&
{\bf -0.012}$\pm${\bf 0.020}&
{\bf 0.502}&
{\bf 0.994}$\pm${\bf 0.050}&
{\bf 1.031}$\pm${\bf 0.019}&
{\bf -0.501}&-&\\


Post-Recon. Multipoles&
$1.015\pm0.016$&
$-0.020\pm0.023$&
0.683&
$0.975\pm0.058$&
$1.037\pm0.018$&
-0.240&
$16/30$&
\\

Post-Recon. Wedges&
$1.020\pm0.015$&
$-0.006\pm0.019$&
0.513&
$1.009\pm0.049$&
$1.027\pm0.018$&
-0.474&
17/30&
\\

 Post-Recon. De-Wiggled&
 $1.015\pm 0.016$&
 $-0.020 \pm 0.023$&
 0.669&
 $0.974\pm 0.057$&
 $1.036\pm 0.018$&
 -0.163&
 16./30&

 \\

Post-Recon. Isotropic&
$1.022\pm 0.013$  &
-- &
--&
--&
--&
--&
$14/17$&
\\


Pre-Recon. Multipoles&
$1.004\pm0.016$&
$-0.024\pm0.025$&
0.439&
$0.956\pm0.057$&
$1.029\pm0.024$&
-0.346&
$36/30$&
\\

Pre-Recon. Wedges&
$1.004\pm0.018$&
$-0.015\pm0.022$&
0.104&
$0.975\pm0.049$&
$1.020\pm0.028$&
-0.482&
30/30&
\\

Pre-Recon. De-Wiggled&
 $1.012\pm 0.018 $&
 $-0.021\pm 0.026 $&
 0.555&
 $0.969 \pm 0.063 $&
 $1.035 \pm 0.023 $&
 -0.237&
 $35/30$&
 \\

Pre-Recon Isotropic&
$1.022\pm 0.017$  &
-- &
--&
--&
--&
-&
$16/17$&
\\
\hline

\end{tabular}
\end{table*}

\begin{table*}
  \caption{CMASS post-reconstruction DR11 results for several variations
    of the fitting models. We can see that the central values of
    $\alpha_\perp$ and $\alpha_\parallel$, and the errors around these
    best-fit values are robust to the changes in methodology
    considered. Were we to extend the range of $\alpha$ and $\epsilon$
    probed, then this would not be the case, and the derived
    errors would change. More details and further tests can be found
    in \protect\citep{Vargas13}}
\label{tab:datatest}

\begin{tabular}{@{}lccccccccc}
\hline\hline
Method&$\alpha$&$\epsilon$&$\rho_{\alpha, \epsilon}$&$\alpha_{||}$&$\alpha_\perp$&$\rho_{||, \perp}$&$\chi^2$/d.o.f&$B_0$&$\beta$\\
\hline
$\mathrm{Fiducial}$&
$  1.017\pm  0.009
$&$ -0.033\pm  0.013
$&$  0.505
$&$  0.952\pm  0.031
$&$  1.051\pm  0.013
$&$ -0.610
$&$    21./30
$&$  1.095
$&$ -0.096$\\
$\mathrm{Fitting} \; 30<r<200$&
$  1.019\pm  0.008
$&$ -0.030\pm  0.012
$&$  0.384
$&$  0.959\pm  0.028
$&$  1.050\pm  0.013
$&$ -0.638
$&$    36./30
$&$  1.113
$&$  0.028$\\
$\mathrm{Only} \; B_0 \; \mathrm{prior}$&
$  1.017\pm  0.010
$&$ -0.031\pm  0.014
$&$  0.580
$&$  0.955\pm  0.034
$&$  1.049\pm  0.013
$&$ -0.607
$&$    20./30
$&$  1.084
$&$ -0.199$\\
$\mathrm{Only} \; \beta\; \mathrm{prior}$&
$  1.016\pm  0.009
$&$ -0.034\pm  0.014
$&$  0.537
$&$  0.949\pm  0.032
$&$  1.052\pm  0.013
$&$ -0.622
$&$    20./30
$&$  1.106
$&$ -0.091$\\
$\mathrm{No \; priors}$&
$  1.016\pm  0.010
$&$ -0.032\pm  0.015
$&$  0.612
$&$  0.953\pm  0.036
$&$  1.049\pm  0.013
$&$ -0.614
$&$    20./30
$&$  1.094
$&$ -0.190$\\
$\mathrm{Fixed \;}\beta=0.0$&
$  1.017\pm  0.008
$&$ -0.034\pm  0.012
$&$  0.447
$&$  0.949\pm  0.029
$&$  1.053\pm  0.012
$&$ -0.608
$&$    21./30
$&$  1.105
$&$  0.000$\\
$\mathrm{2-term} \; A_0(s) \; \& \; A_2(s)$&
$  1.017\pm  0.009
$&$ -0.025\pm  0.013
$&$  0.560
$&$  0.967\pm  0.031
$&$  1.044\pm  0.012
$&$ -0.555
$&$    37./30
$&$  1.048
$&$ -0.210$\\
$\mathrm{4-term} \; A_0(s) \; \& \; A_2(s)$&
$  1.016\pm  0.008
$&$ -0.034\pm  0.013
$&$  0.438
$&$  0.948\pm  0.029
$&$  1.052\pm  0.013
$&$ -0.601
$&$    16./30
$&$  1.094
$&$ -0.039$\\
\hline
\end{tabular}
\end{table*}

\subsection{Robustness Checks on Data}
\label{sec:fitting_data_aniso}

We measure the DR11 post-reconstruction anisotropic BAO scale with various choices of methodology, in order to test the robustness of our anisotropic BAO measurements. Because of the tight correlation between results calculated from fits to either multipoles or wedges (see Section~\ref{sec:aniso_mocktest}), we only present robustness checks for fits to the multipoles. We have only summarized the results of the robustness that are of immediate relevance to this paper here, while the full robustness test of our anisotropic BAO fitting methodology is shown 
in \citet{Vargas13}. 

We vary the choices when fitting to the data in the same way as we did when testing the results on the mock samples in Section~\ref{sec:aniso_mocktest}. The results are summarized in Table \ref{tab:datatest} and we can see that the differences in central fitted values when we consider different choices of fitting parameters are impressively small.  
The central fitted values of $\alpha$ vary less than $0.1$ per cent, while the various fitted errors vary less than $0.2$ per cent. 
For all cases but one, the central fitted values of $\epsilon$ vary less than $0.2$ per cent, while the fitted errors vary less than $0.2$ per cent. 
The largest variation found is on $\epsilon$ when we we change the broadband polynomial such that each component ($A_{\ell}(s)$) is only limited to 2-terms, which is still relatively small, at $0.8$ per cent, which is less than $0.6-\sigma$.  
We can turn our attention to $\alpha_\parallel$ and $\alpha_\perp$, but as expected, since the variations are not large for $\alpha$ and $\epsilon$, the changes in $\alpha_\parallel$ and $\alpha_\perp$ are equally small. 

We also investigate the effects of priors. We refer the reader to the priors listed in Section~\ref{sec:aniso_method_5.1}. 
In both mocks and data of DR11 post-reconstructed, we find that as long as we either limit the $\epsilon$ to reasonable physical intervals when we calculate the error or use priors on both $\beta$ and $B_0$,  the final fitted central values and errors remain relatively unchanged to within $0.1$ per cent. 
We have discussed this further in a companion paper \citep{Vargas13}, which should be consulted for more details. 
Finally, it is also interesting to note that a fixed $\beta$ parameter does not change the error or central values by more than $0.1$ per cent. 
To conclude, the variations of DR11 post-reconstructed data is well within the scatter predicted when the same varying choices are applied to mock galaxy catalogs. 
 
\subsection{DR10 Anisotropic BAO measurement}

Although our default results are for DR11, it is instructive to examine the results from the reduced DR10 data set, which allow us to follow the transition in data quality from our previous DR9 results to our new DR11 results. Consequently we present anisotropic BAO measurements from DR10 alongside the DR11 measurements in Table~\ref{tab:data_res_clustering}. For the mock catalogues, Fig.~\ref{fig:sigdr11_multip} showed that, for fits to the DR10 multipoles, the expected improvement of the measurement in $\alpha_{\perp}$ and $\alpha_{\parallel}$ with reconstruction is from $2.8$ per cent and $5.4$ per cent to $1.9$ per cent and $3.6$ per cent respectively. Using the DR10 data, we measure $\alpha_{\perp}= 1.039 \pm 0.024$ and $\alpha_{\parallel}= 0.956 \pm 0.057$ pre-reconstruction. 
Post-reconstruction, we measure $\alpha_{\perp}= 1.037 \pm 0.018$ and $\alpha_{\parallel}= 0.975 \pm 0.058$, showing remarkable consistency with the mock results. 
The measurement using wedges, also presented in Table~\ref{tab:data_res_clustering}, are similar and consistent. Thus the precision of the BAO measurements from the DR10 data are typical. This can also be seen in Fig.~\ref{fig:sigdr11_multip}, where the orange star representing the data results is within the locus of the blue circles representing results from the mocks. 

It is interesting that, for DR10, the error post-reconstruction is slightly larger for $\alpha_{\parallel}$, compared with the pre-reconstruction error. We can see this more clearly by looking at the $\alpha$ and $\epsilon$ pair pre- and post-reconstruction in DR10 in Table~\ref{tab:data_res_clustering}.  It seems there is no improvement in $\sigma_\alpha$, while there is some slight improvement in $\sigma_\epsilon$ post-reconstruction.  We compare this to the mocks to try and understand this behaviour. Figure~\ref{fig:sigdr11_multip} shows that not all mocks in DR10 have improved constraints on $\alpha_{\parallel}$ after reconstruction, even though it is uncommon: $\approx 20$ out of 600 mocks that do not improve. We do see improvement on nearly all mocks in DR11, which may be due to the fact that DR11 has a more contiguous mask, so that there is less volume close to boundaries. This may contribute to a more successful reconstruction in DR11.

\subsection{Comparison with Isotropic Results}
\label{sec:ani_iso_compare} 

For ease of comparison between our isotropic and anisotropic measurements, we include the results from isotropic fits to the correlation function (presented in Table~\ref{tab:isobaoresults}) in Table~\ref{tab:data_res_clustering}. 
Post-reconstruction, the central values of $\alpha$ measured from isotropic and anisotropic clustering are consistent to well within 1-$\sigma$.  Pre-reconstruction, the central values of $\alpha$ from  the isotropic correlation function are approximately 1-$\sigma$ higher than $\alpha$ from the anisotropic clustering , for both DR10 and DR11. Part of this difference can be explained by the different correlation function templates used for the isotropic and anisotropic analyses. 
The anisotropic fitting uses $P_{\rm pt}(k)$ as described in Eq~\ref{eqn:RPT} which was chosen as it provides less biased measured values of $\alpha$ and $\epsilon$ fitting, while 
the isotropic fitting uses a non-linear power-spectrum ``De-Wiggled template" \cite{And12,And13}. The differences though between the templates are quite small and are further explored in \cite{Vargas13}. For comparison, we provide anisotropic fits made using the same ``De-Wiggled'' power spectrum template used for the isotropic fits in Table~\ref{tab:data_res_clustering}, it is not surprising that the anisotropic results with the same power spectrum template provides more similar fits to  those from the isotropic fits  in both pre-reconstruction and post-reconstruction datasets (this is explored further in \citealt{Vargas13}).

For DR11 post-reconstruction, which is our default choice for making cosmological measurements, we note that  the isotropic power spectrum fits give lower values of $\alpha$ than the isotropic  correlation function fits, pulling the isotropic consensus values of $\alpha$ down (see Table~\ref{tab:isorobust}). 
On the other hand,  the correlation function monopole measurement of $\alpha$ agrees very well with the $\alpha$ values measured from anisotropic fits to both the monopole and the quadrupole. 
They are both higher than the consensus value of the isotropic fits (a combination of both isotropic power-spectrum fit and the correlation function fit), and the effect of this is noticeable when the measurements are combined with the CMB data and turned into cosmological constraints (see Fig.~\ref{fig:DaH_CMB} and Section~\ref{sec:CMB_scale_cmpr}). 

Isotropic fits of $\alpha$ only allows us to measure the spherically averaged distance $D_{V}(z) \propto D_A^2(z)/H(z)$, where z is the median redshift of the sample. This has made the approximation that the clustering of the galaxy sample is isotropic. More importantly, the Hubble parameter $H(z)$ is degenerate with $D_A(z)$ in this isotropic measurement, and thus we cannot directly probe the expansion of the Universe. 
The clustering of galaxies is not truly isotropic due to both large-scale redshift-space distortions and from assuming the wrong cosmology when we calculate the 2-point statistics. Therefore, the fit to the anisotropic clustering provides more information by breaking the degeneracy between $H(z)$ and $D_A(z)$. We are therefore not surprised that the anisotropic clustering measurements provides stronger cosmological constraints as demonstrated by the different contour sizes in Fig.~\ref{fig:DaH_CMB}.  
We further compare and contrast the isotropic and anisotropic fits in  Figure~\ref{fig:DaH}. 
While on average the anisotropic degeneracy direction should lie along
the isotropic ($D_V$) direction, in our data set the orientation is closer to vertical. This slight rotation is driven by the shot-noise differences along the line of sight and perpendicular to the line of sight. 
This is expected, given the comparison of the data and ensemble of
mock constraints on $\alpha_{\perp}$ and $\alpha_{\parallel}$ shown in
Fig.~\ref{fig:sigdr11_multip}.   This figure also illustrates the 0.5 per cent increase in the best-fit $\alpha$ from the anisotropic fits compared with the isotropic ones.  
Anisotropic clustering's constraining power is also amplified depending on the models we explore. 
For example,  variation in dark energy equation of state ($w$) shifts $D_A$ at fixed CMB acoustic scale, and anisotropic clustering measurements provide stronger constraints than isotropic ones in the direction of $D_A$ (Fig.~\ref{fig:DaH}).  

Therefore, we choose the anisotropic clustering measurements to be the default measurement of the CMASS measurement in our cosmological analysis (thus will only be referred to as CMASS in later sections).


\section{The Cosmological Distance Scale}
\label{sec:distance_scale}

\subsection{Systematic Errors on BOSS BAO Measurements}  \label{sec:sys_err}

Sections \ref{sec:fit-2pt} and \ref{sec:aniso_fit} presented the acoustic
scale fits and their statistical errors.  Here we present estimates of 
systematic errors, which we believe to be subdominant by a considerable margin.
We organize the discussion into two separate classes of systematic errors.  
The first set includes possible artifacts from our survey, including the
effects of survey boundaries and observational systematics on the
reconstruction and fitting methodology.  
The second set concerns the possible residual effects of galaxy clustering bias
on the shift of the acoustic scale after one applies our reconstruction
algorithm assuming a large filled survey.

As shown in Table \ref{tab:mockbao}, when run on our mock catalogs,
the estimators for both the spherically averaged correlation function and the power
spectrum return unbiased results in DR11 after reconstruction, with
precision of 0.04 per cent in the mean.  
Table \ref{tab:sigmamethodr11rec} and Table \ref{tab:sigmadr11rec} show that when we run our different estimators (Multipoles and Wedges) for the anisotropic clustering signals on mock catalogs they return unbiased results in DR11 both pre-reconstruction and post-reconstruction. 
This is an extremely sharp test, as it
includes the effects of the survey geometry and ability of reconstruction
to remove the non-linear shifts of the acoustic scale that arise from
Lagrangian perturbation theory as used in our mocks.  It also validates
our fitting methodology, e.g., demonstrating that effects of binning,
interpolation, and integrations in the measurement and fitting procedures
have been handled well.  \citet{Ross13} use two other sets of mocks, using
the same formalism as \citet{Man12} but with different halo mass cuts; 
they find similar unbiased performance after reconstruction, with
precision better than 0.1 per cent.  

The effects of variations in the fitting methodology was discussed
in sections \ref{sec:results_iso} and  \ref{sec:results_aniso},
showing only small offsets, at the level of 0.1-0.2 per cent in $\alpha$
for cases that were expected to agree.  \citet{And12},
\citet{And13}, and \citet{Vargas13} present further tests, again
finding no substantial offsets.

One can also search for systematic errors by comparing different aspects of our
analysis.  Indeed, we do find cases in which different analyses of the same data
return acoustic scales that differ by of order 0.5 per cent,
e.g., the comparison of the $\alpha$ measured from $\xi$ and $P$ in DR11.
However, these discrepancies
occasionally occur in our mock catalogs and hence are not sufficiently unusual
to indicate a systematic error, particularly because we examined a substantial
number of these comparisons, many of which were unremarkable.
If these differences are indeed due to systematic effects, then it must be for reasons 
that are not present in our mock catalogs, as each of our estimators is
unbiased when averaged over many mocks.  

The mocks do not include large-angle observational systematics due to
such things as variations in star-galaxy separation effectiveness or
seeing, as were discussed in \citet{Ross12}.  However, the acoustic
peak measurement is highly robust to such effects, as they tend to
have smooth angular power spectra.  One would expect that
if such effects were present, they would be much more severe if we
omitted the broadband nuisance terms described in \S~\ref{sec:fit-2pt}.
As shown in \S~\ref{sec:results_iso}, performing our correlation function
fits with fewer or even none of the three broadband nuisance terms
produces changes in $\alpha$ of 0.2 per cent or less.  Removing two terms
from the power spectrum fit also changes the answer by only 0.2 per cent.

Although we measure the clustering within a redshift bin
of non-zero thickness, we interpret the fitted scale as measuring
the distance to a single effective redshift.  We base this estimate on
the mean redshift of fully weighted pairs, rounded off for simplicity.
This effective redshift is not formally well defined---for example,
it might depend on scale or differ between line-of-sight and transverse
clustering---but different reasonable choices vary by only 0.01 in redshift.
We then expect the effect of this assumption to be small because any error in the 
effective redshift enters only as the variation with redshift in
the ratio of the true cosmology to the fiducial cosmology.  For example,
we will see in \S~\ref{sec:params} that the ratio of $D_V$ between the 
best-fit $\Lambda$CDM model and a model with $w=-0.7$ that matches the CMB data 
varies by about 1 per cent for each 0.1 in redshift.
This would be a 0.1 per cent shift for an 0.01 change in effective redshift.
Yet this much tilt in the
distance-redshift relation is already disfavored by the BAO Hubble diagram
and by the supernova data.  Hence, we argue that errors in the effective
redshift affect our interpretations at below 0.1 per cent.

Similarly, 
our mocks are based on a single redshift snapshot of the simulations,
rather than light-cone outputs that track the exact structure at each
redshift.  This approach could create errors when we combine a broad
redshift bin into one clustering measurement and interpret the acoustic
peak as arising from a single, effective redshift.  Note that the amplitude
of galaxy clustering changes much more slowly than the predicted variation
in the amplitude of the matter clustering, which limits the mismatch of
combining different redshifts.
Preliminary tests of this approximation with light cone simulations
in a few cases show the effects to be small, but we
intend to extend these tests in the future.

The choice of fiducial cosmology also enters through the linear power spectrum
used in our fitting.  The assumption of our methodology is that the $\alpha$
values recoved from fits with other template spectra would be well predicted
by the ratios of sound horizons computed in these cosmologies to that of the
fiducial model.  Were this not the case, we would simply have to repeat the fit
for each new cosmology, searching for cases of $\alpha=1$.  This assumption
has been investigated in previous papers and found to be a good approximation
\citep{Seo08,Xeaip,Xu12b}, with systematic offsets typically at or below
0.1 per cent in $\alpha$.  One exception was
presented in \citet{Xeaip}, where a case with an extra relativistic neutrino
species created an uncorrected 0.5 per cent shift of $\alpha$ due to template
mismatches.  Hence, more exotic cosmologies may require additional
consideration of whether the sound horizon fully captures the impact
of the variation in the fitting template.

Our conclusion from these tests is that there is no evidence for
systematic errors from the survey effects and fitting above the 0.1 per cent
r.m.s. level from any effect we have considered.  However, there are
several such terms that could accumulate, so we triple this to
adopt a systematic error of 0.3 per cent for our measurements of $D_V$.
We believe that further tests on a more diverse and realistic set
of mock catalogs would boost confidence in the methods at the 0.1 per cent
aggregate level.

The analysis of the anisotropic BAO could be subject to additional 
systematic errors due to the above effects.  The anisotropic fitting
is more complicated because of redshift distortions and the inherent
anisotropy of the survey geometry and light-cone effects.  
Our tests on mock catalogs show the estimators to be unbiased
at the level of 0.2 per cent in $\epsilon$.  
\citet{Vargas13} presents an exhaustive set of tests of the multipole fitting
method; \citet{Xu12b}, \citet{And13}, and \citet{Kaz13} present a wide
variety of tests on earlier data sets.  For the DR11 post-reconstruction case,
\citet{Vargas13} find variations in $\epsilon$ at the 0.1-0.2 per cent level
as the parameters in the fitting method are varied.  We take these results to 
indicate a 0.3 per cent r.m.s. systematic uncertainty in $\epsilon$ due to fitting.
We increase this estimate to 0.5 per cent to include possible errors in the 
anisotropic BAO external to our mocks, e.g., due to light cone effects,
evolution in the sample, inaccuracies in assumptions about peculiar
velocities in the mocks or reconstruction, or mismatches between
our fiducial cosmology and the true one.

Our estimate of statistical error does depend on the assumption that the 
amplitude of clustering in the mocks matches that in the true data, as the
sample variance of the density field depends on its power spectrum.
Our current mocks have about 10 per cent less power than the data,
which might lead to a small underestimate of the sample variance in
the correlation function.
The variance of the power spectrum analysis would actually be slightly
overestimated because the covariance matrix was computed for $\ln P$ and
hence includes only the fractional error on the power.
The fractional error would be somewhat larger because of the increased
importance of shot noise relative to a weaker clustering signal.
The fact that the effects of a mismatch in clustering amplitude have
opposite effects on the estimated errors in $\xi$ and $P$, combined
with the result that the uncertainties in $\alpha$ recovered from each
statistic match closely, further argues that this effect is small.
At present, we make no correction to our statistical error bars for the offset
of clustering amplitude in our mocks, as the mismatch is small and
the exact size of the resulting correction not well known.  We also do
not include a term in our systematic errors for possible mismatches
of the amplitude of clustering, as this does not represent a bias
in the mean, but rather an error on the error.

We next turn to systematic errors from true astrophysical shifts due 
to non-linear structure formation and galaxy clustering bias.
Prior to reconstruction, one can see the small expected shift, of order
0.4 per cent, in the fitting of the mocks.
From perturbation theory \citep{CroSco08,PadWhi09}
and simulations \citep{PadWhi09,Seo10}
we expect shifts in the clustering of matter at 0.2-0.25 per cent at
these redshifts.
Galaxy bias produces additional small shifts \citep{PadWhi09,Meh11}.
As reconstruction improves due to the larger and more contiguous survey
volume, we expect it to remove the shifts due to large-scale velocities.
\citet{Meh11} found no example in their models in which the shift after reconstruction was
non-zero, with errors of about 0.1 per cent r.m.s..
The mock catalogs used here, as well as the two in \citet{Ross13},
also show no offsets at this level.  
Of course, our mock catalogs and the galaxy bias models of \citet{Meh11} 
do not span all possibilities, but there is a good physical reason
why reconstruction is successful at removing shifts:  
in a wide range of bias models, the galaxy density field is proportional 
to the dark matter density field at scales
above $10\,$Mpc.
The shifts in the acoustic scale arise in second-order perturbation theory
due to large-scale flows, which are well predicted by the galaxy maps.
Reconstruction substantially reduces the flows and hence the source of
the acoustic scale shifts.
To be conservative, we triple the level of uncertainty implied by our current
mocks and adopt a systematic error of 0.3 per cent in $\alpha$ for shifts
{}from galaxy bias that are not corrected by reconstruction.

Our systematic error budget for galaxy clustering bias does not
encompass offsets that could result from the
effects of relative streaming velocities between baryons and dark
matter in the earliest collapse of proto-galaxies \citep{Tse10}.
Although this
effect is large at the cosmological Jeans scale of $10^6\ {\rm M}_\odot$
halos, the galaxies we measure in BOSS occupy halos over a million
times larger and one might imagine that the impact of the early
streaming velocities have been significantly diluted.  Empirically, a recent
paper by \cite{Yoo13} limited the acoustic scale shifts from this
effect through its impact on the large-scale DR9 power spectrum;
they found a remaining r.m.s. uncertainty of 0.6 per cent.  While we look
forward to more work on the possible effects of relative streaming
velocities, we do not inflate our systematic errors by this much,
as theories often predict the effect to be negligible at mass scales
well above the cosmological Jeans scale \citep[see e.g.][]{McQOLe12}.

To summarize, for our isotropic analysis, we adopt systematic errors of
0.3 per cent for fitting and survey effects and 0.3 per cent for unmodeled
astrophysical shifts.  These are applied in quadrature.  
These systematic errors increase the error on the CMASS consensus $D_V$
value from 0.9 per cent to 1.0 per cent and the error on the LOWZ consensus
value $D_V$ from 2.0 per cent to 2.1 per cent.
For the anisotropic analysis, we apply the above effects in quadrature to
$\alpha$ and then add an additional independent systematic error of 0.5 per cent
in quadrature to $\epsilon$.  The impact on the measurement of $D_A$
and $H$ is subdominant to the statistical errors.

\begin{table*}
\begin{center}
\caption{Comparison between the different CMASS-DR11 results. While
our study focuses on the BAO information in the clustering signal,
all other studies model the anisotropic broadband clustering in order
to measure the cosmological distortion \citep{AP} and
redshift-space distortions.  In addition to the differences in 
modeling, only the results of this paper use reconstruction.
The $\alpha$ values from some of the other papers have been corrected 
to match our fiducial cosmological values.}
\begin{tabular}{llccc}
  \hline       
\multicolumn{5}{c}{Comparison between different CMASS-DR11 results}\\                 
source & method & $\alpha$ & $\alpha_{\parallel}$ & $\alpha_{\perp}$\\
\hline
this analysis & consensus & $1.019\pm 0.010$ & $0.968\pm 0.033$ & $1.045\pm 0.015$\\
\cite{Beutler13} & $P(k)$-multipoles & $1.023\pm 0.013$ & $1.005\pm 0.036$ & $1.021\pm 0.016$\\
\cite{Sam13} & $\xi(s)$-multipoles & $1.020\pm 0.013$ & $1.013\pm 0.035$ & $1.019\pm 0.017$\\
\cite{Chuang13b} & $\xi(s)$-multipoles & $1.025\pm 0.013$ & $0.996\pm 0.031$ & $1.039\pm 0.019$\\
\cite{Sanchez13} & $\xi(s)$-wedges & $1.011\pm 0.013$ & $1.001\pm 0.031$ &$1.016\pm 0.019$ \\
  \hline  
\end{tabular}
\label{tab:cmp}
\end{center}
\end{table*}

\subsection{The Distance Scale from BOSS BAO}

\newcommand{\zeff}{\ensuremath{z_{\rm eff}}}
\newcommand{\fid}{\ensuremath{{\rm fid}}}

As described in \citet{And12} and \citet{And13}, the value of $\alpha$ is directly related
to the ratio of the quantity $D_V(z)/r_d$ to its value in our fiducial model:
\begin{equation}
D_V/r_d = \alpha \left(D_V/r_d\right)_\fid.
\end{equation}
Similarly, $\alpha_\perp$ and $\alpha_\parallel$ measure the ratios of
$D_A/r_d$ and $r_d/H$, respectively, to their values in our fiducial model.

We opt to quote our results by writing these quantities as 
\begin{eqnarray}
D_V(\zeff) &=& \alpha D_{V,\fid}(\zeff) \left(r_d\over r_{d,\fid}\right), \\
D_A(\zeff) &=& \alpha_\perp D_{A,\fid}(\zeff) \left(r_d\over r_{d,\fid}\right), \\
H(\zeff)   &=& \alpha_\parallel H_\fid(\zeff) \left(r_{d,\fid}\over r_d\right).
\end{eqnarray}
With this form, we emphasize that only the ratio of $r_d$ between the adopted
and fiducial cosmology matters.  There are a variety of possible conventions and
fitting formulae available for $r_d$; any of these can be used so long as one
is consistent.  Moreover, within the usual class of CDM cosmologies, the CMB
data sets tightly constrain $r_d$.  For example, the \citet{PlanckXVI}
results imply $r_d$ to 0.4 per cent r.m.s. precision for the minimal
$\Lambda$CDM model and extensions to spatial curvature and low-redshift dark energy.
As this is somewhat tighter than our statistical errors on the $\alpha$'s,
it is reasonable to choose a form of the results that emphasizes the absolute
measurement of the distance scale.

The effective redshift of CMASS is $z_{\rm eff}=0.57$, while that of LOWZ
is $z_{\rm eff}=0.32$.
Our fiducial cosmology is $\Omega_m = 0.274$, $H_0=70\,$km\;s$^{-1}$\;Mpc$^{-1}$, 
$\Omega_bh^2=0.0224$, $n_s=0.95$, $m_\nu=0\,$eV, $w=-1$, $\Omega_K=0$, and $\sigma_8=0.8$.
Using this cosmology, we obtain $D_{V,\fid}(0.57) = 2026.49\,$Mpc,
$D_{A,\fid}(0.57)=1359.72\,$Mpc, and $H_\fid(0.57)=93.558\,$km\;s$^{-1}$\;Mpc$^{-1}$
for CMASS.
For LOWZ, we have $D_{V,\fid}(0.32) = 1241.47\,$Mpc,  
$D_{A,\fid}(0.32)=966.05\,$Mpc, 
and $H_\fid(0.32)=81.519\,$km\;s$^{-1}$\;Mpc$^{-1}$.

Inserting the constraints on $\alpha$, we find the primary isotropic results 
of this paper:
\begin{eqnarray}
D_V(0.57) &=& \left(2056 \pm 20 {\rm\;Mpc}\right)\left(r_{d}\over r_{d,\rm fid}\right) \\
D_V(0.32) &=& \left(1264 \pm 25 {\rm\;Mpc}\right)\left(r_{d}\over r_{d,\rm fid}\right)
\end{eqnarray}
for the post-reconstruction DR11 consensus values.  
For the anisotropic CMASS fit, we find
\begin{eqnarray}
D_A(0.57) &=& \left(1421 \pm 20 {\rm\ Mpc}\right) \left(r_d\over r_{d,\fid}\right), \\
H(0.57)   &=& \left(96.8 \pm 3.4 {\rm\ km\;s^{-1}\;Mpc^{-1}}\right) \left(r_{d,\fid}\over r_d\right),
\end{eqnarray}
with a correlation coefficient between $D_A$ and $H$ of 0.539 (in the sense that
higher $H$ favors higher $D_A$).   As described in Section~\ref{sec:ani_iso_compare},
we recommend the anisotropic values as our primary result at $z=0.57$
when fitting cosmological models.

When applying these constraints to test cosmology, one must of course consider
the variation in the sound horizon in the models.
Our fiducial cosmology has a sound horizon $r_{d,\rm fid}= 153.19\,$Mpc if
one adopts the definition in Eqs.~4 through 6 of
\citet[][hereafter, EH98]{Eis98}.
Alternatively, if one adopts the definition of the sound horizon in {\sc Camb},
one finds $r_{d,\rm fid}=149.28\,$Mpc, which is 2.6 per cent less. 
Much of the past BAO literature uses the EH98 convention, but we now recommend
using {\sc Camb} as it provides a transparent generalization to models with
massive neutrinos or other variations from vanilla CDM.
As discussed in \citet{Meh12}, the ratio of the EH98 and {\sc Camb} sound
horizons is very stable as a function of $\Omega_m h^2$ and $\Omega_bh^2$,
varying by only 0.03 per cent for the range $0.10<\Omega_ch^2<0.13$ and 
$0.020<\Omega_b h^2<0.023$.  Thus in evaluating the ratios that appear in
our expressions for $D_V$, $D_A$, and $H$, the choice is largely irrelevant.
We further find that for $0.113<\Omega_ch^2<0.126$, $0.021<\Omega_bh^2<0.023$ and
$m_\nu<1\,$eV, the approximation of
\begin{equation}
  r_d = \frac{55.234\,{\rm Mpc}}{(\Omega_ch^2+\Omega_bh^2)^{0.2538}(\Omega_bh^2)^{0.1278}
        (1+\Omega_\nu h^2)^{0.3794}}
\end{equation}
matches {\sc Camb} to better than 0.1 per cent, whatever the mass hierarchy.
One can use any of these conventions for the sound horizon in applying our 
results, so long as one is consistent in evaluating $r_d$ and $r_{d,\fid}$.

For comparison to past work, using the EH98 sound horizon, 
we find $D_V(0.57)/r_d = 13.42\pm 0.13$ and $D_V(0.32)/r_d = 8.25 \pm 0.16$. 
Using the {\sc Camb} sound horizon instead,
this shifts to $D_V(0.57)/r_d = 13.77\pm 0.13$ and $D_V(0.32)/r_d = 8.47 \pm 0.17$.

Finally, for the DR10 consensus values, we find
\begin{eqnarray}
D_V(0.57) &=& \left(2055 \pm 28 {\rm\;Mpc}\right)\left(r_{d}\over r_{d,\rm fid}\right) \\
D_V(0.32) &=& \left(1275 \pm 36 {\rm\;Mpc}\right)\left(r_{d}\over r_{d,\rm fid}\right), \\
D_A(0.57) &=& \left(1386 \pm 26 {\rm\ Mpc}\right) \left(r_d\over r_{d,\fid}\right), \\
H(0.57)   &=& \left(94.1 \pm 4.7 {\rm\ km\;s^{-1}\;Mpc^{-1}}\right) \left(r_{d,\fid}\over r_d\right).
\end{eqnarray}

\subsection{Comparison with other DR11 Studies and Past Work}

We next compare these distance measurements to prior results in the literature.
First, we note that the CMASS results from DR9, DR10, and DR11 are in close
agreement.  DR10 and DR11 are double and triple the survey volume of DR9,
respectively, and the survey geometry has become substantially more contiguous.
For the consensus values for DR9 after reconstruction, \citet{And12} found
$\alpha=1.033\pm 0.017$, in good agreement with the DR10 value of
$\alpha = 1.014\pm 0.014$ and DR11 value of $\alpha = 1.0144\pm 0.0098$.
The DR9 anisotropic analysis of \citet{And13} found $\alpha= 1.024\pm 0.029$,
also in good agreement with our results.

Similarly, the new values are in good agreement with DR9 analyses that
utilized the whole broadband correlation function and power spectrum,
without the broadband marginalization of the BAO-only analysis.
In particular, by fitting the full anisotropic clustering, these 
analyses are sensitive to the \citet{AP} distortion of the 
broadband clustering, which gives additional information on the 
product $D_A(z)H(z)$.  This requires modeling to separate from
the redshift-space distortions.
\citet{Rei12} model the monopole and quadrupole moments of the redshift-space
DR9 correlation function above $25\,h^{-1}$Mpc and find $D_V(0.57) =(2070\pm46)\,$Mpc
when allowing $f\,\sigma_8$, $D_A$ and $H$ as free parameters in the fit.
\citet{Kaz13} also use the correlation function, but fit to clustering
wedges rather than the multipoles.  They found consistent values.
\citet{Sanchez13} also analyzed the correlation function of the DR9 CMASS
sample using clustering wedges, fitting to the data above $44\,h^{-1}$Mpc,
but combined their constraints with those derived from other BAO measurements,
CMB and SNe data.  Their inferences are entirely consistent with the other DR9
measurements.
Finally, \citet{Chuang13} also constrained cosmology from the DR9 CMASS
correlation function, finding $D_V(0.57) =(2072\pm53)\,$Mpc.
These analyses are all clearly consistent with each other and with the more
precise values we find for DR11.

Similar analyses of the additional cosmological information residing
in the anisotropic broadband clustering have again been performed 
for the CMASS DR11 sample.  These are presented in a series 
of companion papers. 
\cite{Beutler13} analyses the power spectrum multipoles to measure
the BAO signal as well as redshift-space distortions using the
clustering model of \cite{Taruya10}. \cite{Sam13} and \cite{Chuang13b}
use correlation function multipoles, also including additional
information from redshift-space distortions. While \cite{Sam13}
uses the model suggested by \cite{Rei11}, \cite{Chuang13b} uses a
model suggested by \cite{Eis06}, \cite{CroSco06} and \cite{Mat08b}.
\cite{Sanchez13} analyses the correlation function wedges together
with external datasets to constrain a wide variaty of cosmological
parameters.
We compare the various results in Table~\ref{tab:cmp}, finding 
good agreement with those of this paper.  The agreement on $\alpha$
is close in most cases, while our BAO results differ by about 1~$\sigma$
when split anisotropically.  Perfect agreement is 
not expected: these analyses are gaining additional information 
on $D_A(z)H(z)$ from anisotropies in the broadband shape, but none
of them use reconstruction.  Given the difference in these 
treatments and the range of clustering statistics and template
modeling, we are encouraged by this level of agreement.

\citet{And12} compared the DR9 CMASS distance measurement to that from the
acoustic scale measured by 6dFGS \citep{Beutler11}, WiggleZ \citep{Bla11a}
and from the BAO detections in SDSS-III imaging data \citep{Pad07,Car12,Seo12}.
Our DR11 measurement remains in good agreement, within $1\,\sigma$, with these
studies.


The LOWZ measurements may be compared to previous work on the SDSS-II Luminous
Red Galaxy sample, which covered a similar area of sky but with fewer galaxies.
We find very close agreement with the results of
\citet{Per10} and \citet{Pad12}.
The survey footprints of these studies overlap substantially, but not entirely,
with those of DR11 LOWZ.
Moreover, \citet{Per10} included substantial volume at lower redshift through
the SDSS-II MAIN sample \citep{Str02} and 2dFGRS data sets \citep{colless03};
this resulted in an effective redshift of $z=0.275$.
Both \citet{Per10} and \citet{Pad12} used the SDSS-II LRG sample out to
$z=0.47$.  \citet{Pad12} used density-field reconstruction, while
\citet{Per10} did not.
However, the results are all similar, with differences that are well
within $1\sigma$.
For example, \citet{Pad12} measure $D_V(0.35)/r_d= 8.88\pm 0.17$;
if we adjust this to $z=0.32$ using the best-fit $\Lambda$CDM model and
convert to $\alpha$, we find $\alpha = 1.012\pm 0.019$, very similar to
the DR11 LOWZ value of $\alpha = 1.018\pm0.021$.

Previous analyses of the SDSS-II LRG sample have measured the anisotropic BAO
to determine $D_A$ and $H$ separately \citep{Oku08,Gaz09,CW11,Xu12b}.  As we have
not yet done an anisotropic analysis with LOWZ, we cannot directly compare to
these works.  However, all of these works inferred cosmological
parameters in good agreement with what we find in \S\ref{sec:params}, 
indicating that the
distance scales are compatible.

\section{Cosmological Parameters}
\label{sec:params}

\subsection{Data Sets and Methodology}
\label{sec:datasets}
We next consider the cosmological implications of our distance scale 
measurements.  From BOSS, we consider several different measurements.
First, we have the $D_V(0.57)$ measurement from CMASS galaxy clustering
in each of DR9, DR10, and DR11.  Second, we have the $D_V(0.32)$ measurement
from LOWZ clustering in DR10 and DR11.  Finally, we have the $D_A(0.57)$
and $H(0.57)$ joint measurement from CMASS in DR11.  In all cases, we 
use the post-reconstruction consensus values.  When not stated, 
we refer to the DR11 measurement.  We adopt the CMASS anisotropic values
as our best cosmological data set, labeling this as ``CMASS'', but 
also show results for the isotropic fit, labeling this as ``CMASS-iso''.

At points, we combine our CMASS and LOWZ measurements with two other BAO detections 
at different redshifts: the measurement of $D_V$ at $z=0.10$ from 
the 6dFGS \citep{Beutler11} and the measurement of $D_A$ and $H$ at 
$z=2.3$ in the Lyman $\alpha$ forest in BOSS \citep{Busca13,Slosar13,Kirkby13}.
These will be labeled as ``6dF'' and ``Ly$\alpha$F'', and the union the BAO data 
sets will be labelled in plots as ``BAO.''

As discussed in the previous section, our BOSS galaxy BAO measurements
are consistent with those from the WiggleZ survey \citep{Bla11a,Kaz14}
at $z=0.44$, 0.60, and 0.73 and with earlier SDSS-II LRG analyses \citep{Per10,Pad12,Xeaip,Meh12}.
We do not include these in our data compilations because of the
overlap in survey volume and redshift.

The anisotropies of the cosmic microwave background are an important 
part of our BAO analysis.  We consider three different CMB data sets.
The first is the Planck temperature anisotropy data set,
excluding lensing information from the 4-point correlations in the CMB
\citep{PlanckI},
supplemented by Wilkinson Microwave Anisotropy Probe (WMAP) 9-year 
polarization data \citep{Bennett13} to control the optical depth
to last scattering.  
This
is the so-called ``Planck+WP'' data set in \citet{PlanckXVI}; we will
abbreviate it as ``Planck''.  This is our primary CMB data set.

Our second CMB data set is the WMAP 9-year temperature and polarization
data set \citep{Bennett13}.  We abbreviate this as ``WMAP''.  We also
consider a third option, in which we combine WMAP 9-year data with 
the temperature power spectra from the finer scale and deeper data
from the South Pole Telescope \citep[SPT;][]{Story13} and Atacama
Cosmology Telescope \citep[ACT;][]{Das13}.  We abbreviate this as 
``WMAP+SPT/ACT'' or more briefly as 
``\textit{e}WMAP''. The likelihood code used is the publicly available \textsc{actlite} \citep{{Dunkley13},{Calabrese13}}.

As has been widely discussed \citep[e.g.,][]{PlanckXVI}, the
cosmological fits to these CMB data sets mildly disagree.  This issue can
be easily characterized by comparing the fitted ranges for $\Omega_m h^2$ in the
vanilla flat $\Lambda$CDM model.  The values range from $\Omega_m h^2 = 0.1427 \pm
0.0024$ for Planck \citep{PlanckXVI}, to $0.1371 \pm 0.0044$ for WMAP,
and then to $0.1353 \pm 0.0035$ for WMAP+SPT/ACT.  Note these
numbers shift slightly from others in the literature because, following the
Planck collaboration, we include a total of 0.06 eV in neutrino masses in all
our chains.
The 5 per cent shift in $\Omega_m h^2$ is $2\,\sigma$ between the central values of Planck and
WMAP+SPT/ACT and hence can produce noticeable variations in
parameters when combining our BAO results with those from the CMB.

We include cosmological distance measurements from Type Ia supernovae
by using the ``Union 2'' compilation by the Supernova Cosmology Project from
\citet{suzuki12}.
Supernova data are an important complement to our BAO data because they 
offer a precise measurement of the relative distance scale at low
redshifts.  We refer to this data set as ``SN''.  However, we note that
the recent recalibration of the SDSS-II and Supernova Legacy Survey photometric
zeropoints \citep{Bet13} will imply a minor adjustment, not yet available,
to the SNe distance constraints.

We use CosmoMC \citep{lewis02} Markov Chain Monte Carlo sampler 
to map the posterior distributions of these parameters. 
In most cases, we opt to compute chains using the CMASS DR9 data and then 
reweight those chains by the ratio of the DR10 or DR11 BAO likelihood
to the CMASS DR9 likelihood.  For each choice of cosmological model, CMB data
set, and inclusion of SNe, we ran a new chain. Using these chains, the
variations over choices of the BAO results could be produced quickly.
This approach is feasible because the new BAO distance measurements are well
contained within the allowed regions of the DR9 CMASS measurements.

We explore a variety of cosmological models, starting from the minimal
$\Lambda$CDM model.  We considered dark energy models of constant $w$ and varying
$w = w_0+(1-a)w_a$, which we notate as ``$w$CDM'' and ``$w_0w_a$CDM'', respectively.
In each case, we consider variations in spatial curvature, labeled 
as ``oCDM'', ``o$w$CDM'', and ``o$w_0w_a$CDM''.  
Following \citet{PlanckXVI}, we assume a minimal-mass normal hierarchy
approximated as a single massive eigenstate with $m_{\nu}=0.06\,$eV.  This is
consistent with recent oscillation data \citep{Forero12}.
We note this since even in this minimal neutrino mass case, the contribution 
to the expansion history is becoming noticeable in cosmological analyses.

\subsection{Comparison of BAO and CMB Distance Scales in $\Lambda$CDM}   \label{sec:CMB_scale_cmpr}
\begin{figure}
  \centering
  \resizebox{0.9\columnwidth}{!}{\includegraphics{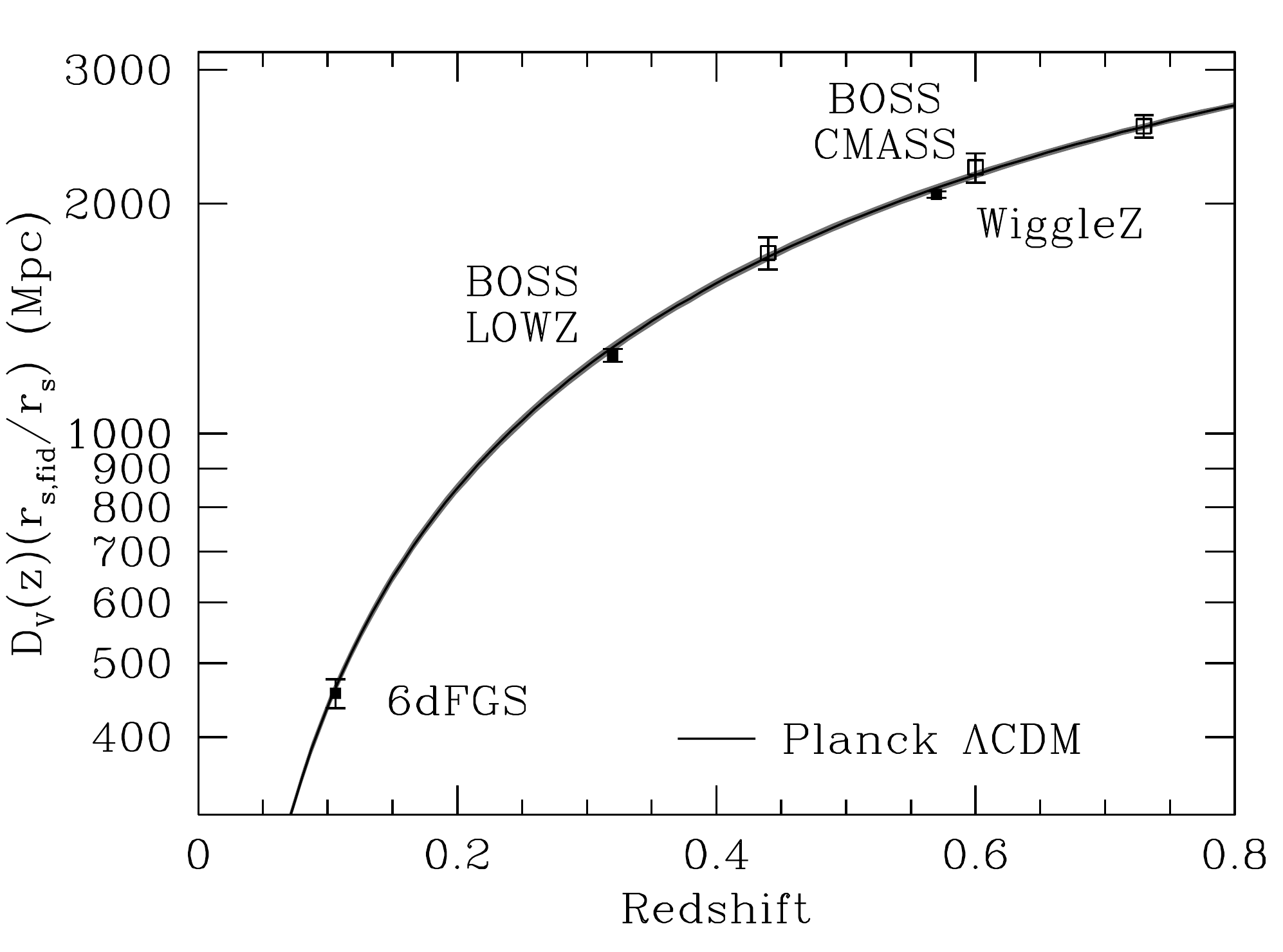}}
\caption{\label{fig:DV_vs_z}%
The distance-redshift relation from the BAO method on galaxy surveys. 
This plot shows $D_V(z)(r_{s,{\rm fid}}/r_d)$ versus $z$ from the DR11 CMASS and LOWZ 
consensus values from this paper, along with those from the acoustic
peak detection from the 6dFGS \citep{Beutler11} and WiggleZ survey \citep{Bla11a,Kaz14}.
The grey region shows the $1\sigma$  prediction for $D_V(z)$ from the Planck 2013 results,
assuming flat $\Lambda$CDM and using the Planck data without lensing combined with smaller-scale CMB observations
and WMAP polarization \citep{PlanckXVI}.  One can see the superb agreement in these cosmological
measurements. 
}
\end{figure}

\begin{figure}
  \centering
  \resizebox{0.9\columnwidth}{!}{\includegraphics{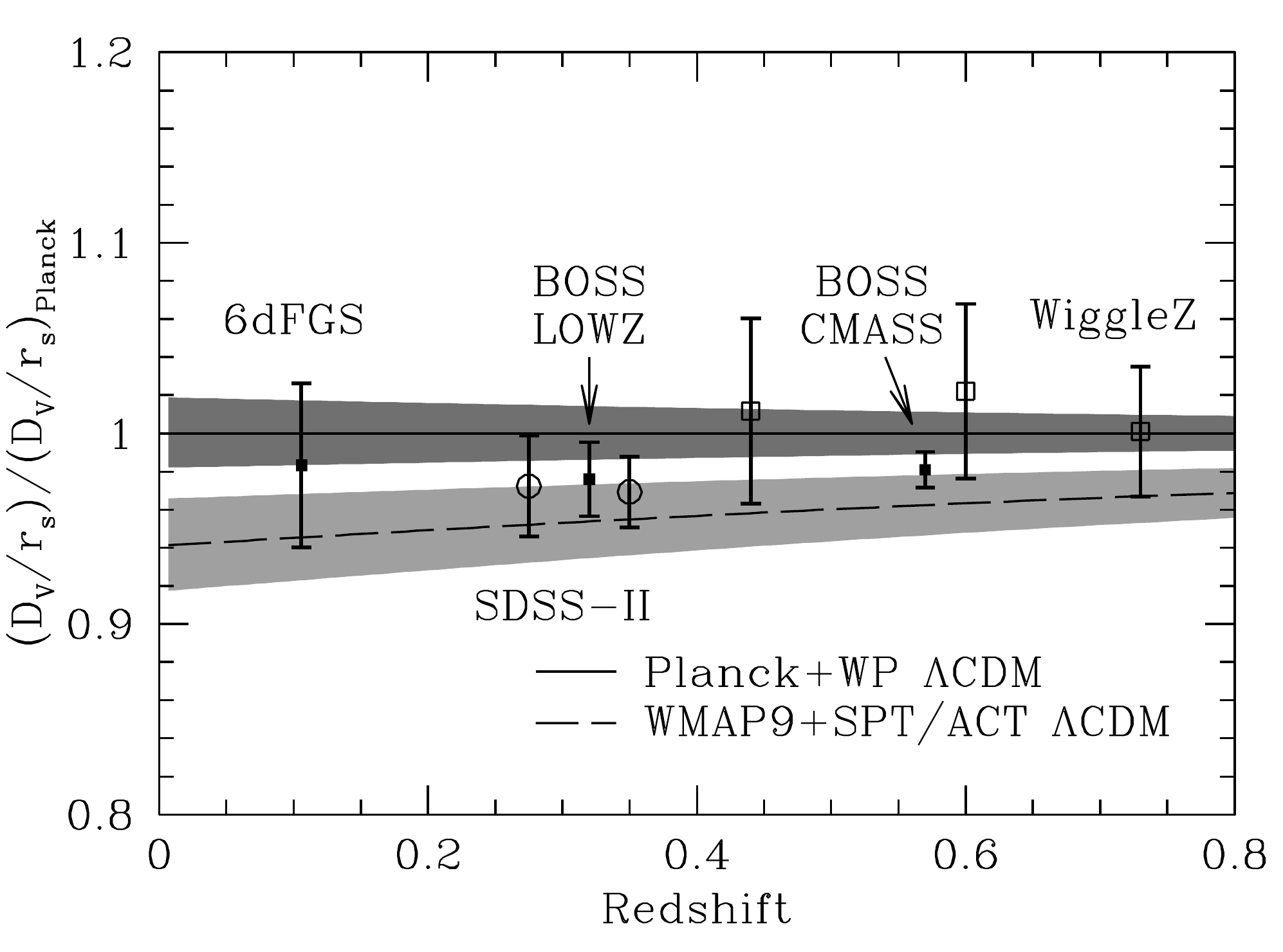}}
\caption{\label{fig:DVplanck}%
The $D_V(z)/r_d$ measured from galaxy surveys, divided by the best-fit flat $\Lambda$CDM
prediction from the Planck data.  All error bars are $1\,\sigma$.
The Planck prediction
is a horizontal line at unity, by construction.  The dashed line shows the best-fit flat $\Lambda$CDM 
prediction from the WMAP+SPT/ACT results, including their smaller-scale CMB compilation
\citep{Bennett13}.  In both cases, the grey region shows the 1~$\sigma$ variation in
the predictions for $D_V(z)$ (at a particular redshift, as opposed to the whole redshift range), which are dominated by uncertainties in $\Omega_m h^2$.  As the
value of $\Omega_m h^2$ varies, the prediction will move coherently up or down, with
amplitude indicated by the grey region.
One can see the mild tension between the two sets of CMB results, 
as discussed in \citet{PlanckXVI}.
The current galaxy BAO data fall in between the two predictions and are clearly consistent
with both.  
As we describe in Sec.~\ref{sec:ani_iso_compare}, the anisotropic CMASS fit would yield a prediction
for this plot that is 0.5 per cent higher than the isotropic CMASS fit; this value 
would fall somewhat closer to the Planck prediction.
In addition to the BOSS data points, we plot SDSS-II results as open circles,
that from \citet{Per10} at $z=0.275$ and from \citet{Pad12} at $z=0.35$.
These data sets have a high level of overlap with BOSS LOWZ and with each other, so one
should not include more than one in statistical fitting.  However, the results are 
highly consistent despite 
variations in the exact data sets and differences in methodology.  We also 
plot results from WiggleZ from \citet{Kaz14} as open squares; however, we note that the 
distance measurements from these three redshift bins are substantially correlated.
}
\end{figure}

\begin{table*}
\centering
\caption{Comparison of CMB flat $\Lambda$CDM predictions for the BAO distance scale
to our BOSS DR11 measurements.  We translate the CMB predictions to our
observables of $\alpha$, $\epsilon$, $\alpha_\parallel$, and $\alpha_\perp$.
As the CMB data sets vary notably in the value of $\Omega_m h^2$, we report these quantities.
We also translate our BOSS distance measurements to the constraints they
imply on $\Omega_mh^2$, assuming the flat $\Lambda$CDM model and using
the CMB measurements of $\Omega_b h^2$ and the angular acoustic scale.
We stress that this inference of $\Omega_m h^2$ is entirely model-dependent
and should not be used as a more general result of this paper.  However,
it does allow an easy comparison of the CMB and BOSS data sets in the 
context of $\Lambda$CDM.
}
\begin{tabular}{llccccc}
\hline
dataset & $z_{\rm eff}$ & $\alpha$ & $\epsilon$ & $\alpha_{\parallel}$ & $\alpha_{\perp}$ & $\Omega_m h^2$\\
\hline
Planck & 0.32 & $1.040 \pm 0.016$ & $-0.0033 \pm 0.0013$ & $1.033 \pm 0.014$ & $1.043 \pm 0.018$ & $0.1427 \pm 0.0024$\\
WMAP & 0.32 & $1.008 \pm 0.029$ & $-0.0007 \pm 0.0021$ & $1.007 \pm 0.025$ & $1.009 \pm 0.031$ & $0.1371 \pm 0.0044$ \\
eWMAP & 0.32 & $0.987 \pm 0.023$ & $0.0006 \pm 0.0016$ & $0.988 \pm 0.020$ & $0.986 \pm 0.025$ & $0.1353 \pm 0.0035$ \\
LOWZ & 0.32 & $1.018 \pm 0.021$ & - & - & - & $0.1387 \pm 0.0036$ \\
\hline
Planck & 0.57 & $1.031 \pm 0.013$ & $-0.0053 \pm 0.0020$ & $1.020 \pm 0.009$ & $1.037 \pm 0.015$ & $0.1427 \pm 0.0024$\\
WMAP & 0.57 & $1.006 \pm 0.023$ & $-0.0012 \pm 0.0034$ & $1.004 \pm 0.017$ & $1.007 \pm 0.027$ & $0.1371 \pm 0.0044$ \\
eWMAP & 0.57 & $0.988 \pm 0.019$ & $0.0010 \pm 0.0027$ & $0.990 \pm 0.013$ & $0.987 \pm 0.021$ & $0.1353 \pm 0.0035$\\ 
CMASS-iso & 0.57 & $1.0144 \pm 0.0098$ & - & - & - & $0.1389 \pm 0.0022$ \\
CMASS & 0.57 & $1.019 \pm 0.010$ & $-0.025 \pm 0.014$ & $0.968 \pm 0.033$ & $1.045 \pm 0.015$ & $0.1416 \pm 0.0018$\\
\hline
\label{tab:CMBpredictions}
\end{tabular}
\end{table*}

Results from the BAO method have improved substantially in the last decade
and we have now achieved measurements at a wide range of redshifts.
In Fig.~\ref{fig:DV_vs_z} we plot the distance-redshift relation obtained from
isotropic acoustic scale fits in the latest galaxy surveys.  In addition to
the values from this paper, we include the acoustic scale measurement from
the 6dFGS \citep{Beutler11} and WiggleZ survey \citep{Bla11a,Kaz14}.
As the BAO method actually measures $D_V/r_d$, we plot this quantity multiplied
by $r_{d,{\rm fid}}$.
The very narrow grey band here is the prediction from the Planck CMB dataset
detailed in Sec.~\ref{sec:datasets}.
In vanilla flat $\Lambda$CDM, the CMB acoustic peaks imply precise measurements
of $\Omega_m h^2$ and $\Omega_b h^2$, which in turn imply the acoustic scale.
The angular acoustic scale in the CMB then determines the distance to $z=1089$,
which breaks the degeneracy between $\Omega_m$ and $h$ once the low-redshift
expansion history is otherwise specified  (e.g., given $\Omega_K$, $w$, and
$w_a$).  The comparison between low-redshift BAO measurements and the
predictions from the CMB assuming a flat $\Lambda$CDM cosmology therefore
allows percent-level checks on the expansion history in this model over a
large lever arm in redshift.  One sees remarkably good agreement between the
BAO measurements and the flat $\Lambda$CDM predictions from CMB observations.

Fig.~\ref{fig:DVplanck} divides by the best-fit prediction from
\citet{PlanckXVI} to allow one to focus on a percent-level comparison.
In addition to the BAO data from the previous figure, we also plot older BAO
measurements based primarily on SDSS-II LRG data \citep{Per10,Pad12}.
This figure also shows the flat $\Lambda$CDM prediction from the WMAP+SPT/ACT
data set.
The predictions from these two data sets are in mild conflict due to the
$\sim 5$ per cent difference in their $\Omega_m h^2$ values, discussed
in Section \ref{sec:datasets}.
One can see that the isotropic BAO data, and the BOSS measurements in
particular, fall between the two predictions and are consistent with both. {Note that the recent revision of Planck data by \citet{Spe13} results in a value of $\Omega_m h^2$ that is in excellent agreement with our isotropic BAO measurements, which brings Planck predictions of the distance scale at $z=0.32$ and $z=0.57$ much closer to BOSS measurements.}

\begin{figure}
{\includegraphics[trim=0cm 0cm 0cm 0cm, clip=true,width=3.5in]{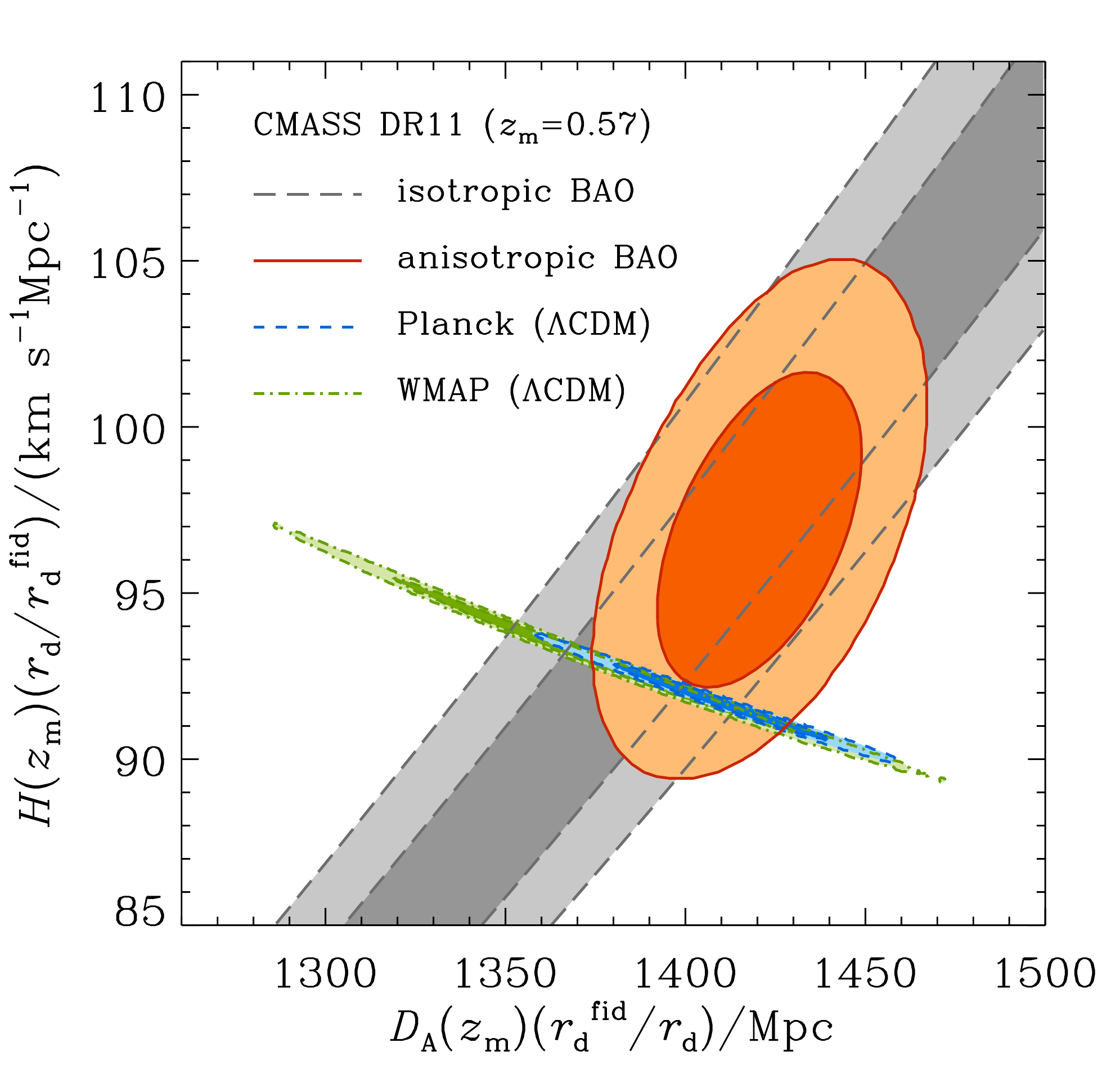}}
\caption{Comparison of the 68 and 95 per cent constraints in the
$D_A(0.57)(r_d^{\rm fid}/r_d)-H(0.57)(r_d^{\rm fid}/r_d)$ plane from
CMASS consensus anisotropic (orange) and isotropic (grey) BAO constraints.
The Planck contours correspond to Planck+WMAP polarization (WP) and no lensing.
The green contours show the constraints from WMAP9.}
\label{fig:DaH_CMB}
\end{figure}

Our 68 and 95 per cent  constraints in the
$D_A(0.57)(r_d^{\rm fid}/r_d)-H(0.57)(r_d/r_d^{\rm fid})$
plane from CMASS consensus anisotropic measurements are highlighted in
orange in Fig.~\ref{fig:DaH_CMB}.   In grey we overplot one-dimensional
1- and $2\,\sigma$ contours of our consensus isotropic BAO fit. 
Also shown in Fig.~\ref{fig:DaH_CMB} are the flat $\Lambda$CDM predictions
from the Planck and WMAP CMB data sets detailed in Section \ref{sec:datasets}.
The CMB constraints occupy a narrow ellipse defined by the extremely precise
measurement of the angular acoustic scale of 0.06 per cent \citep{PlanckXVI}.
The extent of the ellipse arises primarily from the remaining uncertainty on
the physical cold dark matter density, $\Omega_c h^2$;
Planck narrows the allowed range by nearly a factor of two compared with WMAP.
The CMASS isotropic BAO constraints are consistent with both CMB predictions
shown here.  The anisotropic constraints in particular prefer larger values
of $\Omega_c h^2$ (right edge of the WMAP contour) also favored by Planck.
Also evident in this plot is the offset between the best fit anisotropic
constraint on $H(0.57)(r_d/r_d^{\rm fid})$ (or $\epsilon$) and the flat
$\Lambda$CDM predictions from the CMB.

To make the flat $\Lambda$CDM comparison between the CMB and our BAO
measurements more quantitative, we report in Table~\ref{tab:CMBpredictions} 
the Planck, WMAP, and \textit{e}WMAP $\Lambda$CDM predictions for our isotropic 
and anistropic BAO observables at $z=0.32$ and $z=0.57$.
All three predictions are in good agreement with our isotropic measurements.  
The largest discrepancy between the Planck $\Lambda$CDM predictions and BOSS
measurements is about $1.5\,\sigma$ for the anisotropic parameter $\epsilon$
(or the closely related $\alpha_{\parallel}$) at $z=0.57$.
\textit{e}WMAP and BOSS disagree at about $1.8\,\sigma$ in $\epsilon$, which leads to
an approximately $2.2\,\sigma$ offset in $\alpha_{\perp}$.

Our measurements therefore provide no indication that additional parameters
are needed to describe the expansion history beyond those in flat $\Lambda$CDM.
However, it is also clear from Fig.~\ref{fig:DVplanck} and
Table~\ref{tab:CMBpredictions} that the disagreement between the WMAP+SPT/ACT
and Planck $\Lambda$CDM BAO predictions is comparable to the error on the BOSS
acoustic scale measurement.  
Under the assumption of a flat $\Lambda$CDM model, our anisotropic measurements
show a mild preference for the Planck parameter space over WMAP+SPT/ACT.
We are optimistic that the further analysis of the CMB data sets will resolve
the apparent difference.

Since the uncertainties in the $\Lambda$CDM prediction of the BAO
observables from the CMB are dominated by the uncertainty in $\Omega_c h^2$,
another way to summarize and compare the BAO measurements across redshift is
as a constraint on $\Omega_m h^2$ from the flat $\Lambda$CDM model holding
the CMB acoustic scale, $\ell_A$ \citep[Eq.~10 of][]{PlanckXVI},
and physical baryon density, $\Omega_b h^2$ fixed.
These values are given in the $\Omega_m h^2$ column of
Table~\ref{tab:CMBpredictions}.
We stress that these inferences depend critically on the assumption of a
flat $\Lambda$CDM expansion history.
Using this method, the BOSS inferences are more precise than the CMB and
fall between the WMAP and Planck constraints.
The isotropic CMASS analysis yields $\Omega_m h^2 = 0.1389 \pm 0.0022$,
in close agreement with the LOWZ result of $0.1387 \pm 0.0036$.
Our anisotropic analysis shifts to a notably larger value,
$\Omega_m h^2 = 0.1416 \pm 0.0018$, closer to the Planck measurement.
This shift in $\Omega_m h^2$ between the isotropic and anisotropic CMASS
fits is simply a restatement of the half sigma shift in $\alpha$ between our
isotropic and anistropic fits, discussed in Sec.~\ref{sec:ani_iso_compare}.

For our cosmological parameter estimation, we present Planck in most cases
but show the results for WMAP and WMAP+SPT/ACT in some cases so that the 
reader can assess the differences.  
For most combinations, the agreement is good.  This is because the BAO data
fall between the two CMB results and hence tend to pull towards reconciliation,
and because the low-redshift data sets dominate the measurements of dark energy
in cosmologies more complicated than the vanilla flat $\Lambda$CDM model.

Fig.~\ref{fig:DaH_CMB} and Table~\ref{tab:CMBpredictions} illustrate many of
the features of the $\Lambda$CDM model fits we present in
Table~\ref{tab:bigcos}.
For instance, the addition of a CMASS BAO measurement to the CMB improves the
constraint on $\Omega_m h^2$ by 40 per cent for Planck (with similar
improvements for the other CMB choices).
The central values for all three reported $\Lambda$CDM parameters
shift by one sigma between isotropic and anisotropic CMASS fits.
There are also one sigma shifts between Planck and WMAP/\textit{e}WMAP central parameter
values at fixed BAO measurements; taken together, WMAP+CMASS-iso or
\textit{e}WMAP+CMASS-iso and Planck+CMASS differ in their central values of
$\Omega_m$ and $H_0$ by about $2\,\sigma$.
Additionally combining with other BAO and SN measurements relaxes this
tension to about $1\,\sigma$.
Within the context of the $\Lambda$CDM model, the combination of CMB and BAO
provides 1 per cent (3 per cent) constraints on $H_0$ and $\Omega_m$,
respectively.  These constraints relax by a factor of 3 (2) in the most
general expansion history model, o$w_0w_a$CDM.

In \citet{And12} we showed that the BAO distance-redshift relation is
consistent with that measured by Type Ia supernovae.
This remains true with these DR11 results.

\subsection{Cosmological parameter estimates in extended models}

\begin{figure}
  \centering
  \resizebox{0.9\columnwidth}{!}{\includegraphics{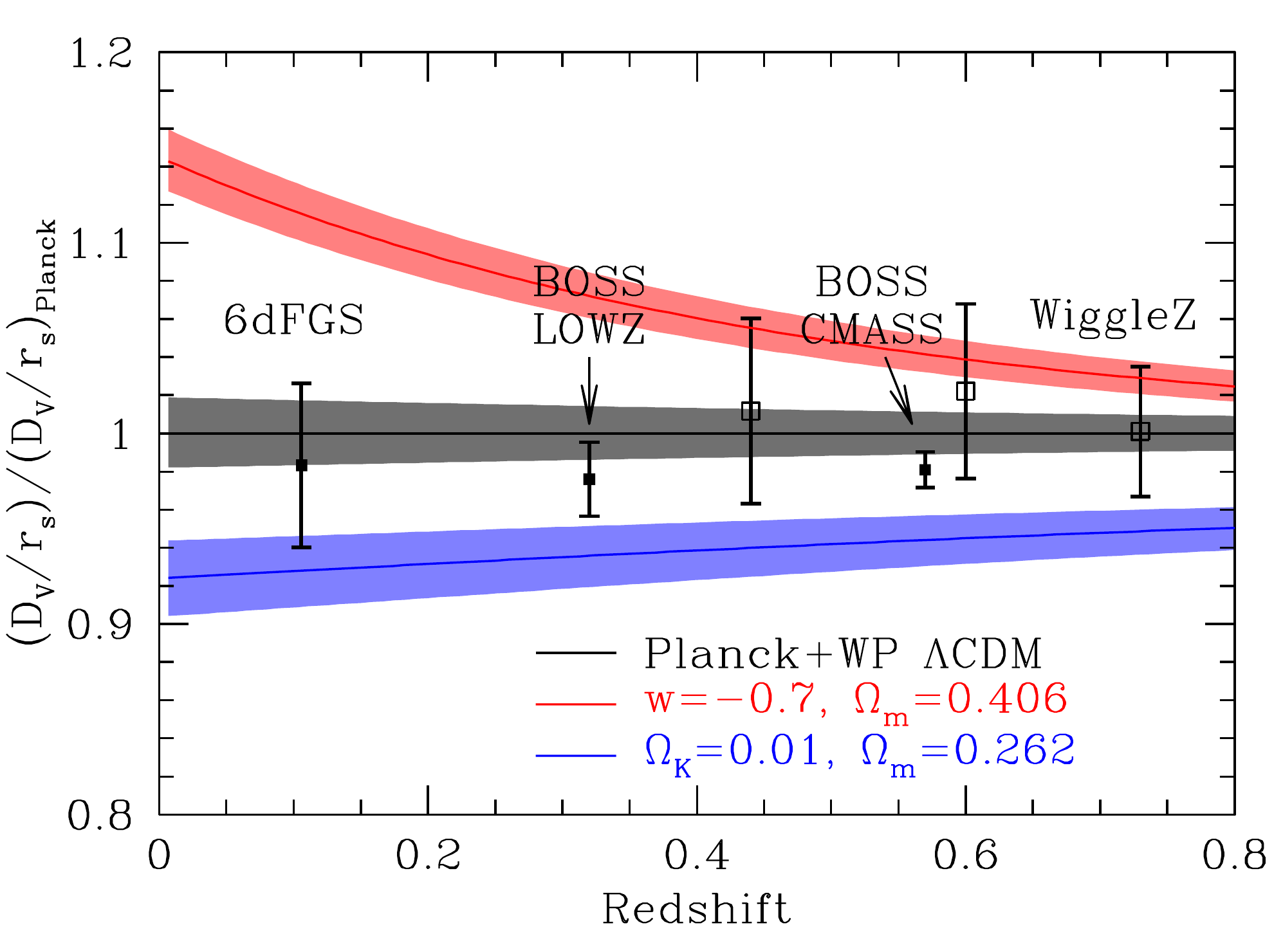}}
\caption{\label{fig:DVplanck_cosm}%
The $D_V(z)/r_d$ measured from galaxy surveys, divided by the best-fit flat $\Lambda$CDM
prediction from the Planck data.  All error bars are 1~$\sigma$.  We now vary the 
cosmological model for the Planck prediction.  Red shows the prediction assuming
a flat Universe with $w=-0.7$; blue shows the prediction assuming a closed Universe
with $\Omega_K = -0.01$ and a cosmological constant.  
}
\end{figure}

\begin{figure*}
  \centering
  \resizebox{0.9\textwidth}{!}{\includegraphics{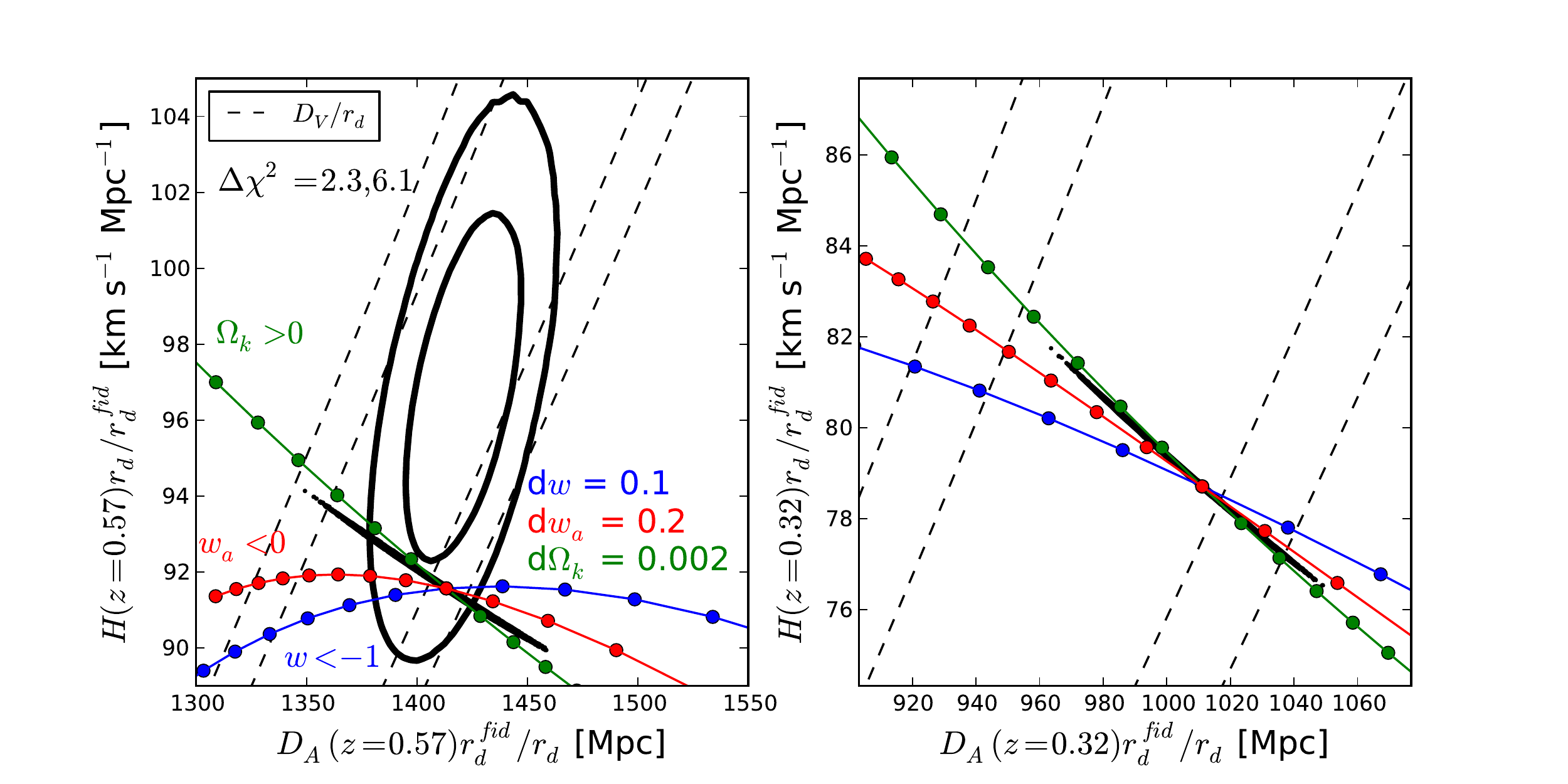}}
\caption{\label{fig:cosmoDAHz}%
 $\Delta
\chi^2 = 2.3, 6.1$ contours for both the isotropic (dashed) and anistropic
(solid) fits for the BAO observables at $z=0.57$ (left panel) and $z=0.32$ (right panel).  Overplotted are
the Planck flat $\Lambda$CDM predictions (narrow black band), where the
uncertainty is dominated by the uncertainty on $\Omega_c h^2$.  We overlay
predictions for the BAO observables for three one-parameter extensions
($\Omega_K$, $w$, or $w_a$) at fixed $\Omega_{c,b} h^2$ and CMB acoustic scale.
Given our relative errors on $D_A$ and $H$ at $z=0.57$, 
we can see that for the models of interest, the
improved constraint on $D_A$ is driving the improvement of our results from the isotropic to anisotropic analysis.
Also note that none of these models move along the long-axis of our anisotropic
constraints towards our best-fit values.
}
\end{figure*}

While the flat $\Lambda$CDM expansion history is sufficient to explain current CMB and
BAO measurements, the addition of precise low-redshift BAO distances greatly
improves constraints on parameters that generalize the flat $\Lambda$CDM expansion
history.  In this section we allow for non-zero spatial curvature ($\Omega_K$), a fixed equation
of state for dark energy ($w$), and a time-varying dark energy equation of state
($w_0$ and $w_a$).

Fig.~\ref{fig:DVplanck_cosm} illustrates the utility of BAO measurements for
constraining these additional parameters.  As one changes the model of the
spatial curvature or dark energy equation of state, the $\Omega_m$ and $H_0$
values required to simultaneously match the CMB measurement of $\Omega_m h^2$
and the distance to $z=1089$ change.  Here, we show the result assuming $w=-0.7$
for a flat cosmology, as well as that for a closed Universe with
$\Omega_K=-0.01$ and a cosmological constant.  One can see that these
predictions are sharply different from flat $\Lambda$CDM at low redshift.

In Fig.~\ref{fig:cosmoDAHz} we focus instead on the two effective redshifts of
our BAO observables, now examining how variations in the new parameters alter
predictions for both $D_A$ and $H$. For ease of comparison, we plot $\Delta
\chi^2 = 2.3, 6.1$ contours for both the isotropic (dashed) and anistropic
(solid) fits; these values correspond to 68 and 95 per cent confidence regions when
fitting two parameters.  The extremely narrow black ellipse (nearly parallel
with the green curve) shows the predictions from Planck in a flat $\Lambda$CDM model;
the uncertainty in the Planck predictions are dominated by the uncertainty in
cold dark matter density, 
$\Omega_c h^2$.  The three colored curves cross at the Planck best fit
cosmology, and show how the predictions for the BAO observables depend on each
of the extra parameters.  To produce these curves, we held $\Omega_c h^2$,
$\Omega_b h^2$, 
and the CMB acoustic scale fixed; the reader should keep in mind that
marginalizing over $\Omega_c h^2$ (the width of the Planck flat $\Lambda$CDM prediction)
will allow a larger range of parameter values to be consistent with both the CMB
and BAO observables compared with the fixed $\Omega_c h^2$ case.  

Fig.~\ref{fig:cosmoDAHz} already anticipates many of the results from detailed
joint parameter fitting reported in Tables \ref{tab:bigcos} and \ref{tab:bigcos2}.
For instance, by comparing the model variations to the isotropic BAO measurement
uncertainties, the constraint on $\Omega_K$ should be about 30 per cent better from
the $z=0.57$ isotropic BAO feature than the $z=0.32$ measurement.  For the case
combining CMASS isotropic and Planck constraints, the uncertainty on $\Omega_c
h^2$ (e.g., the extent of the flat $\Lambda$CDM Planck contour) degrades the constraint
on $\Omega_K$ from $\sim 0.002$ to 0.003.  For the wCDM model, the situation is
reversed: the lower redshift isotropic BAO measurement is more constraining even
though the fractional measurement errors are larger.  The wCDM model curves also
help explain why the Planck + CMASS-iso constraint, $w = -1.34 \pm
0.25$, does not improve the error on $w$ over our DR9 result, $w = -0.87 \pm
0.25$ \citep{And12}, even though our error on the BAO scale has improved from
1.7 per cent to 1 per cent: models with $w < -1$, favored by our CMASS isotropic BAO
measurement, produce smaller changes in the BAO observables at $z=0.57$ per
unit change in $w$ than models close to $w = -0.7$.  Moreover, the best-fit parameters for
both the CMB and BAO datasets have shifted between DR9 and DR11.  In fact,
combining CMASS-DR9 with Planck instead of WMAP7 yields $w=-1.18 \pm 0.25$.  In
that case, the BAO and CMB flat $\Lambda$CDM constraints have closer best fit $\alpha$
values.

The left panel of Fig.~\ref{fig:cosmoDAHz} also demonstrates why the CMASS
anisotropic constraints are more constraining than the isotropic ones,
particularly for dark energy parameters.  Variation in $w$ at fixed CMB acoustic
scale primarily shifts $D_A(0.57)$, and the anisotropic measurements provide
tighter constraints in that direction.  
Note that none of these extra parameters drive the expansion
rate as high as our anisotropic best fit to $H(0.57)$.

\begin{figure*}
  \centering
  \resizebox{0.9\columnwidth}{!}{\includegraphics{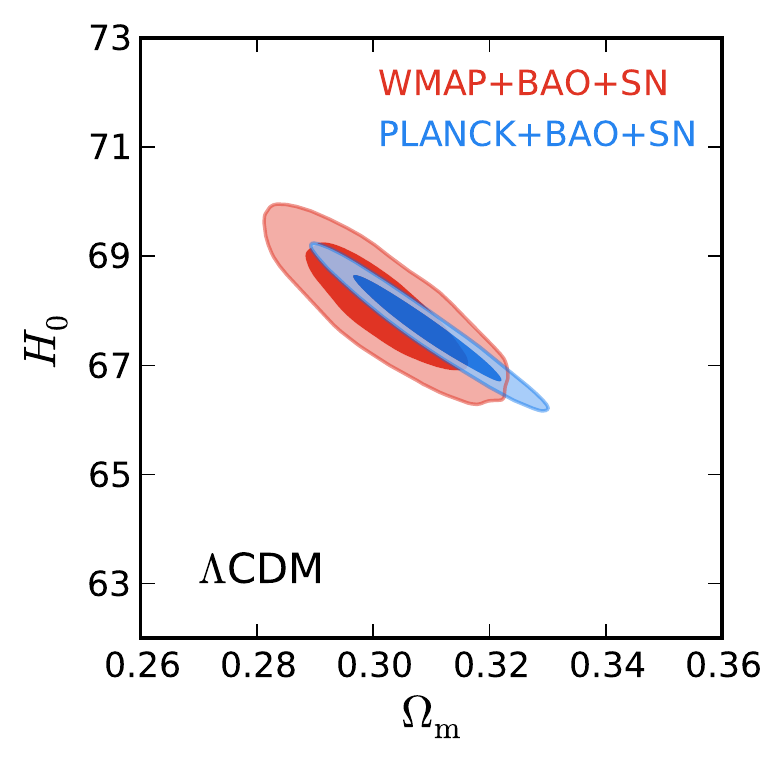}}
  \hspace{0.1\columnwidth}
  \resizebox{0.9\columnwidth}{!}{\includegraphics{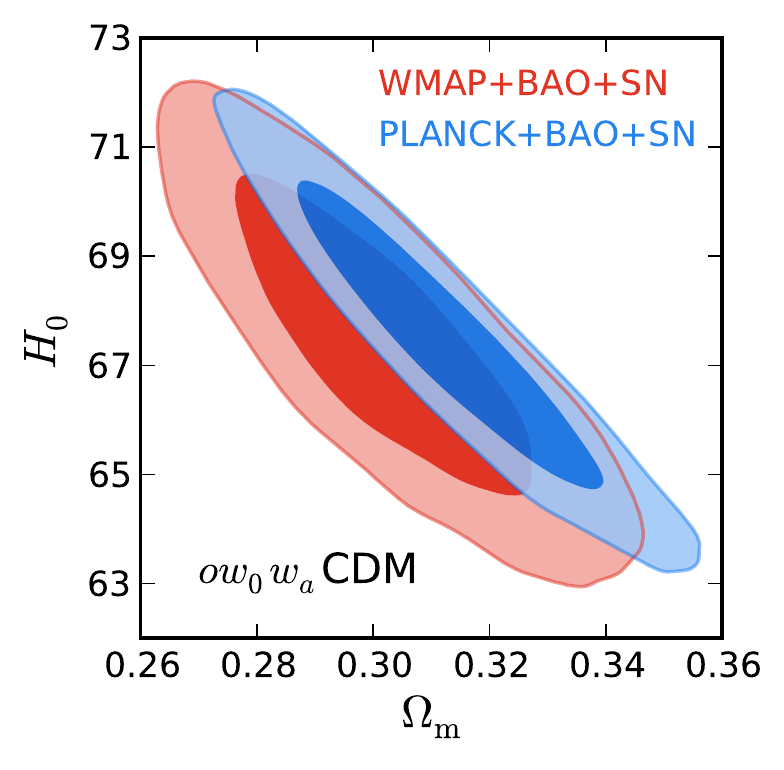}}
\caption{Constraints in the $\Omega_m$--$H_0$ plane for the combination CMB+BAO+SN, in the $\Lambda$CDM (left) and o$w_0w_a$CDM (right) cosmological models.
Here we show the degeneracy direction in this plane and we compare the allowed regions in this parameter space when the CMB dataset used is WMAP9 (red) or Planck (blue). The allowed regions open up when adding more degrees of freedom to the cosmological model; however, they still exclude values of 73 km s$^{-1}$ Mpc$^{-1}$ and above. The BAO and SN datasets make the $H_0$ values from WMAP9 and Planck agree with each other.  The best-fit value of $\Omega_m$ is slightly different between the two, but still consistent within 1 sigma.}
\label{fig:omh0_cmbbaosn}
\end{figure*}

\begin{figure}
  \centering
  \resizebox{0.9\columnwidth}{!}{\includegraphics{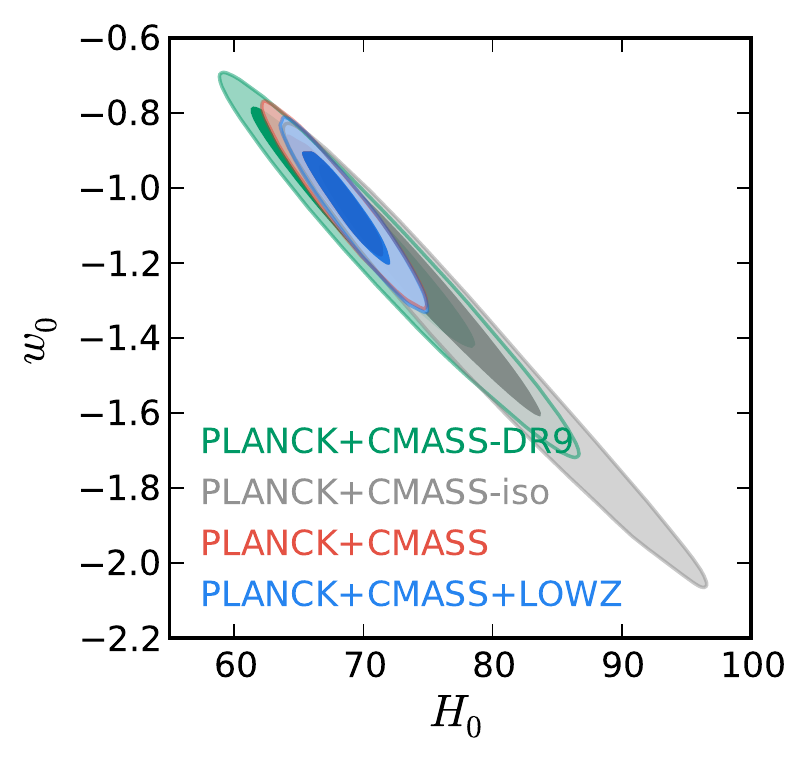}}
\caption{Constraints in the $H_0$--$w$ plane for Planck+DR9, Planck+CMASS-isotropic, 
Planck+CMASS (anisotropic), and Planck+CMASS+LOWZ. This figure shows the degeneracy 
between the Hubble constant and the dark energy equation of state, assumed constant in time.
Comparing with the Planck+CMASS-DR9 results (green contours), we note that the additional volume in CMASS-DR11 did not
help that much (dark contours). However performing an anisotropic BAO analysis of the same data really improves the constraints (red contours). The addition of the LOWZ isotropic BAO measurement at lower redshift (blue contours) has a marginal improvement over the CMASS anisotropic constraints, but it is a significant improvement over CMASS isotropic (see Table~\ref{tab:bigcos}).
}
\label{fig:h0w_cmbbao}
\end{figure}

In order to explore our results on the full multi-dimensional parameter space in
which we derive our cosmological constraints, we now describe the results of our
MCMC chains. Here we use our BAO measurements in combination with CMB results, 
and supplemented at times by SN
data and other BAO measurements, doing the analysis in the context of
different cosmological models. We first start by comparing constraints on the
parameters $\Omega_{\rm m} h^{2}$, $\Omega_{\rm m}$, and $H_{0}$ from our
different BAO datasets in Table~\ref{tab:bigcos}. In this case we combine BAO
with different CMB datasets: Planck, WMAP9, or \textit{e}WMAP, in the
$\Lambda$CDM, oCDM, or \textit{w}CDM cosmological models.  We
find that all CMB+BAO combinations return similar cosmological fits 
in $\Lambda$CDM and oCDM models, with $H_0$ around 68 km s$^{-1}$ Mpc$^{-1}$,
$\Omega_m$ around 0.30, and negligible spatial curvature.
Somewhat more variation is seen in the \textit{w}CDM case, because
of a degeneracy between $w$ and $H_0$ that is described later in this
section.  However, these variations are accompanied by larger formal 
errors and are highly consistent with the $\Lambda$CDM fit.
In our best constrained case
(Planck+CMASS in $\Lambda$CDM), we find a 1 per cent measurement of $\Omega_{\rm m} h^{2}$,
a 1 per cent measurement of $H_{0}$, and a 3 per cent measurement of
$\Omega_{\rm m}$.  These broaden only slightly in oCDM, to 2 per cent in 
$\Omega_{\rm m} h^2$.  We find a tight measurement of curvature, consistent
with a flat Universe with 0.003 error.

The degeneracy between $\Omega_{\rm m}$ and $H_{0}$ is shown in
Fig.~\ref{fig:omh0_cmbbaosn}. Here we compare the allowed parameter space in
the case of Planck and WMAP9, for the minimal $\Lambda$CDM model (left panel)
and the o$w_0w_a$CDM model \citep{{Chev01},{linder03}} (right panel) 
The latter was recommended by the Dark Energy Task Force
(hereafter DETF; \citealt{albrecht06}) for dark energy
Figure of Merit comparisons.
This model contains three more degrees of freedom (curvature and a
time-dependent equation of state for dark energy).
As was discussed in \citet{Meh12} and \citet{And12}, the combination
of CMB, BAO, and SNe data results produces a reverse distance
ladder that results in tight constraints on $H_0$ and $\Omega_m$ 
despite this flexibility in the cosmological model.
The CMB determines the acoustic scale, which the BAO uses to measure
the distance to intermediate redshift.  The SNe then transfer that
distance standard to low redshift, which implies $H_0$.  Combining this
with the CMB measurement of $\Omega_mh^2$ yields $\Omega_m$.
As shown in the figure, changing between Planck and WMAP data does not 
significantly shift these constraints.

As has been discussed before \citep{Meh12,And12,PlanckXVI}, 
the $H_0$ value inferred from this reverse distance ladder, 
$67.5\pm1.8$ km~s$^{-1}$~Mpc$^{-1}$,
is notably lower than some recent local measurements.  For example,
\citet{riess11} finds $H_0 = 73.8 \pm 2.4$ km s$^{-1}$ Mpc$^{-1}$ and  
\citet{Fre12} finds $H_0 = 74.3 \pm 2.1$ km s$^{-1}$ Mpc$^{-1}$.
The \citet{riess11} value would be decreased by a small recalibration 
of the water maser distance to NGC~4258 \citep{Hum13}.
\citet{Efs13} warns about possible
biases in the period-luminosity relation fits due to low-metallicity
Cepheids and finds a lower value of 
$H_0 = 70.6 \pm 3.3$ km s$^{-1}$ Mpc$^{-1}$ using only NGC~4258 as the
primary distance standard, including the maser recalibration,
or 
$H_0 = 72.5 \pm 2.5$ km s$^{-1}$ Mpc$^{-1}$ using three sets of 
primary standards.
While we believe that the comparison of these direct measurements to
our BAO results is important, the results are also affected by
the ongoing photometric recalibration of the SDSS and SNLS SNe
data \citep{Bet13}.  We have therefore not pursued a more quantitative
assessment at this time.

\begin{figure}
  \centering
  \resizebox{0.9\columnwidth}{!}{\includegraphics{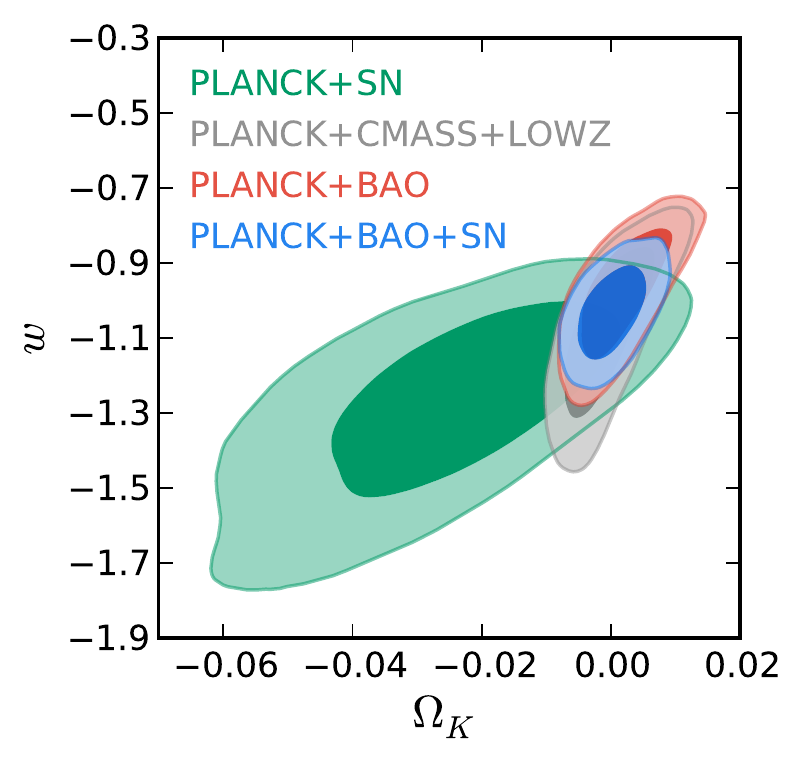}}
\caption{Constraints in the $\Omega_K$--$w$ plane for Planck+CMASS+LOWZ, 
Planck+BAO, Planck+BAO+SN, and Planck+SN.  The combination of CMB and SNe 
(green contours) 
has a substantial statistical degeneracy in this parameter space;
however, combining CMB and BAO 
strongly constrains the curvature
(grey contours for the LOWZ+CMASS results presented in this paper, 
and red contours when adding low and high redshift BAO measurements).
This makes the combination of CMB, BAO, and SNe (blue contours) a powerful 
one in this parameter space, yielding a fit 
centered around the $\Lambda$CDM values of $\Omega_K=0$ and $w=-1$.
}
\label{fig:okw_cmbbaosn}
\end{figure}

\begin{figure*}
  \centering
  \resizebox{0.9\columnwidth}{!}{\includegraphics{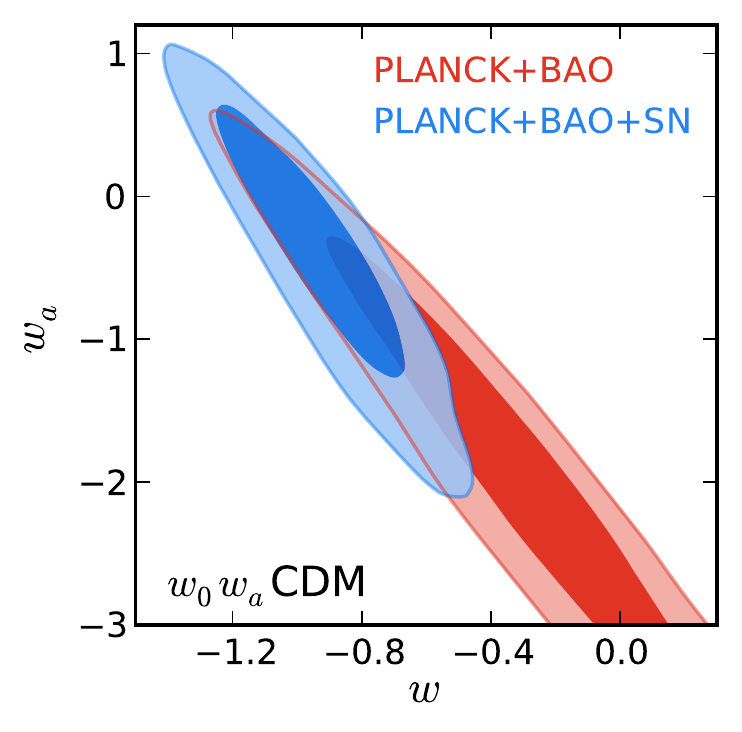}}
  \hspace{0.1\columnwidth}
  \resizebox{0.9\columnwidth}{!}{\includegraphics{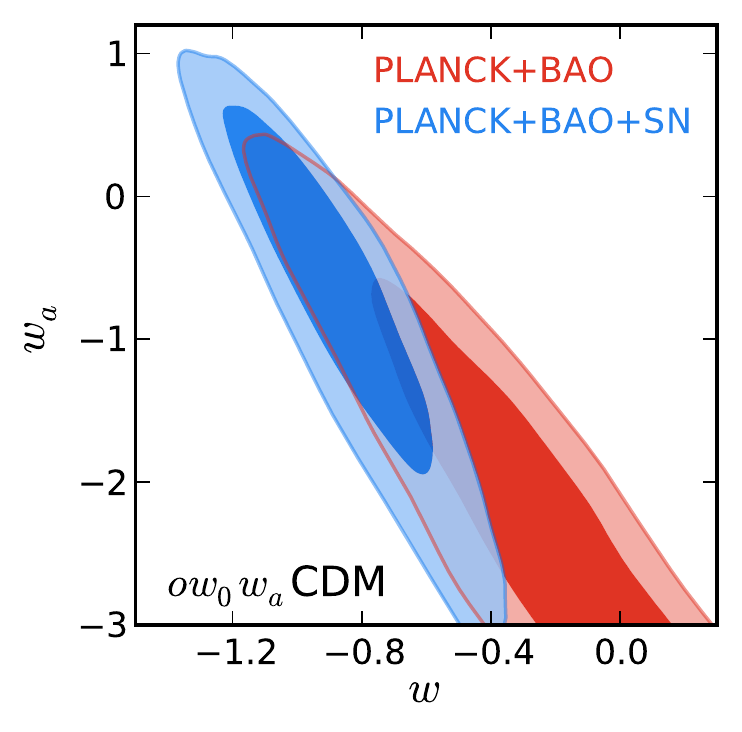}}
\caption{Constraints in the $w_a$--$w_0$ plane for Planck+BAO (red contours), and Planck+BAO+SN (blue contours), 
for both $w_0w_a$CDM (left panel) and o$w_0w_a$CDM (right panel). Note that the area of the 95 per cent contour in the right panel is
related to the dark energy Figure of Merit, as recommended by the Dark Energy Task Force. 
The degeneracy direction in clear in both panels, but the addition of SN data helps rule out very negative values of $w_a$. 
Furthermore, the best fit values for these two parameters in this case are closer to those of a $\Lambda$CDM cosmology
($w_0=-1$, $w_a=0$) than without SN data, in which case $\Lambda$CDM falls outside of the 68 per cent ellipse.  
}
\label{fig:w0wa_cmbbaosn}
\end{figure*}

We next discuss how BAO can help constrain additional degrees of freedom. In
Table~\ref{tab:bigcos2} we present our results in more general cosmological
models: $\Lambda$CDM, oCDM (adding curvature), $w$CDM (adding a equation of
state parameter for dark energy), o$w$CDM (adding both), $w_0w_a$CDM (allowing
for time-dependence in the e.o.s. of dark energy), and o$w_0w_a$CDM (our most
general model, for DETF comparisons). 
In each case, we begin with the results of combining our CMASS and LOWZ 
data with Planck, showing both isotropic and anisotropic CMASS cases.
We then extend the data combination with anisotropic CMASS to include additional BAO information
from the 6dFGS and Ly$\alpha$ forest, as well as SNe results from the
Union 2 compilation.  Finally, for the full combination of BAO and SNe, 
we vary the CMB measurements between Planck, WMAP, and \textit{e}WMAP to explore any
dependency on the tensions between those data sets.

We find that these datasets can constrain the equation of state of dark energy to
6 per cent and curvature to 0.2 per cent, although the time evolution of dark energy is still
unconstrained. In the DETF cosmology, we find a Figure of Merit value (inverse
square root of the minor of the covariance matrix containing the covariances of
$w_0$ and $w_a$) of 13.5. We find that the
anisotropic BAO measurement from CMASS-DR11 is much more powerful when
constraining the equation of state of dark energy (even when considering
time-evolving dark energy) than its isotropic counterpart.

Fig.~\ref{fig:h0w_cmbbao} shows the constraints in the $H_{0}$--$w$ plane for
different BAO datasets combined with Planck results. The degeneracy between
both parameters is quite evident, showing that a more negative value
for $w$ can result in a higher estimation for the Hubble constant. 
This effect can also be seen in Fig.~\ref{fig:DVplanck_cosm}; 
for the $w$CDM model, variations in the
distance to intermediate redshift produce larger variations in the
local distance scale.
The extent of the error contours as we vary the choice BAO data set
is somewhat complicated, as was illustrated in Fig.~\ref{fig:cosmoDAHz}. 
The efficacy of a given BAO distance precision to constrain $w$ degrades
as the fit shifts to more negative values of $w$; this is because models
with $w\ll-1$ have their dark energy disappear by intermediate redshift,
leaving the BAO and CMB constraints degenerate.  The improvement when
we change from the isotropic CMASS results to the anisotropic ones
is partially due to a shift in $w$ toward 0 and partially because of the
rotation of the contours to favor a $D_A$ constraint.
Overall, this figure also shows the consistency
between the various BOSS results and the tight constraints on $w$ that
the BAO now provides.
\comment{Beth, Ashley, ok with this rewrite?}\comment{AJR:yes}

We turn next to the o$w$CDM case, attempting to measure a constant dark energy 
equation of state in the presence of non-zero spatial curvature. 
These constraints are shown in Fig.~\ref{fig:okw_cmbbaosn} for 
several combinations of datasets. The
allowed region in this parameter space by the combination CMB+SN is large, 
due to a substantial degeneracy between $w$ and curvature.
This degeneracy is lifted by the BAO, which in combination with the CMB
sharply constrains the curvature.
Even without the SNe data, the BAO distance constraints are now 
strong enough to measure $w$ while simultaneously measuring $\Omega_K$.  
With Planck, CMASS, and LOWZ measurements alone, we find $w=-1.08\pm0.15$.
Further combine with the BAO measurement from 6dF and the Lyman-alpha
forest BAO measurement from BOSS, we find $w=-0.98\pm0.11$.  In both
cases, the fitted cosmologies are consistent with a flat Universe.
Hence, the BAO distance scale now provides enough precision, without
additional data beyond the CMB, to 
measure $w$ to 11 per cent even while marginalizing over spatial curvature.
It is remarkable that the BAO data prefers a flat Universe with $w=-1$
despite simultaneously opening two additional degrees
of freedom relative to the flat cosmological constant model.
We note the BAO and SNe constraints remain highly complementary
in their degeneracy directions;
adding the SN data shrinks the allowed region further, to $w=-1.04\pm0.07$
while remaining consistent with a flat Universe.
\comment{Substantial edits here.}

Finally, in Fig.~\ref{fig:w0wa_cmbbaosn} we show our constraints on a
time-dependent dark energy equation of state.
The contours show the allowed parameter space using the 
combination of CMB and BAO data, with and without SNe data,
in a flat $w_0w_a$CDM model (left panel) and an o$w_0w_a$CDM model with
curvature (right panel).  This parameter space is poorly
constrained, with a clear degeneracy between the $w_0$ and $w_a$ parameters, 
such that less
negative values of $w_0$ are related to more negative values of $w_a$. 
The addition of SN data suppress the likelihood of less negative $w_0$ values,
greatly reducing the allowed parameter space.  We note that allowing
non-zero spatial curvature degrades the dark energy constraints, but
not catastrophically.  The area covered by
the 2--$\sigma$ contour in the $ow_0w_a$CDM case 
is the DETF Figure of Merit.

\section{Conclusion}
\label{sec:discuss}

We have presented constraints on cosmology and the distance-redshift
relation from the Data Release 10 and 11 galaxy samples of the Baryon
Oscillation Spectroscopic sample.  These results, based on the largest
volume of the Universe ever surveyed at this high density ($8.4\,{\rm
Gpc}^3$, including both LOWZ and CMASS samples), provide the strongest
constraints on the distance-redshift relation achieved with the BAO
method and the most accurate determination of the distance scale in
the crucial redshift range where the expansion of the Universe begins
to accelerate.

The combination of large survey volume, high sampling density and high
bias of the LOWZ and CMASS galaxies allows detection of the acoustic
oscillation signal at unprecedented significance.
The acoustic signature is seen in both the power spectrum and the
correlation function, before density field reconstruction and after
reconstruction.  The measures are all highly consistent and the values and
errors are in accord with our models and mock catalogs \citep{Man12,Man13}.
Unlike our earlier results based upon DR9, we find density-field reconstruction
significantly improves our measurement of the acoustic scale
(see Fig.~\ref{fig:reconcomcmass}),
with the amount of improvement consistent with expectations if the underlying
cosmology were of the $\Lambda$CDM family.

With the larger volume of data, we now have statistically significant
evidence for variations in the target catalog density that are
correlated with seeing and stellar density.  We correct for these
systematics, along with a correction for redshift failures and
galaxies for which a redshift was not obtained due to fiber
collisions, using weights.  A similar procedure was used in
\citet{And12}, except the weights have been revised to correct for the
effects of seeing.

We fit the acoustic signature to an appropriately scaled template in both
the correlation function and power spectrum, marginalizing over broad-band
shape.  Our results are insensitive to the model of broad-band power and
highly consistent between configuration- and Fourier-space.
As an extension of the work reported in \citet{And12}, we now explicitly
consider the effects of binning in the correlation function and power spectrum
and combine the two methods using several different binning choices.
We measure a spherically averaged distance,
$D_V\equiv [cz(1+z)^2D_A/H]^{1/3}$,
in units of the sound-horizon, $r_d$, at two ``effective'' redshifts:
$z=0.32$ and $z=0.57$.
Our consensus results for the distance, including a budget for systematic
errors, are
$D_V(0.32)=(1264 \pm 25 {\rm\;Mpc})\left(r_{d}/r_{d,\rm fid}\right)$
and 
$D_V(0.57)=(2056 \pm 20 {\rm\;Mpc})\left(r_{d}/r_{d,\rm fid}\right)$.
The measurement at $z=0.57$ is the first ever 1 per cent measurement of a
distance using the BAO method.

As in \citet{And13}, we have used the anisotropy in the measured configuration-space clustering,
induced by redshift-space distortions, to separately constrain the distance
along and across the line-of-sight.
We compress the dependence on the angle to the line-of-sight into two
statistics, either the multipole moments or ``wedges''.
We obtain consistent fits from both methods.
A detailed study of possible systematics in inferences from anisotropic
clustering is presented in \citet{Vargas13}.
Our consensus results for the CMASS sample at $z=0.57$ are
$D_A(0.57) = \left(1421 \pm 20 {\rm\ Mpc}\right) \left(r_d/ r_{d,\fid}\right)$
and 
$H(0.57)   = \left(96.8 \pm 3.4 {\rm\ km/s/Mpc}\right) \left(r_{d,\fid}/ r_d\right)$ with a correlation coefficient between the two of $0.539$.

\citet{Sam13}, \citet{Beutler13} and \citet{Sanchez13}
have used the correlation function and power spectrum over a wide range of
scales, along with a model for the broad-band power, to constrain cosmological
parameters including the distance-redshift relation and $H(z)$.
We find excellent agreement between their results and the pure-BAO measurement
described here, despite differences in the procedure.
This is not unexpected, in that the bulk of the information is contained
in the acoustic signal rather than the broad-band power.

The BOSS results provide the tightest constraints in an reverse distance
ladder that tightly constrains the expansion rate from $z\sim 0$ to $0.6$.
The measurements reported here are in excellent agreement with earlier BAO
results by BOSS \citet{And12} and by other groups
\citep{Per10,Beutler11,Bla11a,Pad12}.

The DR11 $D_V$ distance to $z=0.57$ is approximately $1.8$ per cent
smaller than that reported to the same redshift based on the DR9 data.
This shift is approximately $1\,\sigma$ relative to the DR9 error bars.
As the data set has tripled in size, such a shift is consistent with expectations.
Both the $z\simeq 0.32$ and $z\simeq 0.57$ distance measurements are highly
consistent with expectations from the Planck and WMAP CMB measurements
assuming a $\Lambda$CDM model, lying approximately midway between the
inferences from the two experiments.  Our results for $D_A$ and $H$ are
similarly consistent with both CMB data sets; in detail, the anisotropic
results are slightly closer to the Planck best fit $\Lambda$CDM prediction.
While there are some mild tensions between the CMB data sets, the distance scale
inferred from acoustic oscillations in the distant Universe ($z\simeq 10^3$)
and in the local Universe ($z<1$) are in excellent agreement with the
predictions of a $\Lambda$CDM model, with gravity well described by General
Relativity and with a time-independent and spatially constant dark energy
with equation-of-state $p=-\rho$.

The BOSS will finish data taking within the next year.  Along with the
additional data, constraints at higher $z$ from the Ly$\alpha$ forest,
improvements in the analysis and a full systematic error study, we expect
BOSS to provide the definitive measurement of the absolute distance scale
out to $z\simeq 0.7$ for some time to come.

\section{Acknowledgements}


Funding for SDSS-III has been provided by the Alfred P. Sloan
Foundation, the Participating Institutions, the National Science
Foundation, and the U.S. Department of Energy Office of Science. The
SDSS-III web site is http://www.sdss3.org/.

SDSS-III is managed by the Astrophysical Research Consortium for the
Participating Institutions of the SDSS-III Collaboration including the
University of Arizona,
the Brazilian Participation Group,
Brookhaven National Laboratory,
University of Cambridge,
Carnegie Mellon University,
University of Florida,
the French Participation Group,
the German Participation Group,
Harvard University,
the Instituto de Astrofisica de Canarias,
the Michigan State/Notre Dame/JINA Participation Group,
Johns Hopkins University,
Lawrence Berkeley National Laboratory,
Max Planck Institute for Astrophysics,
Max Planck Institute for Extraterrestrial Physics,
New Mexico State University,
New York University,
Ohio State University,
Pennsylvania State University,
University of Portsmouth,
Princeton University,
the Spanish Participation Group,
University of Tokyo,
University of Utah,
Vanderbilt University,
University of Virginia,
University of Washington,
and Yale University.

We acknowledge the use of the Legacy Archive for Microwave Background
Data Analysis (LAMBDA). Support for LAMBDA is provided by the NASA
Office of Space Science.

This research used resources of the National Energy Research Scientific
Computing Center, which is supported by the Office of Science of the
U.S. Department of Energy under Contract No. DE-AC02-05CH11231.

Some of the CMASS reconstruction and MCMC computations were supported by
facilities and staff of the Yale University Faculty of Arts and
Sciences High Performance Computing Center.

LOWZ reconstruction computations were supported by the facilities and
staff of the UK Sciama High Performance Computing cluster supported by
SEPNet and the University of Portsmouth. Power spectrum calculations,
and fitting made use of the COSMOS/Universe supercomputer, a UK-CCC
facility supported by HEFCE and STFC in cooperation with CGI/Intel.

We thank Christian Reichardt for his help in using the \textsc{actlite}
likelihood code.

\begin{table*}
\centering
\begin{tabular}{lllllll}
\hline
Cosmological & Data Sets & $\Omega_{\rm m} h^{2}$ & $\Omega_{\rm m}$ & $H_{0}$ & $\Omega_{\rm K}$ & $w_{0}$ \\
Model & & & & km s$^{-1}$ Mpc$^{-1}$ & & \\
\hline
$\Lambda$CDM & Planck & 0.1427 (24) & 0.316 (16) & 67.3 (11) & \nodata & \nodata \\
$\Lambda$CDM & WMAP & 0.1371 (44) & 0.284 (25) & 69.6 (21) & \nodata & \nodata \\
$\Lambda$CDM & \textit{e}WMAP & 0.1353 (35) & 0.267 (19) & 71.3 (17) & \nodata & \nodata \\
\hline
$\Lambda$CDM & Planck + CMASS-DR9 & 0.1428 (20) & 0.317 (13) & 67.1 (9) & \nodata & \nodata \\
$\Lambda$CDM & Planck + CMASS-iso & 0.1408 (15) & 0.304 (9) & 68.1 (7) & \nodata & \nodata \\
$\Lambda$CDM & Planck + CMASS & 0.1418 (15) & 0.311 (9) & 67.6 (6) & \nodata & \nodata \\
$\Lambda$CDM & Planck + LOWZ & 0.1416 (20) & 0.309 (13) & 67.7 (9) & \nodata & \nodata \\
$\Lambda$CDM & WMAP + CMASS-DR9 & 0.1403 (30) & 0.305 (16) & 67.9 (12) & \nodata & \nodata \\
$\Lambda$CDM & WMAP + CMASS-iso & 0.1383 (25) & 0.292 (10) & 68.8 (8) & \nodata & \nodata \\
$\Lambda$CDM & WMAP + CMASS & 0.1400 (24) & 0.302 (10) & 68.0 (8) & \nodata & \nodata \\
$\Lambda$CDM & WMAP + LOWZ & 0.1379 (30) & 0.289 (15) & 69.2 (13) & \nodata & \nodata \\
$\Lambda$CDM & \textit{e}WMAP + CMASS-DR9 & 0.1401 (25) & 0.295 (14) & 69.0 (11) & \nodata & \nodata \\
$\Lambda$CDM & \textit{e}WMAP + CMASS-iso & 0.1393 (18) & 0.290 (9) & 69.3 (7) & \nodata & \nodata \\
$\Lambda$CDM & \textit{e}WMAP + CMASS & 0.1409 (17) & 0.300 (9) & 68.6 (7) & \nodata & \nodata \\
$\Lambda$CDM & \textit{e}WMAP + LOWZ & 0.1409 (24) & 0.282 (13) & 69.9 (11) & \nodata & \nodata \\
\hline
oCDM & Planck + CMASS-DR9 & 0.1418 (25) & 0.323 (15) & 66.3 (14) & -0.0029 (42) & \nodata \\
oCDM & Planck + CMASS-iso & 0.1418 (25) & 0.303 (9) & 68.4 (8) & +0.0016 (30) & \nodata \\
oCDM & Planck + CMASS & 0.1420 (24) & 0.311 (9) & 67.5 (8) & +0.0000 (30) & \nodata \\
oCDM & Planck + LOWZ & 0.1418 (25) & 0.307 (14) & 68.0 (15) & +0.0007 (42) & \nodata \\
oCDM & WMAP + CMASS-DR9 & 0.1372 (41) & 0.306 (15) & 67.0 (14) & -0.0050 (48) & \nodata \\
oCDM & WMAP + CMASS-iso & 0.1370 (41) & 0.290 (11) & 68.7 (10) & -0.0017 (41) & \nodata \\
oCDM & WMAP + CMASS & 0.1378 (41) & 0.300 (10) & 67.8 (9) & -0.0027 (41) & \nodata \\
oCDM & WMAP + LOWZ & 0.1371 (41) & 0.291 (15) & 68.7 (16) & -0.0017 (50) & \nodata \\
oCDM & \textit{e}WMAP + CMASS-DR9 & 0.1356 (35) & 0.302 (15) & 67.1 (14) & -0.0079 (44) & \nodata \\
oCDM & \textit{e}WMAP + CMASS-iso & 0.1357 (36) & 0.288 (10) & 68.6 (9) & -0.0045 (37) & \nodata \\
oCDM & \textit{e}WMAP + CMASS & 0.1360 (36) & 0.296 (9) & 67.7 (8) & -0.0061 (38) & \nodata \\
oCDM & \textit{e}WMAP + LOWZ & 0.1357 (35) & 0.288 (15) & 68.6 (16) & -0.0046 (47) & \nodata \\
\hline
$w$CDM & Planck + CMASS-DR9 & 0.1439 (23) & 0.284 (48) & 72.1 (71) & \nodata & -1.19 (26) \\
$w$CDM & Planck + CMASS-iso & 0.1439 (23) & 0.251 (36) & 76.4 (66) & \nodata & -1.33 (24) \\
$w$CDM & Planck + CMASS & 0.1425 (22) & 0.305 (20) & 68.5 (25) & \nodata & -1.04 (11) \\
$w$CDM & Planck + LOWZ & 0.1432 (24) & 0.279 (26) & 72.0 (36) & \nodata & -1.17 (14) \\
$w$CDM & WMAP + CMASS-DR9 & 0.1378 (50) & 0.324 (47) & 65.9 (65) & \nodata & -0.91 (27) \\
$w$CDM & WMAP + CMASS-iso & 0.1380 (47) & 0.288 (37) & 69.9 (61) & \nodata & -1.04 (26) \\
$w$CDM & WMAP + CMASS & 0.1354 (43) & 0.323 (18) & 64.8 (25) & \nodata & -0.84 (12) \\
$w$CDM & WMAP + LOWZ & 0.1373 (47) & 0.292 (25) & 68.8 (36) & \nodata & -0.99 (16) \\
$w$CDM & \textit{e}WMAP + CMASS-DR9 & 0.1366 (35) & 0.341 (34) & 63.5 (36) & \nodata & -0.79 (13) \\
$w$CDM & \textit{e}WMAP + CMASS-iso & 0.1360 (35) & 0.311 (22) & 66.2 (30) & \nodata & -0.87 (12) \\
$w$CDM & \textit{e}WMAP + CMASS & 0.1354 (33) & 0.330 (16) & 64.1 (20) & \nodata & -0.80 (8) \\
$w$CDM & \textit{e}WMAP + LOWZ & 0.1358 (35) & 0.299 (23) & 67.5 (29) & \nodata & -0.90 (10) \\
\hline
\end{tabular}
\caption{Constraints by different CMB+BAO datasets in the cosmological parameters $\Omega_{\rm m} h^{2}$, $\Omega_{\rm m}$, and $H_{0}$ in the $\Lambda$CDM model, $o$CDM model where we also show constraints in $\Omega_{\rm K}$ and $w$CDM model where we also show constraints in $w_0$. Here we compare the constraining power of different BAO measurements at different redshifts (e.g. LOWZ vs. CMASS) as well as different analyses (isotropic vs. anisotropic). We refer to 'CMASS-DR9' as the isotropic measurement presented in Anderson et al. 2012, 'CMASS-iso' as the isotropic measurement from the CMASS sample in this work, and the anisotropic one as simply 'CMASS'. 'LOWZ' is the isotropic measurement of the LOWZ sample also shown here. Given the volume sampled by the CMASS sample, and the constraining power of the anisotropic analysis, we get the best cosmological constraints in this case, especially when combined with Planck.}
\label{tab:bigcos}
\end{table*}

\begin{table*}
\centering
\begin{tabular}{llllllll}
\hline
Cosmological & Data Sets & $\Omega_{\rm m} h^{2}$ & $\Omega_{\rm m}$ & $H_{0}$ & $\Omega_{\rm K}$ & $w_{0}$ & $w_{a}$ \\
Model & & & & km s$^{-1}$ Mpc$^{-1}$ & & & \\
\hline
$\Lambda$CDM & Planck + CMASS-iso + LOWZ & 0.1405 (14) & 0.302 (8) & 68.2 (6) & \nodata & \nodata & \nodata \\
$\Lambda$CDM & Planck + CMASS + LOWZ & 0.1415 (14) & 0.308 (8) & 67.8 (6) & \nodata & \nodata & \nodata \\
$\Lambda$CDM & Planck + BAO & 0.1417 (13) & 0.310 (8) & 67.6 (6) & \nodata & \nodata & \nodata \\
$\Lambda$CDM & Planck + CMASS + LOWZ + SN & 0.1414 (13) & 0.308 (8) & 67.8 (6) & \nodata & \nodata & \nodata \\
$\Lambda$CDM & Planck + BAO + SN & 0.1416 (13) & 0.309 (8) & 67.7 (6) & \nodata & \nodata & \nodata \\
$\Lambda$CDM & WMAP + BAO + SN & 0.1399 (22) & 0.302 (8) & 68.1 (7) & \nodata & \nodata & \nodata \\
$\Lambda$CDM & \textit{e}WMAP + BAO + SN & 0.1411 (16) & 0.301 (8) & 68.5 (6) & \nodata & \nodata & \nodata \\
\hline
oCDM & Planck + CMASS-iso + LOWZ & 0.1417 (25) & 0.302 (8) & 68.5 (8) & +0.0017 (30) & \nodata & \nodata \\
oCDM & Planck + CMASS + LOWZ & 0.1420 (25) & 0.309 (8) & 67.8 (7) & +0.0006 (30) & \nodata & \nodata \\
oCDM & Planck + BAO & 0.1423 (25) & 0.311 (8) & 67.7 (7) & +0.0007 (29) & \nodata & \nodata \\
oCDM & Planck + CMASS + LOWZ + SN & 0.1416 (24) & 0.308 (8) & 67.9 (7) & +0.0003 (30) & \nodata & \nodata \\
oCDM & Planck + BAO + SN & 0.1419 (24) & 0.309 (8) & 67.7 (7) & +0.0004 (29) & \nodata & \nodata \\
oCDM & WMAP + BAO + SN & 0.1384 (40) & 0.300 (9) & 67.9 (8) & -0.0019 (40) & \nodata & \nodata \\
oCDM & \textit{e}WMAP + BAO + SN & 0.1364 (34) & 0.296 (8) & 67.9 (7) & -0.0054 (35) & \nodata & \nodata \\
\hline
$w$CDM & Planck + CMASS-iso + LOWZ & 0.1431 (22) & 0.274 (21) & 72.5 (32) & \nodata & -1.19 (13) & \nodata \\
$w$CDM & Planck + CMASS + LOWZ & 0.1425 (21) & 0.299 (16) & 69.1 (21) & \nodata & -1.07 (9) & \nodata \\
$w$CDM & Planck + BAO & 0.1419 (21) & 0.308 (14) & 67.9 (18) & \nodata & -1.01 (8) & \nodata \\
$w$CDM & Planck + CMASS + LOWZ + SN & 0.1426 (19) & 0.299 (12) & 69.1 (16) & \nodata & -1.07 (7) & \nodata \\
$w$CDM & Planck + BAO + SN & 0.1423 (19) & 0.305 (12) & 68.4 (14) & \nodata & -1.04 (6) & \nodata \\
$w$CDM & WMAP + BAO + SN & 0.1380 (33) & 0.307 (12) & 67.0 (16) & \nodata & -0.94 (8) & \nodata \\
$w$CDM & \textit{e}WMAP + BAO + SN & 0.1379 (28) & 0.312 (11) & 66.5 (15) & \nodata & -0.90 (7) & \nodata \\
\hline
o$w$CDM & Planck + CMASS-iso + LOWZ & 0.1418 (25) & 0.261 (31) & 74.1 (46) & -0.0022 (36) & -1.27 (21) & \nodata \\
o$w$CDM & Planck + CMASS + LOWZ & 0.1420 (24) & 0.298 (23) & 69.2 (27) & -0.0005 (44) & -1.08 (14) & \nodata \\
o$w$CDM & Planck + BAO & 0.1422 (24) & 0.315 (19) & 67.3 (20) & +0.0018 (44) & -0.98 (11) & \nodata \\
o$w$CDM & Planck + CMASS + LOWZ + SN & 0.1421 (25) & 0.298 (14) & 69.1 (16) & -0.0008 (34) & -1.07 (8) & \nodata \\
o$w$CDM & Planck + BAO + SN & 0.1422 (25) & 0.306 (13) & 68.2 (15) & +0.0002 (34) & -1.03 (7) & \nodata \\
o$w$CDM & WMAP + BAO + SN & 0.1374 (42) & 0.306 (13) & 67.1 (16) & -0.0010 (44) & -0.95 (8) & \nodata \\
o$w$CDM & \textit{e}WMAP + BAO + SN & 0.1357 (35) & 0.305 (13) & 66.7 (15) & -0.0039 (40) & -0.93 (8) & \nodata \\
\hline
$w_0w_a$CDM & Planck + CMASS-iso + LOWZ & 0.1434 (22) & 0.302 (53) & 69.8 (66) & \nodata & -0.90 (51) & -0.78 (124) \\
$w_0w_a$CDM & Planck + CMASS + LOWZ & 0.1431 (21) & 0.350 (41) & 64.3 (41) & \nodata & -0.54 (39) & -1.41 (101) \\
$w_0w_a$CDM & Planck + BAO & 0.1428 (21) & 0.361 (32) & 63.1 (29) & \nodata & -0.43 (30) & -1.62 (84) \\
$w_0w_a$CDM & Planck + CMASS + LOWZ + SN & 0.1433 (22) & 0.304 (17) & 68.7 (19) & \nodata & -0.98 (19) & -0.36 (64) \\
$w_0w_a$CDM & Planck + BAO + SN & 0.1431 (22) & 0.311 (16) & 67.8 (18) & \nodata & -0.93 (18) & -0.41 (62) \\
$w_0w_a$CDM & WMAP + BAO + SN & 0.1372 (43) & 0.302 (16) & 67.5 (17) & \nodata & -1.00 (16) & 0.16 (59) \\
$w_0w_a$CDM & \textit{e}WMAP + BAO + SN & 0.1366 (31) & 0.300 (15) & 67.5 (17) & \nodata & -1.04 (14) & 0.41 (40) \\
\hline
o$w_0w_a$CDM & Planck + CMASS-iso + LOWZ & 0.1419 (25) & 0.296 (50) & 70.0 (62) & -0.0042 (40) & -0.83 (45) & -1.41 (115) \\
o$w_0w_a$CDM & Planck + CMASS + LOWZ & 0.1417 (25) & 0.347 (38) & 64.2 (37) & -0.0039 (47) & -0.50 (34) & -1.79 (91) \\
o$w_0w_a$CDM & Planck + BAO & 0.1420 (24) & 0.361 (30) & 62.9 (27) & -0.0020 (47) & -0.40 (28) & -1.82 (82) \\
o$w_0w_a$CDM & Planck + CMASS + LOWZ + SN & 0.1419 (25) & 0.306 (16) & 68.1 (19) & -0.0042 (44) & -0.87 (20) & -0.98 (89) \\
o$w_0w_a$CDM & Planck + BAO + SN & 0.1423 (25) & 0.313 (16) & 67.5 (17) & -0.0023 (43) & -0.87 (20) & -0.74 (83) \\
o$w_0w_a$CDM & WMAP + BAO + SN & 0.1371 (44) & 0.302 (16) & 67.4 (18) & +0.0018 (68) & -1.00 (18) & 0.22 (73) \\
o$w_0w_a$CDM & \textit{e}WMAP + BAO + SN & 0.1358 (35) & 0.301 (15) & 67.2 (17) & -0.0023 (55) & -0.99 (16) & 0.18 (60) \\
\hline
\end{tabular}
\caption{Cosmological constraints by different datasets in the cosmological models $\Lambda$CDM, oCDM, $w$CDM, o$w$CDM, $w_0w_a$CDM, and o$w_0w_a$CDM. We compare the cosmological constraints from combining Planck with acoustic scale from BOSS galaxies as well as lower and higher redshift BAO measurements from the 6-degree field galaxy redshift survey (6DF) and the BOSS-Lyman alpha forest (Ly$\alpha$F), respectively. We also compare how these combinations benefit from the constraining power of type-Ia Supernovae from the Union 2 compilation by the Supernovae Cosmology Project (SN). The WMAP and \textit{e}WMAP cases have been added for comparison. As in Table~\ref{tab:bigcos}, 'CMASS-iso' means the isotropic measurement from the CMASS sample, whereas the anisotropic one is referred to simply as 'CMASS'. 'LOWZ' is the isotropic measurement from the LOWZ sample. 'BAO' stands for the combination CMASS + LOWZ + 6DF + Ly$\alpha$F.}
\label{tab:bigcos2}
\end{table*}

\label{lastpage}

\end{document}